\documentclass[journal]{IEEEtran}

\ifCLASSINFOpdf

\else

\fi
\usepackage{mathrsfs}
\usepackage{amsmath}
\usepackage{ntheorem}

\usepackage{makeidx}  
\usepackage{algorithm}
\usepackage{algpseudocode}
\usepackage{graphicx}
\usepackage{epstopdf}
\usepackage{bm}
\usepackage{cite}
\usepackage{stfloats}
\usepackage{setspace}
 
\usepackage{subcaption}
\usepackage{float}

\usepackage{amssymb}
\usepackage{graphicx}

\usepackage{url}

\usepackage{tabularx,booktabs}
\newcolumntype{C}{>{\centering\arraybackslash}X} 
\usepackage{lipsum}

\usepackage{makecell} 

\usepackage{graphicx}
\usepackage{color}

\usepackage{multirow}

\definecolor{dblue}{RGB}{15,89,164}

\usepackage{mathbbol}

\newcommand{\av}{{\bf a}}

\newcommand{\nv}{{\bf n}}

\newcommand{\pv}{{\bf p}}

\newcommand{\uv}{{\bf u}}

\newcommand{\yv}{{\bf y}}

\newcommand{\Am}{{\bf A}}

\newcommand{\Id}{{\bf I}}

\newcommand{\Rm}{{\bf R}}
\newcommand{\Sm}{{\bf S}}

\newcommand{\Um}{{\bf U}}

\newcommand{\gammav}{\hbox{\boldmath$\gamma$}}

\newcommand{\muv}{\hbox{\boldmath$\mu$}}

\newcommand{\rhov}{\hbox{\boldmath$\rho$}}

\newcommand{\Gammam}{\hbox{\boldmath$\Gamma$}}

\newcommand{\Sigmam}{\hbox{\boldmath$\Sigma$}}
\newcommand{\Phim}{\hbox{\boldmath$\Phi$}}

\newcommand{\Psim}{\hbox{\boldmath$\Psi$}}

\def\bC{{\mathbb C}}

\def\bE{{\mathbb E}}

\def\bR{{\mathbb R}}

\def\diag{{\text{diag}}}
\newcommand{\transp}{{\sf T}}
\newcommand{\herm}{{\sf H}}
\renewcommand{\vec}{{\rm vec}}
\newcommand{\trace}{{\hbox{tr}}}

\begin{document}
 
\title{ Near-Field Integrated Imaging and Communication in Distributed MIMO Networks}

\author{ Kangda Zhi, Tianyu Yang, Shuangyang Li,  Yi Song, Amir Rezaei, and Giuseppe Caire, {\itshape Fellow, IEEE \upshape}		
	\thanks{ The authors are with Communications and Information Theory Group (CommIT), Technische Universit\"{a}t Berlin, 10587 Berlin, Germany (e-mail: \{k.zhi, tianyu.yang, shuangyang.li, yi.song, amir.rezaei, caire\}@tu-berlin.de).}
}

\IEEEspecialpapernotice{(Invited Paper)}

\maketitle

\begin{abstract}
 In this work, we propose a general framework for wireless imaging in distributed MIMO wideband communication systems, considering multi-view non-isotropic targets and near-field propagation effects. For indoor scenarios where the objective is to image small-scale objects with high resolution, we propose a range migration algorithm (RMA)-based scheme using three kinds of array architectures: the full array, boundary array, and distributed boundary array. With  non-isotropic near-field channels, we establish the Fourier transformation (FT)-based relationship between the imaging reflectivity and the distributed spatial-domain signals and discuss the corresponding theoretical properties. Next, for outdoor scenarios where the objective is to reconstruct the large-scale three-dimensional (3D) environment with coarse resolution, we propose a sparse Bayesian learning (SBL)-based algorithm to solve the multiple measurement vector (MMV) problem, which further addresses the non-isotropic reflectivity across different subcarriers. Numerical results demonstrate the effectiveness of the proposed algorithms in acquiring high-resolution small objects and accurately reconstructing large-scale environments.
\end{abstract}

\begin{IEEEkeywords}
Integrated imaging and communication, radio imaging,  network sensing, environmental sensing, near field, wavenumber domain, sparse Bayesian learning, wideband communications.
\end{IEEEkeywords}

%
\IEEEpeerreviewmaketitle

\section{Introduction}
Integrated sensing and communications (ISAC) is listed as one of the six usage scenarios in the sixth generation (6G) recommendation~\cite{wp5d2023m}. In the ISAC paradigm, the same hardware and time-frequency resource is shared between communication and sensing functionalities, enabling attractive usage scenarios such as autonomous vehicles, location-enhanced communications, and low-altitude economy\cite{FengLAE2025,TangLAE2025,lyu2025empowering,lyu2022joint}. In this context, ISAC has been  extensively studied from the perspective of theoretical analysis\cite{xiong2023fundamental,hua20243d,10769538,yu2023active,Meng2024SG}, network and cooperative sensing\cite{songperformance,yang2025cooperative,babu2024precoding,10726912}, wideband sensing\cite{wang2025wideband,10791452,tao2025survey}, wireless localization\cite{wu2024exploit,pan2022overview,wu2024employing}, and precoding design\cite{zhou2024fluid,chen2025multi,hua2023secure,yang2025towards,10596930,liu2024joint,hua2024integrated}.  

Current work of ISAC mainly focused on the sensing and detection of targets' ranges, speeds, locations, and angles. However, as the rapid emergence of versatile application requirements in 6G networks, the imaging capability\cite{wp5d2023m}, i.e., the detection of objects’ shapes and ``radio signatures’’,  will be more crucial in many usage cases, such as security inspection, industrial screening, digital twin, augmented reality, and environmental reconstruction. Imaging can be realized by light or radio, wherein light-based imaging struggles in dark environments and bad weather and has raised growing concerns about privacy exposure, and above all it requires explicit deployment of dedicated hardware, incompatible with wireless communications. In contrast, radio-based imaging does not require good light and a good field of view to work effectively, which motivates the exploration of using communication signal to realize the radio-frequency (RF) imaging function. This technology is called integrated imaging and communication (IIAC) and been investigated in \cite{manzoni2024wavefield,li2021lightweight,yang2025illumination,huang2024fourier,huang2024ris,li2024networked,zheng2024random,jiang2024electromagnetic}. Similar techniques have been also referred to as  environmental sensing/reconstruction\cite{tong2022environment,lu2024deep}, computational imaging\cite{tong2025computational}, and holographic imaging\cite{torcolacci2024holographic}. 

Inspired by the attractive and broad prospects of IIAC, this paper will build the general framework for multi-view RF imaging in wideband distributed MIMO networks, by exploring the distinguished usage scenarios in indoor and outdoor environments, with the requirement of different-resolution image reconstruction.

\subsection{Prior Works}
\subsubsection{Studies on Radio Imaging}
The acquisition of high-resolution image of objects based on RF signal is a non-trivial task. In past decades, this important topic has been widely studied by researchers working in the scope of microwave imaging\cite{ahmed2012advanced,shao2020advances,wang2019review}, diffraction tomography\cite{wu1987diffraction,bolomey1990microwave,ren20183}, synthetic aperture radar (SAR)\cite{moreira2013tutorial,krieger2013mimo,fang2013fast}, and inverse synthetic aperture radar (ISAR)\cite{xu2011bayesian,vehmas2021inverse}.

The state-of-the-art algorithms used in radio imaging can be divided into four main categories. Perhaps the most well-known and widely used method is the back projection algorithm  (BPA)\cite{yanik2020development,soumekh1998wide,grebner2023probabilistic,desai1992convolution}. As a classic algorithm, BPA is a conjugate matched filter method that compensates for phase and amplitude factors for echo signal received by each transceiver pair  and accumulates them coherently over the whole (virtual) aperture and frequency band to obtain the imaging result\cite{ren2019fast,ge2024efficient}. This approach is flexible and can be used for arbitrary array configurations. However, its 
resolution and imaging quality rely on the large array dimension, which could lead to high hardware and computational complexity.

Another well-known approach is the range migration algorithm (RMA) \cite{sheen2001three,lopez20003,zhu2016frequency}. This method leverages the property of echo signal and reshapes spatial-domain signal  as the Fourier transform (FT) of wavenumber-domain signal, enabling computationally efficient approach for short-range microwave image reconstruction. A monostatic imaging setup with a co-located transmit/receive antenna scanning mechanically in a plane was considered in \cite{sheen2001three}.  However, the monostatic case needs a large number of antennas (if operating electronically) or needs long-time scanning (if operating mechanically). To reduce the hardware cost and delay, a multistatic array design where transmitters and receivers are not co-located was proposed in \cite{moulder2016development}, which can effectively improve the diversity of spatial sampling. The proposed boundary array, a topology that employs four linear arrays, can form a large virtual array (by phase center of each transceiver pair) with a small number of antennas on the rectangular boundaries. To implement fast FT (FFT), the signals received by transceivers  with different locations in \cite{moulder2016development} were approximated to co-located phase centers.
A 3D short-range imaging algorithm using a scanning 1D linear MIMO array was proposed in \cite{gao2018novel}. Work \cite{zhu2019sequential} further considered the arbitrary scanning path for 1D linear MIMO.
Without scanning, a 2D planar MIMO array-based 3D near-field imaging algorithm was proposed in \cite{zhuge2012three} which performed the image reconstruction procedure in the frequency-wavenumber domain.
To further improve efficiency, a transverse spectrum deconvolution technique was proposed in \cite{fromenteze2019transverse} to implement the 5D to 3D mapping of the MIMO-RMA. 
Considering a general array setup where antennas could be sparsely located in arbitrary positions, near-field mmWave 2D imaging and 3D imaging algorithms were designed in \cite{yanik2019near} and \cite{wang20203}, respectively. Besides, methods of chirp scaling algorithm (SCA) and  fractional FT in SAR imaging have been proposed in\cite{zhang2022fast} and \cite{hu2018beyond}, respectively.

As a third general approach, the imaging problem can be formulated as a regularized inverse problem, which can be solved by compressed sensing (CS) algorithms\cite{zhang2015generalized,wang2017fast,wang2022efficient,gurbuz2009compressive}. The main idea is to discretize the imaging area as  pixels where  the interval of the grids could be smaller than the imaging resolution. By approximating the received echo as the signal reflected from the pixel points/voxels, the imaging reconstruction problem can be expressed as a classical CS problem. This problem can be solved by iterative thresholding algorithm (ITA)\cite{fang2013fast}, conjugate gradient descent with different regularization\cite{li2018compressive,ma2014mimo}, and complex approximated message passing (CAMP)\cite{bi2017l_}. Before CS reconstruction, work \cite{yang2013sparse} further considered to effectively suppress the azimuth clutter and short-range clutter outside the reconstruction region. Besides, some works have integrate the forward sensing operator of RMA into CS to reduce the computation and storage burdens, by exploiting the feature of FFT operators\cite{li2025compressive,ge2024efficient,chen2023adaptive}.

As a final and more recent approach,  imaging aided by artificial intelligence has emerged as a promising direction. Recent studies have shown the potential of deep learning-based techniques in inverse problems\cite{ongie2020deep}. In \cite{xiong2020spb}, a real-time processing deep network, namely SPB-Net, was proposed to implement imaging and despeckling simultaneously. The authors in \cite{wang2022efficient} first proposed an alternating direction method of multipliers (ADMM) imaging framework which dubbed single-frequency holographic (SFH)-ADMM, by introducing the SFH method imaging theory into the ADMM optimization cycles. Then, to accelerate the convergence, the proposed  SFH-ADMM was mapped into a feed-forward deep network structure with learnable parameters and fixed layers, i.e., SFH-ADMM-Net. In network training, the dataset is randomly generated based on the signal model in the promise of knowing system parameters and imaging geometry.

\subsubsection{Study on Integrated Imaging and Communication} Although the above-mentioned microwave imaging and SAR algorithms well addressed the imaging problems, they cannot be directly applied to wireless communication networks. The reasons are three-fold. First, many algorithms are designed to work in a short range, i.e, in centimeter-level, which is not consistent for wireless communication scenarios. Second, most of the realizations are based on the virtual array and SAR formed by moving sensors, which is time-consuming and cannot represent the fixed array used for communications. Furthermore, their algorithms do not involve the design of beamforming and precoding, which is crucial in communications.

Currently, only a few works have investigated the integrated imaging and communication problem\cite{manzoni2024wavefield,li2021lightweight,yang2025illumination,huang2024fourier,huang2024ris,li2024networked,zheng2024random,jiang2024electromagnetic,tong2022environment,lu2024deep,tong2025computational,torcolacci2024holographic}. A wavenumber-domain back projection algorithm with data fusion from multiple base stations (BSs) was proposed in \cite{manzoni2024wavefield}. A low-complexity zero-forcing (ZF)-like solution was proposed in \cite{li2021lightweight}. Least-square (LS)-based approach was utilized in \cite{yang2025illumination,zheng2024random}. The authors in \cite{jiang2024electromagnetic} employed the diffusion model to generate the point cloud of targets while the authors in \cite{lu2024deep} used deep learning to reconstruct the environment by enhancing the density of sparse point clouds obtained from FT and multiple signal classification (MUSIC). Some reconfigurable intelligent surface (RIS)-aided imaging algorithms based on CS were proposed in \cite{huang2024fourier,huang2024ris,torcolacci2024holographic}, taking advantages of the passive large apertures of RISs. Multi-view imaging considering the obstruction effect has been studied in \cite{tong2022environment} with the generalized approximate message passing (GAMP) method. Multi-view imaging reconstruction aided by moving users has been considered in \cite{huang2024ris,shi20256g}. Besides, the transmission and detection of images with RIS have been studied in \cite{huang2024image,tahira2024irs}.  

\subsection{Motivation and Contribution}
We begin with the observation that RF imaging of 2D/3D non-isotropic objects using distributed MIMO communication networks has not yet been investigated. For non-isotropic objects, distributed MIMO can provide multi-view observations, which improve diversity,  overcome blockages, and mitigate aliasing. Meanwhile, even though each array is not large, distributed arrays can form an equivalent near-field propagation environment, bring high imaging resolutions.
 Besides, none of the previous works has considered establishing a framework for both indoor and outdoor scenarios with different resolution requirements. For indoor scenarios, the imaging targets could be small and high resolution may be desired. For outdoor scenarios, the goal could be the reconstruction of large-scale environment and thus the imaging target may be large and a coarse resolution may be sufficient. Furthermore, the frequency dependency of the reflectivity across different carrier frequencies has been scarcely considered.  
 
 Therefore, in this work, we propose a general framework for RF imaging in distributed MIMO systems, considering the multi-view non-isotropic targets, in scenarios of both indoor target imaging and outdoor environment reconstruction. The main contributions of this paper are listed as follows.

\begin{itemize}
	\item We propose a wideband wireless imaging system empowered by distributed MIMO communication networks. We establish a general model with considering the near-field propagation effect and the multi-view non-isotropic signal reflection. We also model the received signal in the special case of SAR.

	\item For indoor scenario targeting high-resolution imaging of small objects, we proposed RMA-based algorithms under three kinds of arrays, i.e., the conventional full array, the low-cost boundary array, and the flexible distributed-boundary array. We establish the FT relationship between the distributed spatial-domain signal, the wavenumber-domain signal, and the imaging reflectivity, under the near-field non-isotropic channel models. We also discuss the theoretical properties of the algorithms, including resolution, aliasing, and key performance indicators.
	
	\item For outdoor scenario aiming at reconstructing large-scale 3D environments, we propose an SBL-based compressed sensing algorithm by discretizing the area into grids and exploiting the sparsity of the targets. We further tackle the dependence of the reflectivity of targets on the carrier frequencies in the wideband systems by solving the multiple measurement vector (MMV)  problem, which can obtain the correlation of images at different frequencies.
	
	\item Extensive numerical results are presented to validate the effectiveness of the two proposed algorithms. It showcases the quality of high-resolution imaging  of small objects in indoor environments and the accurate reconstruction of large-scale objects in outdoor environments. The superiority of the proposed schemes over several benchmark algorithms is also verified.
	
\end{itemize}

\subsection{Organization and Notations}

The rest of the paper is organized as follows. In Section \ref{section_model}, we establish the receiving signal model for the considered distributed MIMO systems.  Section \ref{section_rma} presents the RMA-based method for indoor imaging. Section \ref{section_sbl} presents the SBL-based method for outdoor environment reconstruction. Section \ref{section_simulation} validates the effectiveness of the algorithms based on numerical simulations. Section \ref{section_summary} summarizes the work.

\emph{Notations}: Vectors and matrices are denoted by boldface lower case and upper case letters, respectively. The transpose, conjugate transpose, and inverse of matrix $\bf X$ are denoted by ${\bf X}^\transp $, ${\bf X}^\herm$, and ${\bf X}^{-1}$, respectively. $\left[{\bf X}\right]_{(a,b)}$ denotes the $(a,b)$-th element of matrix $\bf X$. $[\mathbf{x}]_a$ denotes the $a$-th entry of the vector $\mathbf{x}$. $|\mathcal{X}|$ denotes the cardinality of set $\mathcal{X}$. Besides, $\mathbf{I}_a$ denotes the identity matrix with a dimension of $a\times a$.   The space of $a\times b$ complex matrices is denoted by $\mathbb{C}^{a\times b}$. The $l_2$ norm of a vector $\mathbf{x}$ and the absolute value of a scalar $x$ are denoted by $\left\|\mathbf{x}\right\|$ and $\left|x\right|$, respectively. $j$ denotes the square root of $-1$.  $a\overset{!}{=}b$ means that $a$ shall be equal to $b$.

\section{System Model}\label{section_model}

\begin{figure}
	\centering
	\includegraphics[width= 0.5\textwidth]{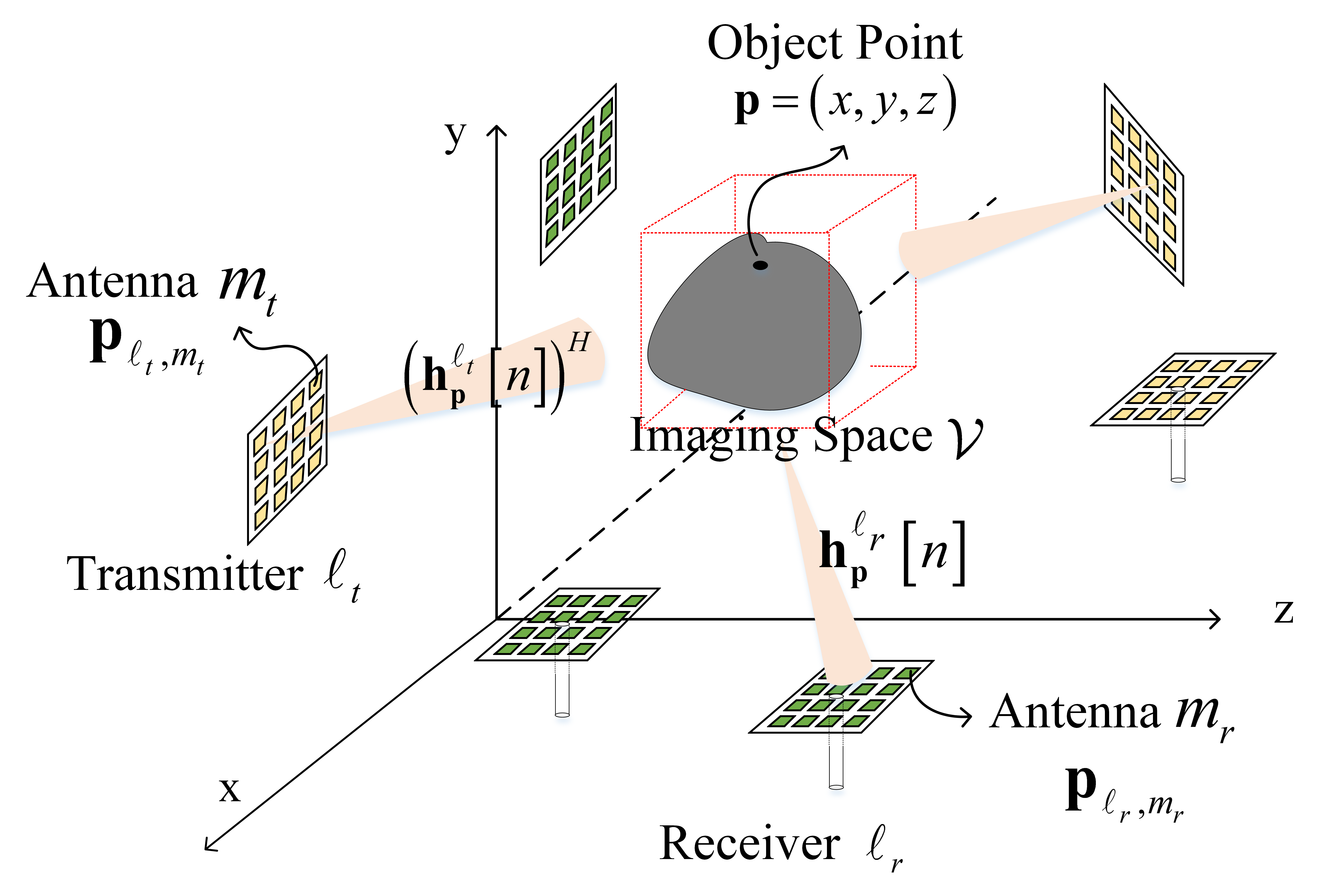}
	\caption{The general scenario of multi-view 3D imaging in distributed MIMO networks.}
	\label{figure1}
\end{figure}

As illustrated in Fig. \ref{figure1}, we investigate the RF imaging of an object in a distributed MIMO network comprising $L$ radio units (RUs). 
Following a common assumption in cell-free and distributed MIMO networks,  all RUs are assumed to be perfectly synchronized in timing and frequency, which can be realized through the fronthaul connecting all the RUs.

Each RU is equipped with a uniform planar array (UPA),  functioning as either a transmitter or a receiver. 
We define $\mathcal{L} = [1,\ldots,L]$ as the index set of $L$ UPAs (i.e., RUs) and define by $\mathcal{L}_{s,n}^t \subset \mathcal{L}$ and $\mathcal{L}_{s,n}^r  \subset \mathcal{L}$ the subset of UPAs that transmit and receive signals, respectively, at time slot $s$ and subcarrier $n$. The number of antennas on each transmitting array and receiving array is denoted by $M_t$ and $M_r$, respectively.

\subsection{Transmit Signal}
We consider the wideband communication systems where RF imaging is conducted based on orthogonal frequency division multiplexing (OFDM) signals\footnote{The OFDM communication signals, which could be sensing-oriented pilots or data signals, are assumed to be known since they are transmitted and received by the same infrastructure.}. The number of subcarriers is assumed to be $N$.  At time slot $s\in[1:S]$ and subcarrer $n \in [1:N]$, the signal transmitted by the $\ell_t$-th array, $  \ell_t  \in  \mathcal{L}_{s,n}^t  $, is denoted by $\mathbf{x}_{\ell_t}[s,n]$. Considering fully digital arrays, the baseband complex signal can be constructed by
\begin{align}
\mathbf{x}_{\ell_t}[s,n] = \mathbf{F}_{\ell_t,s,n} \mathbf{s}_{\ell_t}[s,n],
\end{align}
where $ \mathbf{s}_{\ell_t}[s,n]   \in \mathbb{C}^{U_t \times 1}$, $U_t \le M_t$, denotes the symbol sent by array $\ell_t$ at time slot $s$ in subcarrier $n$ and $\mathbf{F}_{\ell_t,s,n}  \in \mathbb{C}^{M_t \times U_t}$  represents the precoding matrix of array $\ell_t$. 

Given the transmit power budget $P$ of each transmitter, the power constraint can be expressed as
\begin{align}\label{power_constraint}
 \left\| \mathbf{x}_{\ell_t}[s,n]   \right\|^2     \le \frac{P}{N}, \forall \ell_t, s, n,
\end{align}
where the power is equally allocated to each subcarrier.

\subsection{General Channel Model}
In this paper, we consider the non-isotropic reflectivity of the object. To characterize this property, we model the impact of incident directions on the effective channel strength\cite{hu2018beyond} and also consider the variation of visible region from different views\cite{huang2024ris,tong2022environment}. We assume the channel to be purely line-of-sight (LoS) due to the high carrier frequency in mmWave band, since the path gain of LoS path could be $25$ dB larger than the non-LoS (NLoS) path. Then, we model the view-aware effective LoS channel by integrating the incident-angle-aware {\itshape near-field}\footnote{The near-field effect is exhibited via two aspects in our networks. First, it could be caused due to the large aperture of a single compact/sparse UPA. Second, the overall channel formed by all distributed UPAs constructs an {\itshape equivalent} near-field channel.} channel  \cite{hu2018beyond,zhi2024performance,bjornson2020power} with  binary blockage \cite{tong2022environment}. 

Specifically, assume that the arbitrarily shaped object occupies a 3D space $\mathcal{V}$ which is comprised of points (also referred to as a pixel or voxel) located at $\mathbf{p}=(x,y,z) \in \mathcal{V}$. We first introduce the binary vector $\mathbf{v}_{\mathbf{p}}^{\ell_c} \in \mathbb{Z}^{M_c\times 1}$, $c \in \{ t, r\}$, whose elements are binary $\{0,1\}$, characterizing whether pixel $\mathbf{p}$ is visible to the antennas of array $\ell_c$ or not. The values of $\mathbf{v}_{\mathbf{p}}^{\ell_c}$ can be decided according to the situation of whether the LoS path between the pixel and the antennas is blocked by obstacles or other pixels or not. 

Secondly, in order to model the non-isotropic features of the pixel and the UPA antennas, we introduce the incident angle-dependent effective channel strength. On the one hand, let $\mathbf{n}_{\ell_c}$ denote the unit normal vector of UPA $\ell_c$, $c\in\{t,r\}$. On the other hand, assuming that the object is composed of some regular surface, as buildings, vehicles, and furniture exhibit in practice, we assign a unit normal vector $\mathbf{n}_{\mathbf{p}}$ to each pixel $\mathbf{p}$ on its surface\footnote{This modeling holds for all pixels, since pixels inside the object are assumed to be invisible and thus their normal vector can be arbitrary. Besides, if the discretized pixel/voxel is large, it may have multiple normal vectors for different views.}. 

Then, the frequency-domain LoS channel from the $\ell_t$-th transmitter to the point $\mathbf{p}$  at the $n$-th subcarrier is denoted as $(\mathbf{h}_{\mathbf{p}}^{\ell_t} [n] )^\herm  \in \mathbb{C}^{1 \times M_t}$, whose $m_t$-th element, $1\le m_t \le M_t$ is given by
\begin{align}\label{ht}
\begin{aligned}
& \left[  \left(  \mathbf{h}_{\mathbf{p}}^{\ell_t} [n]   \right)^\herm  \right]_{m_t }\\
&=   \underbrace{    \left[ \mathbf{v}_{\mathbf{p}}^{\ell_t}\right]_{m_t}      }_{\text{visibility indicator}} \underbrace{ \frac{1}{  \sqrt{4\pi}  \left\|  \mathbf{p} - \mathbf{p}_{\ell_t,m_t}\right\|        } }_{\text{free-space pathloss}}
\underbrace{       \sqrt{  \frac{      \left( \mathbf{p} - \mathbf{p}_{\ell_t,m_t}  \right)^\transp   \mathbf{n}_{\ell_t}            }{    \left\|  \mathbf{p} - \mathbf{p}_{\ell_t,m_t}\right\|   }      }}_{\text{non-isotropic UPA}, \sqrt{\cos\theta_{\mathbf{p}}^{\ell_t,m_t}} } \\
&\times 
\underbrace{    \sqrt{    \frac{      \left( \mathbf{p}_{\ell_t,m_t}  - \mathbf{p}   \right)^\transp   \mathbf{n}_{\mathbf{p}}            }{    \left\|  \mathbf{p} - \mathbf{p}_{\ell_t,m_t}\right\|    }     } }_{\text{non-isotropic pixel}, \sqrt{\cos\phi_{\mathbf{p}}^{\ell_t,m_t}  }   }
\underbrace{e^{ -j k_n    \left\|  \mathbf{p} - \mathbf{p}_{\ell_t,m_t}\right\|    }}_{\text{frequency-dependent phase}} ,
\end{aligned}
\end{align}
where $\mathbf{p}_{\ell_t,m_t}$ denotes the 3D location of the $m_t$-th antenna on the UPA $\ell_t$, $k_n = \frac{2 \pi f_n}{c} $ denotes the wavenumber at the subcarrier $n$,  $c$ is the speed of light, $f_n = f_c + \frac{B}{N}\left(   n-1 - \frac{N-1}{2}  \right)$, $n\in \left[1:N\right]$, $B$ is the bandwidth, and $f_c$ is the central carrier frequency. The modeling of free-space pathloss, projection-angle-based non-isotropic factors, and transmission phases is based on the electromagnetic  near-field channel model \cite{zhi2024performance,bjornson2020power}. From (\ref{ht}), it can be concluded that the proposed channel model is more general due to the following reasons:
\begin{itemize}
	\item It will degrade to conventional isotropic (i.e., incident-angle-independent) channel model \cite{wang20203} if the transmitting electromagnetic wave from antenna $m_t$ is perpendicular to the UPA's surface and incident electromagnetic wave at point $\mathbf{p}$  is perpendicular to its surface, i.e., $\cos\theta_{\mathbf{p}}^{\ell_t,m_t} = \cos\phi_{\mathbf{p}}^{\ell_t,m_t} = 1$. 
	
	\item  It is a general near-field sphere-wave channel that distinguishes the variation of pathloss and phase of the channel across the whole arrays, which degrades to far-field planar-wave channel \cite{yang2025cooperative} when the array aperture is small and when the imaging distance is large.
	
	\item It characterizes the impact of visible region from different views based on $\mathbf{v}_{\mathbf{p}}^{\ell_t}$. By letting $  \left[ \mathbf{v}_{\mathbf{p}}^{\ell_t}\right]_{m_t} =1$, $\forall m_t$,  pixel $\mathbf{p}$ will be seen from all antennas on array $\ell_t$.
	
	\item It models the beam squint effect caused by $k_n$ of steering vectors in the wideband systems \cite{wang2019beam}. This effect can be neglected if the difference of time delays is negligible between different antennas on the UPA, degrading to frequency-independent steering vectors\cite{cui2022channel,yang2025cooperative}.
	
\end{itemize}

Similarly, the channel from the point $\mathbf{p}$ to the $\ell_r$-th receiver at the $n$-th subcarrier is denoted as  $\mathbf{h}_{ \mathbf{p}}^{\ell_r} [n] \in \mathbb{C}^{M_r \times 1}$ whose $m_r$-th element, $1\le m_r \le M_r$ is given by
\begin{align}
\begin{aligned}
&	\left[   \mathbf{h}_{\mathbf{p}}^{\ell_r} [n]   \right]_{m_r }\\
	&=  \left[ \mathbf{v}_{\mathbf{p}}^{\ell_r}\right]_{m_r}      \frac{1}{ \sqrt{4\pi}   \left\|  \mathbf{p}_{\ell_r,m_r}  -  \mathbf{p} \right\|        }  
\underbrace{    \sqrt{             \frac{      \left(  \mathbf{p}_{\ell_r,m_r} - \mathbf{p}  \right)^\transp        \mathbf{n}_{\mathbf{p}}        }{    \left\|  \mathbf{p}_{\ell_r,m_r}    -   \mathbf{p} \right\|     }    } }_{       \sqrt{ \cos\phi_{\mathbf{p}}^{\ell_r,m_r}  }  }  \\
	&\times 
\underbrace{       \sqrt{           \frac{       \left(     \mathbf{p} - \mathbf{p}_{\ell_r,m_r}   \right)^\transp    \mathbf{n}_{\ell_r}           }{    \left\|    \mathbf{p}_{\ell_r,m_r}  -  \mathbf{p}  \right\|    }     } }_{      \sqrt{  \cos\theta_{\mathbf{p}}^{\ell_r,m_r}  }  } 
\;\;	{e^{ -j k_n    \left\|    \mathbf{p}_{\ell_r,m_r}  -   \mathbf{p}    \right\|    }} .
\end{aligned}
\end{align}

\subsection{Receive Signal}
For each receiving array $\ell_r$, it receives echo signals  reflected from the 3D object. Each point $\mathbf{p}$ of the 3D target is characterized by a complex reflective function $\rho_n(x,y,z) \in \mathbb{C}$ that models the interaction of reflected electromagnetic field with incident electromagnetic field under carrier frequency $f_n$\cite{sheen2001three,manzoni2024wavefield}. We next formulate the receiving signal based on whether different arrays transmit signals cooperatively or not.

Based on the Born approximation  (i.e., neglecting the double bounce reflections)\cite{manzoni2024wavefield}, if the transmission of different arrays are orthogonal by time, frequency, or spatial resources, we   obtain the signal received by array $\ell_r$ that was transmitted from array $\ell_t$ as
\begin{align}\label{y_lt_lr}
	\begin{aligned}
		&\mathbf{y}_{\ell_r,\ell_t}[s,n]= \mathbf{w}_{\ell_r}[s,n]  \\
		&+ \iiint_{\mathcal{V}}  {\rho_n(x, y, z)}  \mathbf{h}_{ \mathbf{p}}^{\ell_r} [n]    \left(\mathbf{h}_{\mathbf{p}}^{\ell_t} [n]\right)^\herm \mathbf{x}_{\ell_t}[s,n]      \mathrm{~d} x  \mathrm{~d} y  \mathrm{~d} z ,
	\end{aligned}
\end{align}
which is the superposition of signals reflected from all infinitesimal points $\mathbf{p}=(x,y,z)$ composing the space $\mathcal{V}$. $\mathbf{w}_{\ell_r}[s,n] \sim \mathcal{C N}\left(\mathbf{0}, \sigma^2 \mathbf{I}_{ M_r}\right)$ denotes the  additive white Gaussian noise with power $\sigma^2$.

For cooperative cases,
    $  \ell_t  \in   \mathcal{L}_{s,n}^t    $ arrays   transmit signal together and the echo signal received by array $\ell_r  \in  \mathcal{L}_{s,n}^r   $ at time slot $s$ and subcarrier $n$ can be formulated as follows
\begin{align}\label{receiving_signal_cooperative}
	\begin{aligned}
		&\mathbf{y}_{\ell_r}[s,n]=\mathbf{w}_{\ell_r}[s,n]  \\
		& + \! \! \sum_{\ell_t    \in   \mathcal{L}_{s,n}^t       } \!\!    \iiint_{\mathcal{V}}  {\rho_n(x, y, z)}  \mathbf{h}_{ \mathbf{p}}^{\ell_r} [n]    \left(\mathbf{h}_{\mathbf{p}}^{\ell_t} [n]\right)^\herm \! \mathbf{x}_{\ell_t}[s,n]    \!   \mathrm{~d} x  \mathrm{~d} y  \mathrm{~d} z . 
	\end{aligned}
\end{align}

\begin{figure}
	\centering
	\includegraphics[width= 0.45\textwidth]{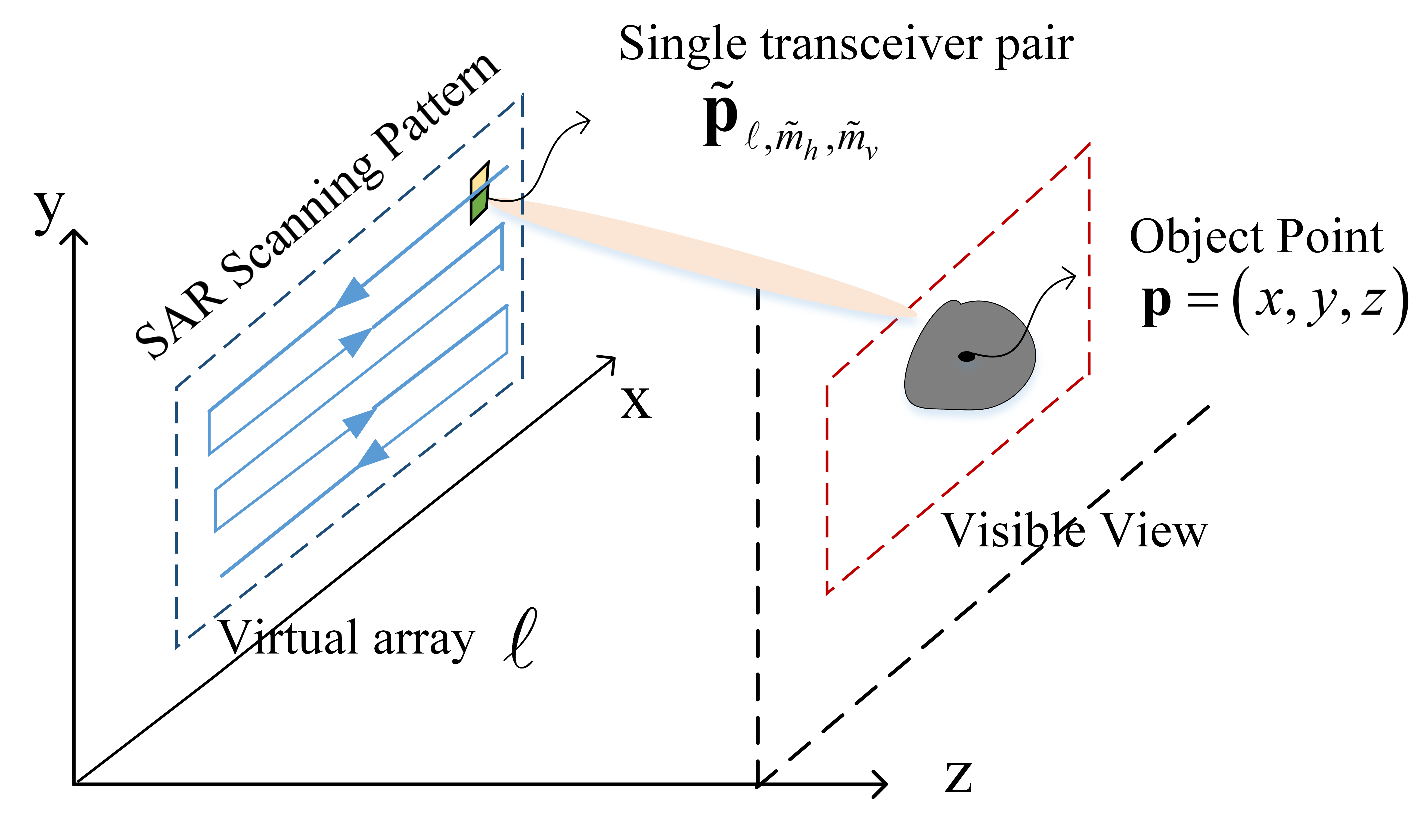}
	\caption{Illustration of SAR imaging.}
	\label{figure3}
\end{figure}
Furthermore, for more insights, we also discuss the received signal obtained in a special case where a single virtual array $\ell$  is formed based on moving antennas, as shown in Fig. \ref{figure3}. This situation possesses a similar principle as scanning array (i.e., SAR) \cite{sheen2001three}, moving sensors\cite{li2021lightweight}, and fluid antenna systems\cite{wu2024fluid,dong2024movable,yao2025framework,zhu2023movable}.  With slight modifications, it is also equivalent to the single-input multiple-output (SIMO) case where the receiving signal at a UPA is reflected from a single transmitting antenna\cite{zhu2016frequency}. To be specific, consider a co-located transceiver pair $\tilde{\mathbf{p}}_{\ell, \tilde{m}_h, \tilde{m}_v}$ monostatically sensing the object by unit pilot and moving to form a virtual $\tilde{M}_h \times \tilde{M}_v$ array $\ell$ after $\tilde{M} = \tilde{M}_h \tilde{M}_v$ time slots. Then, the receiving signal  $\tilde{ \mathbf{Y}}_{\ell}[n]  \in \mathbb{C}^{ \tilde{M}_h \times \tilde{M}_v  }$ of the virtual array can be given by
\begin{align}\label{mono}
	\begin{aligned}
		& \left[     \tilde{ \mathbf{Y}}_{\ell}[n]  \right]_{ \left(  \tilde{m}_h, \tilde{m}_v \right)    }  =  w_{\ell, \tilde{m}_h, \tilde{m}_v}[n] \\
		&+ \iiint_{\mathcal{V}}  {\rho_n(x, y, z)}     
		  v_{\mathbf{p}}^{\ell, \tilde{m}_h, \tilde{m}_v} 
		    \cos\theta_{\mathbf{p}}^{\ell, \tilde{m}_h, \tilde{m}_v}   \cos\phi_{\mathbf{p}}^{\ell, \tilde{m}_h, \tilde{m}_v}    \\
		 & \times \frac{1}{  4\pi \left\|   \mathbf{p} -   \tilde{\mathbf{p}}_{\ell, \tilde{m}_h, \tilde{m}_v} \right\|^2}
		 e^{-j 2 k_n    \left\|   \mathbf{p} -   \tilde{\mathbf{p}}_{\ell, \tilde{m}_h, \tilde{m}_v} \right\|     }
		      \mathrm{~d} x  \mathrm{~d} y  \mathrm{~d} z ,
	\end{aligned}
\end{align}
where $ v_{\mathbf{p}}^{\ell, \tilde{m}_h, \tilde{m}_v} \in \{0,1\} $,   $  \cos\theta_{\mathbf{p}}^{\ell, \tilde{m}_h, \tilde{m}_v}  $, and $ \cos\phi_{\mathbf{p}}^{\ell, \tilde{m}_h, \tilde{m}_v}  $ characterize the impact of visibility and incident angles, respectively. $ w_{\ell, \tilde{m}_h, \tilde{m}_v}[n] $ denote the thermal noise.

\subsection{Objective}
In the sequel of this paper, we will propose two algorithms, based on RMA and SBL, to solve the imaging problem, i.e., estimating $\rho_n(\mathbf{p})$, $\forall \mathbf{p} \in \mathcal{V}$ based on the received communication signals. The two algorithms proposed have their own advantages and disadvantages, respectively.

\section{RMA-Based Design}\label{section_rma}
In this section, we consider applying the RMA to tackle the imaging problem using OFDM signals. This method possesses the advantages of low computational complexity and high imaging resolution by exploiting the features of the structures of receiving signals.

\subsection{MIMO Case}
We now propose the algorithm based on the distributed MIMO arrays. A simple case with SAR  will be discussed later. 
Here, we consider multiple RUs transmitting signals over orthogonal resources as (\ref{y_lt_lr}). Then, we stack the signal that received at RU $\ell_r$ from RU $\ell_t$   after $S$ time slots as
\begin{align}\label{recieved_Y_lr_lt}
	\begin{aligned}
		&\mathbf{Y}_{\ell_r,\ell_t}[n]= \mathbf{W}_{\ell_r}[n]  \\
		&+ \iiint_{\mathcal{V}}  {\rho_n(x, y, z)}  \mathbf{h}_{ \mathbf{p}}^{\ell_r} [n]    \left(\mathbf{h}_{\mathbf{p}}^{\ell_t} [n]\right)^\herm    \mathrm{~d} x  \mathrm{~d} y  \mathrm{~d} z  \; \mathbf{X}_{\ell_t}[n]    ,
	\end{aligned}
\end{align}
where $ \mathbf{Y}_{\ell_r,\ell_t}[n] = [     \mathbf{y}_{\ell_r,\ell_t}[1,n],      \mathbf{y}_{\ell_r,\ell_t}[2,n], \ldots,  \mathbf{y}_{\ell_r,\ell_t}[S,n]      ]$ and 
$ \mathbf{X}_{\ell_t}[n] = [     \mathbf{x}_{\ell_t}[1,n] ,      \mathbf{x}_{\ell_t}[2,n] , \ldots,  \mathbf{x}_{\ell_t}[S,n]     ] $.

To maximize the observation dimensions, we consider $S\ge M_t$ so that space-time matrix $\mathbf{X}_{\ell_t}[n]  \in \mathbb{C}^{M_t \times S}$ could be row-orthogonal. This means that the sequences transmitted from each antenna of RU $\ell_t$ over $ S $ time slots are mutually orthogonal.
Specifically, based on power constraint (\ref{power_constraint}), a DFT-based space-time precoder $\mathbf{X}_{\ell_t}[n]  $ could be designed as  
 \begin{align}
\begin{aligned}
&\left[  \mathbf{X}_{\ell_t}[n]     \right]_{\left(   m_t,s  \right)}  \\
&=\sqrt{ \frac{{P}}{{NM_t}}}e^{-j \frac{2 \pi({m_t}-1)(s-1)}{S}}, {m_t}\in[1:M_t], s\in[1:S] ,
\end{aligned}
 \end{align}
where $\mathbf{X}_{\ell_t}[n]    \mathbf{X}_{\ell_t}^\herm [n]   ={ \frac{{PS}}{{NM_t}}}  \mathbf{I}_{M_t} $. 
Multiplying (\ref{recieved_Y_lr_lt}) on the right by $\mathbf{X}_{\ell_t}^\herm [n] $, we obtain the $M_r$-by-$M_t$ observation matrix as follows
\begin{align}
	\begin{aligned}
		& \mathbf{D}_{\ell_r,\ell_t}[n] \\
		 &  =   \mathbf{Y}_{\ell_r,\ell_t}[n]   \mathbf{X}_{\ell_t}^\herm [n]   \\
		 & = \mathbf{W}_{\ell_r}[n]  \mathbf{X}_{\ell_t}^\herm [n] \\
		&+  { \frac{{PS}}{{NM_t}}}   \iiint_{\mathcal{V}}  {\rho_n(x, y, z)}  \mathbf{h}_{ \mathbf{p}}^{\ell_r} [n]    \left(\mathbf{h}_{\mathbf{p}}^{\ell_t} [n]\right)^\herm    \mathrm{~d} x  \mathrm{~d} y  \mathrm{~d} z   ,
	\end{aligned}
\end{align}
which measures the echo signal between every antenna pair between array $\ell_t$ and $\ell_r$. Based on matrix $  \mathbf{D}_{\ell_r,\ell_t}[n]   $, we can perform the 3D imaging using wideband signals or conducting 2D imaging with single sub-carrier signal, by treating the 3D imaging as the collection of 2D imaging planes\footnote{For brevity, we focus on 2D imaging algorithm in this section. The 3D imaging algorithm can be extended readily by conducting (inverse) FT of 3D space-domain and wavenumber-domain wideband signals. Meanwhile, the construction of 3D image based on multi-view 2D images can be done based on conventional methods using machine learning.}.

 \begin{figure}
 	\centering
 	\includegraphics[width= 0.48\textwidth]{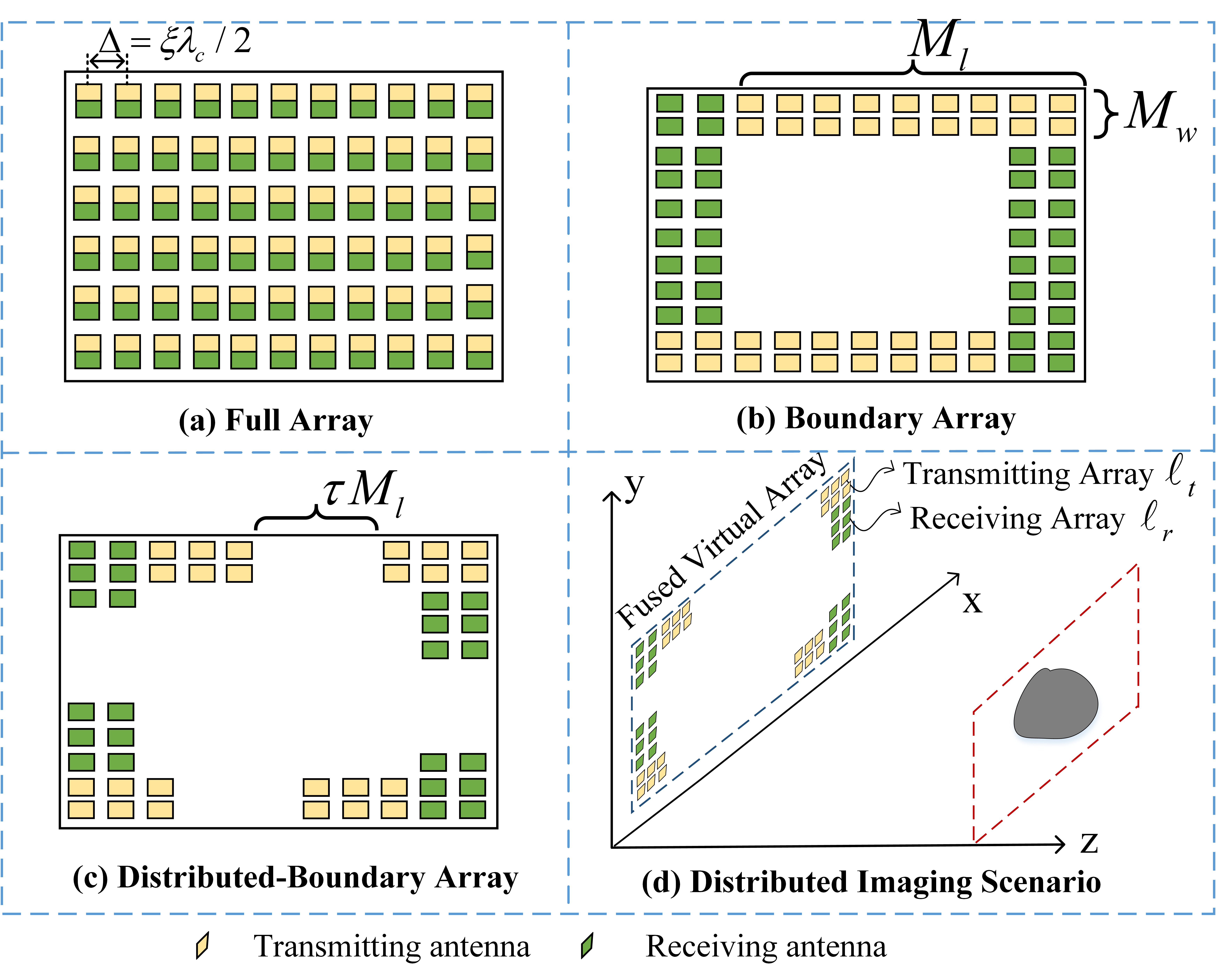}
 	\caption{Different array architectures (a) - (c) and an example scenario of indoor imaging (d).}
 	\label{figure4}
 \end{figure}
We propose to image the objects using three array architectures {\itshape with the same aperture}, as shown in Fig. \ref{figure4}. The first one, which is traditional in radio imaging, is a full array composed of co-located uniform-distributed transmitting/receiving antennas. To improve the imaging quality, the array aperture should be large so that near-field effect appears and brings rich parameter information. In this case, the number of antennas required by full array could be large, which may  not be favorable. The second one, the boundary array, is constructed by removing most of the central antennas of the full array and leaving four $M_l$-by-$M_w$ arrays on the boundary. By maintaining the array aperture, the boundary array could have the same imaging resolution and near-field region as the full array, but with a much smaller number of antennas, effectively reducing the cost and power consumption. Finally, we propose a distributed-boundary array to further reduce the number of antennas and make the array distributed, by removing $\tau M_l$, $\tau\in (0,1)$, antennas on the boundary UPAs. This distributed architecture, comprising eight $(\frac{1-\tau}{2}M_l)$-by-$M_w$ distributed arrays,  could be more flexible in the deployment.

When using distributed arrays, we need to fuse their  signals to form a centralized data matrix corresponding to a virtual full array.   With a slight abuse of notations, we consider a transmitting antenna $m_t \in [1,M_t]$ and a receiving antenna $m_r \in[1,M_r]$ on the full array and denote their locations as $(x_m^t, y_m^t,0)$ and $(x_m^r,y_m^r, 0)$, respectively. Then, denote $s\left(x_m^t, y_m^t, x_m^r, y_m^r, n\right)$ as the signal transmitted from full array's antenna $m_t$ and received by its antenna $m_r$ at subcarrier $n$.
Our objective is to form the data matrix $s\left(x_m^t, y_m^t, x_m^r, y_m^r, n\right)$, $\forall x_m^t, y_m^t, x_m^r, y_m^r$, based on the limited signals, $ \mathbf{D}_{\ell_r,\ell_t}[n]   $, $\forall \ell_t, \ell_r$, that are received by boundary arrays or distributed-boundary arrays.
The specific fusion rule is given as follows
\begin{align}\label{fusion}
\left\{\begin{array}{l}
	s\left(x_m^t, y_m^t, x_m^r, y_m^r, n\right)=  \left\{  \mathbf{D}_{\ell_r, \ell_t}[n]   \right\}_{m_t \rightarrow m_r}, \\
	\qquad \text { if } \left\{x_m^t, y_m^t, x_m^r, y_m^r\right\} \in\left\{\mathbf{p}_{\ell_r}, \mathbf{p}_{\ell_t}, \forall \ell_r, \forall \ell_t\right\}, \\
	s\left(x_m^t, y_m^t, x_m^r, y_m^r, n\right)=0, \text { otherwise },
\end{array}\right.
\end{align}
where $  \left\{  \mathbf{D}_{\ell_r,\ell_t}[n]   \right\}_{m_t \rightarrow m_r} $ represents the signal sent from antenna $m_t$ to antenna $m_r$ that can be obtained by the considered boundary or distributed-boundary  arrays.  If the transmitting-receiving pair $ \left\{x_m^t, y_m^t, x_m^r, y_m^r\right\} $ exists in the considered boundary or distributed-boundary arrays, we extract the corresponding received signal in $  \mathbf{D}_{\ell_r, \ell_t}[n]  $ and  fuse it into the data matrix; otherwise, it is padded by zeros.

Next, we consider producing the object’s image by processing the signal $s\left(x_m^t, y_m^t, x_m^r, y_m^r, n\right)$ in the frequency-wavenumber domain. 
 For 2D imaging, we focus on the imaging of a slice $\mathcal{V}(z)$ of 3D object $\mathcal{V}$ at distance $z$, i.e., $\forall [\mathbf{p}]_3 = z$, parallel to the observation arrays. 
By taking the 4-D Fourier transform for frequency-spatial-domain signal $s(	x_m^t, y_m^t,  x_{m}^r, y_{m}^r, n )$ and neglecting the impact of noise, we obtain signals in the  frequency-wavenumber domain as follows\footnote{Coefficient $ \frac{{PS}}{{NM_t}}$ is omitted for brevity of expression.}
\begin{align}\label{4D_FT}
	\begin{aligned}
		&S(		k_{x_m^t}  , k_{y_m^t},  k_{x_{m}^r} , k_{y_{m}^r}, n ) = {\rm FT}_{4D}\left[           s(	x_m^t, y_m^t,  x_{m}^r, y_{m}^r, n )      \right]   \\
		&=\iint_{\left(    x_m^t ,  y_m^t  \right)}
		\iint_{\left(  x_{m}^r ,  y_{m}^r     \right)}
		s(	x_m^t, y_m^t,  x_{m}^r, y_{m}^r, n) \times \\
		& e^{-j \left(   
			k_{x_m^t}   x_m^t 
			+ k_{y_m^t}   y_m^t
			\right)}
		e^{-j \left(   
			k_{x_{m}^r}   x_{m}^r 
			+ k_{y_{m}^r}   y_{m}^r 
			\right)}
		\mathrm{d} x_m^t  \mathrm{d} y_m^t
		\mathrm{d} x_{m}^r  \mathrm{d} y_{m}^r \\
		& \overset{(a)}{\approx} \iint_{\mathcal{V}(z)}   \frac{ \rho_n(x, y, z)}{4\pi }  
		   \sqrt{ \cos  {\theta}_{\mathbf{p},m_t,m_r }^c   \cos {\phi}_{\mathbf{p},m_t,m_r }^c       }       \\
		&\times \left[ 
		\underbrace{\iint_{\left(    x_m^t ,  y_m^t  \right)}
			\frac{e^{-j k_n R_t} }{ R_t}
			e^{-j \left(   
				k_{x_m^t}   x_m^t 
				+ k_{y_m^t}   y_m^t
				\right)} \mathrm{~d} x_m^t  \mathrm{~d} y_m^t }_{E_t}
		\right]    \\
		&\times\left[ 
		\underbrace{\iint_{\left(  x_{m}^r ,  y_{m}^r     \right)} \frac{e^{-j k_n R_r}  }{ R_r} e^{-j \left(   
				k_{x_{m}^r}   x_{m}^r 
				+ k_{y_{m}^r}   y_{m}^r 
				\right)}\mathrm{~d} x_{m}^r  \mathrm{~d} y_{m}^r}_{E_r} 
		\right]
		\\
		&\times   \mathrm{d} x \mathrm{~d} y 
	\end{aligned}
\end{align}
where 
\begin{align}
\begin{aligned}
	& R_t=\sqrt{\left(x-x_m^t\right)^2+\left(y-y_m^t\right)^2+z^2}, \\
	& R_r=\sqrt{\left(x-x_m^r\right)^2+\left(y-y_m^r\right)^2+z^2}.
\end{aligned}
\end{align}
$(a)$ substitutes the signals $   s(	x_m^t, y_m^t,  x_{m}^r, y_{m}^r, n )   $ with the detailed expressions of virtual full arrays, ignoring the imperfection of zero-padding and obstruction; then, it approximates the incident projection of all pixels, i.e., $\forall \mathbf{p} \in  \mathcal{V}(z)$,  as the same value corresponding to the central pixel of $\mathcal{V}(z)$, i.e., $\mathbf{p}^c$, with
\begin{align}
 \cos  {\theta}_{\mathbf{p},m_t,m_r }^c  \triangleq  \cos\theta_{\mathbf{p}^c}^{\ell_t,m_t}    \cos\theta_{\mathbf{p}^c}^{\ell_r,m_r}  ,\\
 \cos {\phi}_{\mathbf{p},m_t,m_r }^c  \triangleq     \cos\phi_{\mathbf{p}^c}^{\ell_t,m_t}    \cos\phi_{\mathbf{p}^c}^{\ell_r,m_r}     .
\end{align} 

Next, the expression of $E_t$ and $E_r$ can be approximated by the method of stationary phase (MSP) for two-dimensional integrals \cite[(4-4), Page 241]{papoulis1968systems}, using
\begin{align}\label{MSP}
	\iint g(x, y) e^{j k \mu(x, y)} d x d y \approx \frac{2 \pi j g\left(x_o, y_o\right)}{k \sqrt{\mu_{x x} \mu_{y y}-\mu_{x y}{ }^2}} e^{j k \mu\left(x_o, y_o\right)}
\end{align}
where $x_o$ and $y_o$ are points of the stationary phase making 
\begin{align}
	\begin{aligned}
		& \left.\frac{\partial \mu(x, y)}{\partial x}\right|_{\left(x=x_0, y=y_0\right)}=0, \\
		& \left.\frac{\partial \mu(x, y)}{\partial y}\right|_{\left(x=x_0, y=y_0\right)}=0,
	\end{aligned}
\end{align}
and $ \mu_{x x}$, $\mu_{y y}$, and $\mu_{x y} $ denote the second-order partial derivatives of $\mu(x, y)$ evaluated at the stationary points, respectively. To apply the MSP, we can rewrite $E_r$ and $E_t$ as $ E_t= \iint_{\left(    x_m^t ,  y_m^t  \right)}  g_t(x_m^t ,  y_m^t ) e^{j k_n \mu_t(x_m^t ,  y_m^t )} d x_m^t  d  y_m^t  $ and $ E_r= \iint_{\left(  x_{m}^r, y_{m}^r \right)}  g_r(x_{m}^r, y_{m}^r) e^{j k_n \mu_r(x_{m}^r, y_{m}^r)} d x_{m}^r  d y_{m}^r$ where
\begin{align}
\begin{aligned}
	g_t(x_m^t, y_m^t) &= \frac{1}{\sqrt{( x- x_m^t )^2   +   (y-   y_m^t)^2   + z^2}},\\
\mu_t(x_m^t, y_m^t) &= -\sqrt{( x- x_m^t )^2   +   (y-   y_m^t)^2   + z^2} \\
&- \frac{    k_{x_m^t}   x_m^t 
	+ k_{y_m^t}   y_m^t         }{k_n}  ,\\
g_r(x_{m}^r, y_{m}^r) &= \frac{1}{\sqrt{(  x- x_{m}^r)^2   +   (y-y_{m}^r  )^2   + z^2}},\\
\mu_r(x_{m}^r,y_{m}^r) &=- \sqrt{(  x- x_{m}^r)^2   +   (y-y_{m}^r  )^2   + z^2} \\
&- \frac{      k_{x_{m}^r}   x_{m}^r 
	+ k_{y_{m}^r}   y_{m}^r        }{k_n} .
\end{aligned}
\end{align}

Following (\ref{MSP}), we then need to calculate the stationary points and the corresponding second-order  partial derivatives. After some algebraic manipulations\cite{fromenteze2019transverse}, we obtain
\begin{align}
\begin{aligned}
	E_t &= 
\frac{-j 2 \pi}{\sqrt{k_n^2-k_{x_m^t}^2-k_{y_m^t}^2}} \exp \left(-j z\sqrt{k_n^2-k_{x_m^t}^2-k_{y_m^t}^2} \right) \\
&\times 	\exp \left(-j k_{x_m^t} x\right) \exp \left(-j k_{y_m^t} y\right) \\
E_r &=  \frac{-j 2 \pi}{\sqrt{k_n^2-k_{x_{m}^r}^2-k_{y_{m}^r}^2}} \exp \left(-j z\sqrt{k_n^2-k_{x_{m}^r}^2-k_{y_{m}^r}^2} \right) \\
&	\times \exp \left(-j k_{x_{m}^r} x\right) \exp \left(-j k_{y_{m}^r} y\right)
\end{aligned}
\end{align}

To fuse the transmitting and receiving wavenumber domains, define
\begin{align}
	k_{x_{m}}&  \triangleq  k_{x_m^t}  + k_{x_{m}^r},\\
	k_{y_{m}}&  \triangleq   k_{y_m^t}  + k_{y_{m}^r},\\
	k_{z_{m}}&  \triangleq   k_{z_m^t}  + k_{z_{m}^r},
\end{align}
where  $	k_{z_m^t}  \triangleq    \sqrt{k^2 -    (k_{x_m^t})^2  - (k_{y_m^t})^2 }$ and $k_{z_{m}^r}  \triangleq    \sqrt{k^2 -  (k_{x_{m}^r})^2 - (k_{y_{m}^r})^2   }$.
Then, we obtain
\begin{align}
	E_t &= 
	\frac{-j 2 \pi}{k_{z_m^t} } \exp \left(-j k_{z_m^t}  z \right) 
	\exp \left(-j k_{x_m^t} x\right) \exp \left(-j k_{y_m^t} y\right) \\
	E_r &=  \frac{-j 2 \pi}{k_{z_{m}^r} } \exp \left(-j k_{z_{m}^r}  z  \right) 
	\exp \left(-j k_{x_{m}^r} x\right) \exp \left(-j k_{y_{m}^r} y\right)
\end{align}
and
\begin{align} \label{E_tE_r}
	E_t  E_r = \frac{-4 \pi^2}{k_{z_m^t}  k_{z_{m}^r}  } \exp \left(    
	-j\left\{
	k_{x_{m}} x + k_{y_{m}}y +      k_{z_m }     z
	\right\}
	\right)
\end{align}

Substituting this simplified expression (\ref{E_tE_r}) into the measured signal (\ref{4D_FT}), we find
\begin{align}\label{S_frequency_wavenumber}
	\begin{aligned}
		&S(		k_{x_m^t}  , k_{y_m^t},  k_{x_{m}^r} , k_{y_{m}^r}, n ) \\
		&=\! \! \iint_{\mathcal{V}(z)}  \frac{ \rho_n(x, y, z)}{4\pi }
  \sqrt{ \cos  {\theta}_{\mathbf{p},m_t,m_r }^c   \cos {\phi}_{\mathbf{p},m_t,m_r }^c       }
		E_t E_r
		\mathrm{~d} x \mathrm{d} y\\
		&= 
		\frac{- \pi  \sqrt{ \cos  {\theta}_{\mathbf{p},m_t,m_r }^c   \cos {\phi}_{\mathbf{p},m_t,m_r }^c       }       }{k_{z_m^t}  k_{z_{m}^r}  }
		\iint_{\mathcal{V}(z) }  \rho_n(x, y, z) \\
		& \times
		\exp \left(    
		-j\left\{
		k_{x_{m}} x + k_{y_{m}}y + k_{z_{m}}z
		\right\}
		\right)
		\mathrm{~d} x \mathrm{~d} y \\
		&=  \!
		{\rm FT}_{2D} \! \Big[   
		\frac{- \pi   e^{-j k_{z_{m}} z}      }{k_{z_m^t}  k_{z_{m}^r}  }
	 \sqrt{ \cos  {\theta}_{\mathbf{p},m_t,m_r }^c   \cos {\phi}_{\mathbf{p},m_t,m_r }^c       }
		\rho_n(x,y,z)
		\! \Big] \! .
	\end{aligned}
\end{align}

Now, it can be seen that the frequency-wavenumber domain signal  $ S(		k_{x_m^t}  , k_{y_m^t},  k_{x_{m}^r} , k_{y_{m}^r}, n)  $ is a 2-D FT of the object's reflectivity coefficient $\rho_n(x,y,z)$. Therefore, the imaging at distance $z$ can be reconstructed by
\begin{align}\label{mapping_result}
\begin{aligned}
	\rho_n(x,y,z)  = {\rm FT}_{2D}^{-1} \Big[   &
\frac{k_{z_m^t}  k_{z_{m}^r}  }{- \pi   e^{ j k_{z_{m}} z}   \sqrt{ \cos  {\theta}_{\mathbf{p},m_t,m_r }^c   \cos {\phi}_{\mathbf{p},m_t,m_r }^c       }  }      \\
&    \times   S(		k_{x_m^t}  , k_{y_m^t},  k_{x_{m}^r} , k_{y_{m}^r}, n ) 
\Big] .
\end{aligned}
\end{align}

As shown in (\ref{mapping_result}), to perform the RF imaging, we first need to calculate frequency-wavenumber domain signal spectrum $   S(		k_{x_m^t}  , k_{y_m^t},  k_{x_{m}^r} , k_{y_{m}^r}, n )  $ based on the frequency-spatial domain measurement $s(	x_m^t, y_m^t,  x_{m}^r, y_{m}^r, n  )$, which requires two 2-D spatial FTs over the transmitting and receiving virtual arrays, respectively. The 2-D FT can be implemented by using 2-D FFT which effectively reduces the computational complexity. Then, we filter the signal and conduct a 2D inverse FT to obtain the imaging result following (\ref{mapping_result}). The overall algorithm is summarized in Algorithm \ref{alg:RMA}.

\begin{algorithm}[t]
	\caption{RMA-Based Imaging}\label{alg:RMA}
	\begin{algorithmic}[1] 
		\State{According to the adopted array architecture, conduct the signal fusion as (\ref{fusion}) and obtain the full-array frequency-spatial-domain data matrix $s\left(x_m^t, y_m^t, x_m^r, y_m^r, n\right)$ }
		\State  Conduct 4D FFT to obtain the frequency-wevenumber-domain data matrix $S(		k_{x_m^t}  , k_{y_m^t},  k_{x_{m}^r} , k_{y_{m}^r}, n ) $ as (\ref{4D_FT})
		\State  Conduct filtering operation to the data matrix with coefficients of  $\frac{ k_{z_m^t}  k_{z_{m}^r}  }{  -\pi e^{ j k_{z_{m}} z}   \sqrt{ \cos  {\theta}_{\mathbf{p},m_t,m_r }^c   \cos {\phi}_{\mathbf{p},m_t,m_r }^c       }  }    $ as (\ref{mapping_result})
		\State Conduct 2D IFFT to obtain the imaging result
	\end{algorithmic}
\end{algorithm}

\subsection{Special Case of SIMO/SAR}

For more insights and to facilitate understanding,  we briefly discuss the design based on monostatic virtual arrays \cite{sheen2001three}.  Based on (\ref{mono}), the signal received by antenna $m$ on the full virtual array can be approximated as
\begin{align} 
\begin{aligned}
	&s(	x_m, y_m , n) \\
	& \approx  \iint_{\mathcal{V}(z)}  { \cos  {\theta}_{\mathbf{p},m }^c   \cos {\phi}_{\mathbf{p},m }^c       }   
	   \frac{\rho_n(x, y, z)}{ 4\pi  R^2}    e^{-j k_n 2R}          \mathrm{d} x \mathrm{d} y,
\end{aligned}
\end{align}
where $ 	R  = \sqrt{( x- x_m  )^2   +   (y- y_m     )^2   + z^2} $.

Based on Weyl's identity\cite[(2.2.27)]{chew1999waves}, we know that
\begin{align}\label{weyl}
	\frac{e^{ {j} k_n r}}{r}=\frac{     {j}}{2 \pi} \iint_{-\infty}^{\infty}  \frac{e^{{j}\left(k_x r_x+k_y r_y+ k_z      \left|r_z\right|\right)}}{  k_z    } \mathrm{d} k_x \mathrm{d} k_y
\end{align}
where  $r^2=r_x^2+r_y^2+r_z^2$ and
\begin{align}
	k_z     = \begin{cases}\sqrt{k_n^2-k_x^2-k_y^2}, & k_x^2+k_y^2 \leq k_n^2, \\  {j} \sqrt{  k_x^2+k_y^2-k_n^2}, & k_x^2+k_y^2>k_n^2.\end{cases}
\end{align}
Then, by defining $ \tilde{s}(	x_m, y_m , n)  \triangleq \frac{ 4\pi s(	x_m, y_m , n)  }{ { \cos  {\theta}_{\mathbf{p},m}^c   \cos {\phi}_{\mathbf{p},m }^c       }  }$, we have
\begin{align} \label{ft_siso}
	\begin{aligned}
& \tilde{s}(	x_m, y_m ,n) \overset{(b)}{\approx}  \iint_V \frac{\rho_n(x, y, z_0)}{  R }       e^{-jk_n 2R}           \mathrm{d} x \mathrm{d} y \\
& =  \iint_{\mathcal{V}(z)}      \rho_n(x, y, z)       \frac{   e^{-j (2k_n)  R}   }{  R }                  \mathrm{d} x \mathrm{d} y \\
&  \overset{(c)}{=}  \iint_{\mathcal{V}(z)}      \rho_n(x, y, z)        \frac{{-j}}{2 \pi}  \\
&\times \iint_{-\infty}^{\infty}  \frac{e^{{-j}\left(k_x      (x-x_m)    +k_y    (y-y_m)     +  k_z      z     \right)}}{   k_z   } \mathrm{d} k_x  \mathrm{d} k_y                  \mathrm{d} x \mathrm{d} y \\
& =    \iint  \frac{{-j   e^{-jk_z z}}}{2 \pi k_z} 
\underbrace{	\left[     \iint      \rho_n(x, y, z)         	 e^{ -  j      \left(k_x       x   +k_y     y       \right)       } {d} x {d} y  \right]}_{     \mathrm{FT}_{2D}^{}\left\{    \rho_n(x,y,z)   \right\}   } \\
& \times	e^{     j    \left(k_x      x_m    +k_y    y_m      \right)   } 
\mathrm{d} k_x \mathrm{d} k_y                      \\
& =  \mathrm{FT}_{2D}^{-1}     \left\{   \frac{{- j   e^{-jk_z z}}}{ 2 \pi  k_z}  \mathrm{FT}_{2D}^{}\left\{    \rho_n(x,y,z)   \right\}        \right\}
	\end{aligned}
\end{align}
where $ (b) $ approximates $1/R^2$ as $1/R$, $(c)$ applies the Weyl's identity in (\ref{weyl}), and
\begin{align}
	k_z     = \sqrt{4k_n^2-k_x^2-k_y^2}  .
\end{align}

From (\ref{ft_siso}), imaging can be readily obtained through 2D FFT, filtering, and 2D IFFT operations.

\subsection{Imaging Resolution and Sampling Criterion}

\subsubsection{Resolution}For an object located at a distance $z$, the 2D imaging resolution of the cases of full array and boundary array is decided by array aperture, frequency, and distance as follows\cite{lopez20003,zhuge2012three}
 \begin{align}\label{reso}
\begin{aligned}
	\delta_x & =\frac{c  z}{    f_c  \left(    L_{T_x}+L_{R_x}    \right)   }, \\
	\delta_y & =\frac{ c  z}{   f_c      \left(   L_{T_y}+L_{R_z}     \right) ,     }
\end{aligned}
 \end{align}
where $  L_{T_x}$ and $L_{T_y}   $ denote the length of transmitting array aperture along the $x$- and $y$-axes, respectively.  $  L_{R_x}$ and $L_{R_y}   $ denote the length of receiving array aperture along the $x$- and $y$-axes, respectively.
It can be seen that for high-resolution imaging of objects placed at a large distance, the array aperture should be large, to ensure the collection of rich information in the near-field channels. This requirement can be fulfilled by increasing the number of antennas and the antenna spacing to form the concept of sparse extremely large-scale MIMO (XL-MIMO)\cite{lu2024tutorial}.

\subsubsection{Aliasing}  Although enlarging antenna spacing can improve imaging resolutions,  large antenna spacing could cause spatial aliasing. To satisfy the Nyquist criterion of spatial signal sampling, the antenna spacing of transmitter and receiver along the $x$-axis should satisfy\cite{lopez20003,zhuge2012three}
\begin{align}\label{alise}
\begin{aligned}
& \Delta_{ {R_x}} \leq \lambda_{  \min } \frac{\sqrt{\left(L_{ {R_x}}+D_x\right)^2 / 4+ z^2}}{L_{ {R_x}}+D_x}, \\
& \Delta_{ {T_x}} \leq \lambda_{  \min } \frac{\sqrt{\left(L_{ {T_x}}+D_x\right)^2 / 4+ z^2}}{L_{ {T_x}}+D_x} ,
\end{aligned}
\end{align}
where $D_x$ denotes the aperture of the imaging target along the $x$-axis. It can be seen that (\ref{alise}) is an increasing function of imaging distance $z$ and a decreasing function of array aperture. This implies that the antenna could be sparser for long-distance imaging. Also, for a smaller number of antennas, the array aperture could be more sparse.

To sum up, the array could be a sparse XL-MIMO so that large aperture and high resolution are realized. However, the antenna spacing can not be too large in order to satisfy the sampling criterion and mitigate the impact of aliasing. In general, the parameters and the setup will depend on the imaging scenarios (object size, distance), array aperture, antenna spacing, and operation frequency.

\subsection{Image Quality Evaluation}
The evaluation of image quality requires specific criteria that are in line with humans' subjective evaluation. Besides traditional assessment standards such as minimizes the mean square error (MSE) and root mean square error (RMSE) that was widely adopted for channel estimation and localization problems, in the following, we consider three effective criteria commonly adopted for evaluating imaging quality\cite{wang2017associations,li2025compressive,zhang2020amp}, i.e., the  peak signal-to-noise ratio (PSNR), structural similarity (SSIM) index, and Pearson’s correlation coefficient (PCC)\footnote{There are also some other standards such as point spread function (PSF) \cite{li2018compressive} and  power spreading ratio \cite{wang20203}.}. 

Note that the number and coordinates of pixels on the obtained imaging results based on RMA depend on the number of antennas and array aperture. However, the image of the true object is continuous. Thus, before quantitative comparison, pre-operations including grid quantization into pixels, image shifting, and scaling are required to align the original image with the imaging results in terms of pixel size and coordinates. Then, for a 2D image, denote the $(x,y)$-th pixel of the true image and estimated image as $\rho_n( {\mathbf{p}(x,y)})  $ and $\hat{\rho}_n ( {\mathbf{p}(x,y)})  $, respectively. The MSE of the 2D image is given by
\begin{align}
\begin{aligned}
&\mathrm{MSE}\left(        \rho_n( {\mathbf{p} }) ,     \hat{\rho}_n ( {\mathbf{p}})      \right) \\
&= \frac{1}{D_x D_y} \sum_{x=1}^{D_x} \sum_{y=1}^{D_y}  \left|    
\rho_n( {\mathbf{p}(x,y)})    -    \hat{\rho}_n ( {\mathbf{p}(x,y)})  
\right|^2.
\end{aligned}
\end{align}

Then, the PSNR is defined as
\begin{align}
\mathrm{PSNR}\left(         \rho_n( {\mathbf{p} }) ,     \hat{\rho}_n ( {\mathbf{p}})         \right) = 10\log_{10}  \frac{ \max_{x,y}  \left|      \hat{\rho}_n ( {\mathbf{p}(x,y)})    \right|^2  }{\mathrm{MSE}\left(            \rho_n( {\mathbf{p} }) ,     \hat{\rho}_n ( {\mathbf{p}})          \right) }.
\end{align}

SSIM characterizes the similarity of luminance, contrast, and structure of two images.  The SSIM is defined as
\begin{align}
\begin{aligned}
&\operatorname{SSIM}(      \rho_n( {\mathbf{p} }) ,     \hat{\rho}_n ( {\mathbf{p}})             ) \\
&=\left[\frac{2 \mu_{\mathbf{p}} \hat{\mu}_{{\mathbf{p}}}+\varepsilon_1}{\mu_{\mathbf{p}}^2+\hat{\mu}_{{\mathbf{p}}}^2+\varepsilon_1}\right]\left[\frac{2 \sigma_{\mathbf{p}} \hat{\sigma}_{{\mathbf{p}}}+\varepsilon_2}{\sigma_{\mathbf{p}}^2+\hat{\sigma}_{{\mathbf{p}}}^2+\varepsilon_2}\right]  \leq 1,
\end{aligned}
\end{align}
where $\mu_{\mathbf{p}}$ and $\hat{\mu}_{{\mathbf{p}}}$ denote the means of amplitudes of images $    \rho_n( {\mathbf{p} })   $ and $ \hat{\rho}_n ( {\mathbf{p}})      $, respectively. $\sigma_{\mathbf{p}}$ and   $\hat{\sigma}_{{\mathbf{p}}}$ denote the standard deviations of amplitudes of images $   \rho_n( {\mathbf{p} })   $ and $  \hat{\rho}_n ( {\mathbf{p}})     $, respectively. $\varepsilon_1$ and $\varepsilon_2$ are two small positive constants.

PCC measures the similarity between images. The PCC is defined as
\begin{align}
\begin{aligned}
&\operatorname{PCC} \left(  \rho_n( {\mathbf{p} }) ,     \hat{\rho}_n ( {\mathbf{p}})       \right) \\
&=\frac{ \left| \sum_{x,y}\left(    \rho_n( {\mathbf{p} } (x,y) )    - \mu_{\mathbf{p}}\right)^*\left(      \hat{\rho}_n ( {\mathbf{p}}(x,y) )      -   \hat{\mu}_{{\mathbf{p}}}     \right) \right|}{
	\sqrt{\sum_{x,y}\left|    \rho_n( {\mathbf{p} }(x,y) )  - {\mu}_{\mathbf{p}}\right|^2} 
	\sqrt{\sum_{x,y}\left|       \hat{\rho}_n ( {\mathbf{p}}(x,y))  	- \hat{\mu}_{{\mathbf{p}}}   \right|^2}}.
\end{aligned}
\end{align}

\section{SBL-Based Design}\label{section_sbl}

In this section, we focus on 3D image reconstruction by exploiting the multi-view of the image region to overcome spatial non-isotropic reflections and occlusion effects, as illustrated in Fig. \ref{figure1}. By discretizing the 3D imaging region,  we formulate the imaging task as a linear inverse problem and apply a CS-based method, where we further exploit the frequency selectivity of the reflection coefficients. Our approach is feasible for any distributed array geometry and has strong cooperation on multi-view imaging via joint and coherent processing of received data from all distributed arrays.

First, the 3D imaging region $\mathcal{V}$ is discretized uniformly into $Q=Q_x Q_y Q_z$ cube units, where each cube unit is called a voxel with volume $\bar{V}$. Each voxel is represented by its central point.  If there exist scatterers located in a voxel, the reflection coefficient of its central point is not zero. We assume that the discretized cube units are small such that we can approximately treat the pathloss (distance) and scattering reflectivity of all points within the cube to be the same as the center point of the cube. 
Then, we can approximate the received signal in (\ref{receiving_signal_cooperative}) by the signal reflected from $Q$ discretized points, as follows
\begin{align}\label{discretize_signal_cooperative}
	\begin{aligned}
		&	\mathbf{y}_{\ell_r}[s,n]   \approx 
		\sum_{\ell_t    \in   \mathcal{L}_{s,n}^t       }          \sum_{q=1}^{Q}       \bar{V} {\rho_n(\mathbf{p}_q)}  \mathbf{h}_{ \mathbf{p}_q}^{\ell_r} [n] \left(\mathbf{h}_{\mathbf{p}_q}^{\ell_t} [n]\right)^\herm \mathbf{x}_{\ell_t}[s,n]    \\
		& + \mathbf{w}_{\ell_r}[s,n]    \\
		&  =   \sum_{\ell_t    \in   \mathcal{L}_{s,n}^t             }       \bar{V}     \mathbf{H}_{\ell_r} [n]  \operatorname{diag}\left\{    \rho_n(\mathbf{p}_1),    \rho_n(\mathbf{p}_2), \ldots,    \rho_n(\mathbf{p}_Q)   \right\}   \\
		& \times   \mathbf{H}_{\ell_t} [n]  \mathbf{x}_{\ell_t}[s,n]  + \mathbf{w}_{\ell_r}[s,n]    \\
		& = \sum_{\ell_t    \in   \mathcal{L}_{s,n}^t             }            \underbrace{\bar{V} \mathbf{H}_{\ell_r} [n]  \operatorname{diag}\left\{     \mathbf{H}_{\ell_t} [n]  \mathbf{x}_{\ell_t}[s,n]    \right\}}_{\triangleq\Phim_{\ell_r, \ell_t} [s, n]}     \rhov_n  + \mathbf{w}_{\ell_r}[s,n]   \\
		& =\sum_{\ell_t    \in   \mathcal{L}_{s,n}^t             }     \Phim_{\ell_r, \ell_t} [s,n]        \rhov_n  + \mathbf{w}_{\ell_r}[s,n],
	\end{aligned}
\end{align}
where $\mathbf{p}_q$ denotes the coordination of the central point of the $q$-th voxel and 
\begin{align}
	\mathbf{H}_{\ell_r} [n] &\triangleq \left[     \mathbf{h}_{ \mathbf{p}_1}^{\ell_r} [n],   \mathbf{h}_{ \mathbf{p}_2}^{\ell_r} [n]  , \ldots, \mathbf{h}_{ \mathbf{p}_Q}^{\ell_r} [n]     \right]  \in \mathbb{C}^{M_r \times Q},\\
	\mathbf{H}_{\ell_t} [n] &\triangleq \left[   \mathbf{h}_{\mathbf{p}_1}^{\ell_t} [n],    \mathbf{h}_{\mathbf{p}_2}^{\ell_t} [n], \ldots,  \mathbf{h}_{\mathbf{p}_Q}^{\ell_t} [n]       \right]^\herm  \in \mathbb{C}^{Q \times M_t}, \\
	\rhov_n & \triangleq \left[        \rho_n(\mathbf{p}_1),    \rho_n(\mathbf{p}_2), \ldots,    \rho_n(\mathbf{p}_Q)         \right]^\transp   \in \mathbb{C}^{Q \times 1}.
\end{align}

To fully leverage the potential of multi-view observation in improving imaging quality, we consider an alternate illumination-receiving scheme to enrich the diversity.  Specifically, at each time slot $s$ and subcarrier $n$, the index sets of RUs that transmit and receive signals are denoted as $\mathcal{L}^{  t}_{s,n} \subset \mathcal{L}$ and $\mathcal{L}^{  r}_{s,n} = \mathcal{L} \setminus \mathcal{L}^{  t}_{s,n}$, respectively. Then, stacking received signal $ 	\mathbf{y}_{\ell_r}[s,n]    $ of RUs $\ell_r\in \mathcal{L}^{  r}_{s,n}$ after $1\leq s \leq S$ time slots, the observation vector at subcarrier $n$ can be constructed as
\begin{align}\label{y_stack}
	\begin{aligned}
		\mathbf{y}_n &\triangleq \begin{bmatrix}
			{\yv}_{[\mathcal{L}_{1,n}^{ {r}}]_1}  [1,n] \\
			\vdots \\
			{\yv}_{[\mathcal{L}_{1,n}^{ {r}}]_{     L^{{r}}_{1,n}} }    [1,n] \\
			{\yv}_{[\mathcal{L}_{2,n}^{ {r}}]_{1}}[2,n]\\
			\vdots\\
			{\yv}_{[\mathcal{L}_{S,n}^{ {r}}]_{   L^{{r}}_{S,n}       }}[S,n]
		\end{bmatrix} 
	\end{aligned}
\end{align}
where $L^{ {t}}_{n,s} \triangleq    |\mathcal{L}^{ {t}}_{n,s}|$, $L^{{r}}_{n,s} \triangleq |\mathcal{L}^{ {r}}_{n,s}|$, and $[\mathcal{L}]_i$ denotes the $i$-th element of set $\mathcal{L}$. Substituting (\ref{y_stack}) with the detailed expression in (\ref{discretize_signal_cooperative}), we have
\begin{align}
	\begin{aligned}
	\mathbf{y}_n \! =\! \underbrace{ \begin{bmatrix}
			\sum_{i\in\mathcal{L}^{ {t}}_{1,n}}     \Phim_{    [\mathcal{L}^{ {r}}_{n,1}]_1  ,  i   } [   1   ,n]    \\
			\vdots \\
			\sum_{i\in\mathcal{L}^{ {t}}_{1,n}}      \Phim_{    [\mathcal{L}^{ {r}}_{n,1}]_{L^{{r}}_{1,n}}     ,  i   } [   1   ,n]      \\
			\sum_{i\in\mathcal{L}^{ {t}}_{2,1}}         \Phim_{    [\mathcal{L}^{ {r}}_{n,2}]_1  ,  i   } [   2   ,n]           \\
			\vdots \\
			\sum_{i\in\mathcal{L}^{ {t}}_{S,n}}          \Phim_{    [\mathcal{L}^{ {r}}_{n,S}]_{ L^{{r}}_{S,n} }  ,  i   } [   S   ,n]  
	\end{bmatrix}}_{\triangleq \Phim_n}   \rhov_n \!+\!
\underbrace{ \begin{bmatrix}
		 \mathbf{w}_{    [\mathcal{L}^{ {r}}_{n,1}]_1     } [   1   ,n]    \\
		\vdots \\
		 \mathbf{w}_{    [\mathcal{L}^{ {r}}_{n,1}]_{L^{{r}}_{1,n}}     } [   1   ,n]      \\
		 \mathbf{w}_{    [\mathcal{L}^{ {r}}_{n,2}]_1   } [   2   ,n]           \\
		\vdots \\
		 \mathbf{w}_{    [\mathcal{L}^{ {r}}_{n,S}]_{ L^{{r}}_{S,n} }   } [   S   ,n]  
\end{bmatrix}}_{\triangleq  \nv_n}  .
\end{aligned}
\end{align}

We consider the more general frequency-selective reflectivity of the imaging object, where the complex reflection coefficients depend on the carrier frequency \cite{manzoni2024wavefield}. Specifically, the reflection coefficients are assumed to be correlated at different subcarriers of the same position while independent at different positions. Mathematically, we have
\begin{align}
	\bE[\rho_n(\pv_q) \rho^*_{n'}(\pv_{q'})] \!\!=\!  \gamma_q[\Psim]_{n,n'} \mathbb{1}_{q=q'}, \; q\in[1:Q], n\in[1:N],
\end{align}
where $\mathbb{1}_{q=q'}$ is the indicator function of the condition $q = q'$ and $\Psim \in \bC^{N \times N}$ is a positive definite matrix capturing the correlation of $N$ subcarriers. $\gamma_q$ is the reflection power at position $\pv_q$. 
Now, the imaging task is to estimate $\Um \triangleq [\rhov_1, \dots, \rhov_N] $ from the noisy observations  $\{\yv_n\}_{n=1}^N$. We notice that the dimensions of observations on different subcarriers, i.e., $\yv_n$ ($\Phim_n$) and $\yv_{n'}$ ($\Phim_{n'}$) with $n\neq n' \in [1:N]$, are not necessarily the same due to the flexible illumination-receiving scheme. Therefore, this is a special MMV problem with different sensing matrices and correlated measurements. We solve this special MMV problem using an SBL approach. The SBL treats $\gamma_q$ as a hyperparameter, and assumes a prior distribution of the unknown variables $\Um$, which is given as  
\begin{align}
	p(\Um; \Gammam, \Psim) = \prod^Q_{i=1}\Big( p(\bar{\uv}_i; \gamma_i, \Psim) \sim \mathcal{CN} (\mathbf{0}, \gamma_i \Psim)\Big),
\end{align}
where $\Gammam = \diag(\gammav) = \diag([\gamma_1, \dots, \gamma_Q]^\transp) \in \bR^{Q\times Q}$ contains all hyperparameters in a diagonal matrix, and  
$\bar{\uv}_i \triangleq ([\Um]_{i,:})^\transp$,  $i \in [1:Q]$, is the vector of the $i$-th row of $\Um$ containing the unknown reflection coefficients of the $i$-th voxel at all subcarriers.

Stacking all observations, we have $\bar{\yv} \triangleq [
\yv_1^\transp,\dots,\yv_N^\transp]^\transp$. 
The SBL approach maximizes the Bayesian evidence $p(\bar{\yv};\Gammam, \Psim)$ to infer the parameters $\gammav$ and $\Psim$, and then estimates $\Um$ via the posterior mean of the posterior distributions $p(\Um | \bar{\yv}; \gammav, \Psim)$. 
Hence, the objective of SBL is the minus log-likelihood function for maximum-likelihood (ML) estimation of the parameter vector $\gammav$ and $\Psim$, which is given as
\begin{align}\label{eq:cost_function}
	-\log p(\bar{\yv}; \Gammam, \Psim) = \log(\det(\Sigmam_{\bar{\yv}})) + \bar{\yv}^\herm \Sigmam_{\bar{\yv}}^{-1}\bar{\yv},
\end{align}
where $\Sigmam_{\bar{\yv}}$ is the covariance matrix of $\bar{\yv}$, and the constant term is ignored. The cost function in \eqref{eq:cost_function} is not a convex function, which is normally solved iteratively using the expectation–maximization (EM) method. 
Denoting the estimated parameters in the $\ell$-th iteration as $\widehat{\Gammam}^{(\ell)} = \diag(\widehat{\gammav}^{(\ell)}) = \diag([\widehat{\gamma}_1^{(\ell)}, \dots, \widehat{\gamma}_Q^{(\ell)}]^\transp) $ and $\widehat{\Psim}^{(\ell)}$, and treating $\Um$ as hidden variables, each E-step of the EM method evaluates the average minus log-likelihood of the complete data set $\{\bar{\yv}, \Um\}$ as\footnote{In this work, we assume that the noise power $\sigma^2$ is known. However, we emphasize that the proposed SBL approach can also be applied when the noise power is unknown, in which the noise power will be jointly estimated with other unknown parameters. } 
\begin{align}
	&\mathcal{L}^{(\ell)}\left(\gammav, \Psim|\widehat{\gammav}^{(\ell)}, \widehat{\Psim}^{(\ell)}\right)\notag  \\ 
	&= 
	\bE_{\Um|\bar{\yv}; \widehat{\gammav}^{(\ell)}, \widehat{\Psim}^{(\ell)}}[-\log p(\bar{\yv}, \Um;\gammav, \Psim) ]  \notag \\
	&=\bE_{\Um|\bar{\yv}; \widehat{\gammav}^{(\ell)}, \widehat{\Psim}^{(\ell)}}\left[\underbrace{-\log p(\bar{\yv}|\Um)}_{\text{constant due to known $\sigma^2$}} \underbrace{-\log p(\Um;\Gammam, \Psim)}_{\text{using prior of $\Um$}}\right]    \notag\\
	&\overset{(d)}{=} \bE_{\Um|\bar{\yv}; \widehat{\gammav}^{(\ell)}, \widehat{\Psim}^{(\ell)}} \left[\sum^Q_{i=1} -\log \left(\frac{\exp\left(-\bar{\uv}_i^\herm (\gamma_i \Psim)^{-1}\bar{\uv}_i\right)}{\pi^N \det( \gamma_i \Psim)} \right)\right]   \notag\\
	&\overset{(e)}{=}\sum^Q_{i=1} \log(\det(\gamma_i\Psim)) + \bE_{\bar{\uv}_i | \bar{\yv}; \widehat{\gamma}_i^{(\ell)}, \widehat{\Psim}^{(\ell)} }\left[\bar{\uv}_i^\herm (\gamma_i \Psim)^{-1}\bar{\uv}_i\right]   \notag \\
	&=\!\!\sum^Q_{i=1} \! N \! \log(\gamma_i) \!+\! \log\det( \! \Psim) \!+\!  \bE_{\bar{\uv}_i | \bar{\yv}; \widehat{\gamma}_i^{(\ell)}, \widehat{\Psim}^{(\ell)}}\left[\bar{\uv}_i^\herm (\gamma_i \Psim)^{-1}\bar{\uv}_i\right] \label{eq:L_ini}
\end{align}
where in $(d)$ and $(e)$ the constant terms that do not contribute to the subsequent optimization of the parameters are dropped.

To calculate the last term in \eqref{eq:L_ini}, we need a posterior density of each $\bar{\uv}_i$, which can be obtained from the posterior density of the full $\Um$. Due to the aforementioned fact that the dimension of different measurements $\yv_n$ could be different, we need to first zero-pad them to the same dimension and later extract the non-zero indices using a selection matrix. 
Concretely, we define $\mathbb{y}_n=[\yv_n^\transp, \mathbf{0}_{l_{\rm max}-l_n,1}^\transp]^\transp$, where $l_n = |\yv_n|$ is the dimension of $\yv_n$ and $l_{\rm max} = \max(\{|\yv_n|\}_{n=1}^N)$ is the maximum dimension among all measurements. We also zero-pad the corresponding noise vectors  and sensing matrices as
\begin{align}
\begin{aligned}
	&\mathbb{n}_n=[\nv_n^\transp, \mathbf{0}_{l_{\rm max}-l_n,1}^\transp]^\transp ,\\
&	\bar{\Phim}_n = \begin{bmatrix}
	\Phim_n \\
	\mathbf{0}_{l_{\rm max}-l_n, Q}
\end{bmatrix}. 
\end{aligned}
\end{align}
With the zero-padding signals, the noisy observations of all subcarriers can be written as
\begin{align}
	\begin{aligned}
		\mathbb{Y} & \triangleq [\mathbb{y}_1, \ldots, \mathbb{y}_N] \\
		& = [\bar{\Phim}_1 \rhov_1, \dots, \bar{\Phim}_N\rhov_N] + [\mathbb{n}_1, \dots, \mathbb{n}_N] \\
		& = [\bar{\Phim}_1 \rhov_1, \dots, \bar{\Phim}_N\rhov_N] + \mathbb{N},
	\end{aligned}
\end{align}
where $\mathbb{N} \triangleq [\mathbb{n}_1,\dots, \mathbb{n}_N]$.
Then, by using the selection matrix to extract the non-zero signals, we can obtain a linear form between the vectorization of the transpose of $\Um$ and the corresponding non-zero measurements, which is given as
\begin{align}
	\widetilde{\yv} &\triangleq \Sm \vec\left(\mathbb{Y}^\transp\right)\\
	&= \underbrace{\Sm\left(\sum_{n=1}^N \bar{\Phim}_n \otimes \boldsymbol{\Upsilon}_n\right)}_{\triangleq \widetilde{\Phim}}\underbrace{\vec\left(\Um^\transp\right)}_{\triangleq \widetilde{\uv}} + \underbrace{\Sm \vec\left(\mathbb{N}^\transp\right)}_{\triangleq \widetilde{\nv}} \\
	&= \widetilde{\Phim} \widetilde{\uv} +\widetilde{\nv},\label{eq:linear_form}
\end{align}
where $\boldsymbol{\Upsilon}_n \in \{0,1\}^{N\times N}$ is a $N$-by-$N$ square matrix whose elements are all zeros except $[\boldsymbol{\Upsilon}_n]_{n,n} = 1$. $\Sm\in \{0,1\}^{l_{\rm all} \times Nl_{\rm max}}$ is a binary selection matrix that extracts the non-zero entries according to the previous zero-padding process with $l_{\rm all} = \prod_{n=1}^N l_n$ being the dimension of all measurements.  It is noticed that $\widetilde{\uv} = [\bar{\uv}_1^\transp, \dots, \bar{\uv}_Q^\transp]^\transp \in\bC^{NQ}$ is the stacking vector of all rows of $\Um$ representing a two level stacking of reflection coefficients, where the outer level follows the grid point and the inner level follows the subcarrier index. 

The prior distribution of $\widetilde{\uv}$ can be rewritten as $p(\widetilde{\uv};\Gammam,\Psim) \sim \mathcal{CN} (\mathbf{0}, \widetilde{\Gammam})$, where $ \widetilde{\Gammam} = \Gammam \otimes \Psim$.
Then, given \eqref{eq:linear_form}, the posterior distribution of  $\widetilde{\uv}$ is $p(\widetilde{\uv}|\widetilde{\yv};\widetilde{\Gammam}) \sim \mathcal{CN}(\muv_{\widetilde{\uv}}, \Sigmam_{\widetilde{\uv}})$, where the posterior mean and posterior covariance at the $\ell$-th iteration using estimated parameters $\widehat{\gammav}^{(\ell)}$ and $\widehat{\Psim}^{(\ell)}$ are given as \cite{zhang2013extension}
\begin{align}
	\Sigmam_{\widetilde{\uv}}^{(\ell)} &= \left(\frac{1}{\sigma^2}\widetilde{\Phim}^\herm \widetilde{\Phim} + \left(\widetilde{\Gammam}^{(\ell)}\right)^{-1}\right)^{-1} \in \bC^{NQ\times NQ}\\
	&=\widetilde{\Gammam}^{(\ell)} \!- \underbrace{\widetilde{\Gammam}^{(\ell)} \widetilde{\Phim}^\herm \left(\sigma^2 \Id_{l_{\rm all}} + \widetilde{\Phim} \widetilde{\Gammam}^{(\ell)} \widetilde{\Phim}^\herm \right)^{-1}}_{\triangleq \widetilde{\Am}^{(\ell)}} \widetilde{\Phim} \widetilde{\Gammam}^{(\ell)}, \label{eq:poster_cov}\\
	\muv_{\widetilde{\uv}}^{(\ell)} &= \frac{1}{\sigma^2} \Sigmam_{\widetilde{\uv}}^{(\ell)}\widetilde{\Phim}^\herm \widetilde{\yv} =\widetilde{\Am}^{(\ell)} \widetilde{\yv}. \label{eq:poster_mean}
\end{align}
Having \eqref{eq:poster_cov} and \eqref{eq:poster_mean}, we can calculate the last term in \eqref{eq:L_ini}, which is given as
\begin{align}
	&\bE_{\bar{\uv}_i|\bar{\yv}; \widehat{\gamma}^{(\ell)}_i, \widehat{\Psim}^{(\ell)}}\left[\bar{\uv}_i^\herm (\gamma_i \Psim)^{-1}\bar{\uv}_i\right] \\     
	&=\bE_{\bar{\uv}_i|\bar{\yv}; \widehat{\gamma}^{(\ell)}_i, \widehat{\Psim}^{(\ell)}}\left[\trace\left(\bar{\uv}_i\bar{\uv}_i^\herm (\gamma_i \Psim)_i^{-1}\right)\right] \\
	&=\trace\left(\bE_{\bar{\uv}_i|\bar{\yv}; \widehat{\gamma}^{(\ell)}_i, \widehat{\Psim}^{(\ell)}}[\bar{\uv}_i \bar{\uv}_i^\herm] (\gamma_i \Psim)^{-1}\right) \\
	&=\trace\left((\underbrace{\bar{\muv}_i^{(\ell)} (\bar{\muv}_i^{(\ell)})^\herm + \bar{\Sigmam}_i^{(\ell)} }_{\triangleq \Rm_i^{(\ell)}})(\gamma_i \Psim)^{-1}\right) \label{eq:E_u_bar},
\end{align}
where using MATLAB index notation, we have
\begin{align}
	\bar{\muv}_i^{(\ell)} &\triangleq [\muv^{(\ell)}_{\widetilde{\uv}}]_{(i-1)N + 1: i N}, \quad i \in [1:Q] \\
	\bar{\Sigmam}^{(\ell)}_i &\triangleq [\Sigmam_{\widetilde{\uv}}^{(\ell)}]_{(i-1)N + 1: i N, (i-1)N + 1: i N}, \quad i \in [1:Q]
\end{align}
Putting \eqref{eq:E_u_bar} back to \eqref{eq:L_ini}, we have the cost function
\begin{align}
	\mathcal{L}^{(\ell)} &= \sum^Q_{i=1} N \log(\gamma_i) + \log \det\left(\Psim\right) + \trace\left(\Rm_i^{(\ell)}(\gamma_i\Psim)^{-1}\right).
\end{align}

In each M-step, the parameters $\gammav$ and $\Psim$ are updated by minimizing the cost function $\mathcal{L}^{(\ell)}$. Specifically, by setting the partial derivative of the cost function with respect to each $\gamma_i$ to zero, we obtain the update rule for $\gammav$, which is given as
\begin{align}
	\frac{\partial \mathcal{L}^{(\ell)}}{\partial \gamma_i} & = \frac{N}{\gamma_i} - \frac{\trace\left(\Rm_i^{(\ell)}\Psim^{-1}\right)}{\gamma_i^2} \overset{!}{=}  0 \\ 
	\Rightarrow \quad \widehat{\gamma}_i^{(\ell + 1)} & = \frac{1}{N}\trace\left(\Rm_i^{(\ell)}(\widehat{\Psim}^{(\ell)})^{-1} \right), \quad i\in[1:Q]. \label{eq:M_step_gamma}
\end{align}
Similarly, by setting the gradient of the cost function over $\Psim$ to a zero matrix, we obtain the update rule for $\Psim$, which is given as
\begin{align}
	\frac{\partial \mathcal{L}^{(\ell)}}{\partial \Psim}
	&= Q \Psim^{-1} - \sum_{i=1}^Q \gamma_i^{-1}\Psim^{-1} \Rm_i^{(\ell)}\Psim^{-1} \overset{!}{=}\mathbf{0}  \\
	&\Rightarrow \quad \widehat{\Psim}^{(\ell +1)}  = \frac{1}{Q} \sum^Q_{i=1} \left(\widehat{\gamma}_i^{(\ell+1)}\right)^{-1}\Rm_i^{(\ell)}. \label{eq:M_step_psi}
\end{align}

The initial point of the hyperparameter vector and correlation matrix can be respectively set to an all-ones vector and identity matrix, i.e., $\widehat{\gammav}^{(0)} = \mathbf{1}$ and $\widehat{\Psim}^{(0)}=\Id_N$.  The E-step and M-step are alternately applied until the predefined maximum number of iterations $G$ is reached, or the stop condition $\|\widehat{\gammav}^{(\ell+1)} - \widehat{\gammav}^{(\ell)}\|/{\|\widehat{\gammav}^{(\ell)}\|} < \epsilon$ is met, where $\epsilon > 0 $ is the predefined positive number that is very close to zero. When the EM algorithm has converged, the estimated reflection coefficients are given in the posterior mean, i.e., $\muv_{\widetilde{\uv}}^{(\ell)}$. To obtain the estimated reflection coefficients at subcarrier $n$, we can extract them from $\muv_{\widetilde{\uv}}^{(\ell)}$ as $\widehat{\rhov}_n = [\text{mat}(\muv_{\widetilde{\uv}}^{(\ell)}, N, Q)]_{n, :}$, where $\text{mat}(\av, A, B)$ is the inverse vectorization function that transforms the vector $\av$ into a matrix with dimension $A\times B$.
The complete SBL-based approach is given in Algorithm~\ref{alg:SBL-EM}.

\begin{algorithm}[t]
	\caption{SBL-EM for 3D Multiview Imaging}\label{alg:SBL-EM}
	\begin{algorithmic}[1] 
		\State{Initialize $\widehat{\gammav}^{(0)} = \mathbf{1}$, $\widehat{\Psim}^{(0)} = \Id_N$}
		\For{$\ell=0,\dots,G$} 
		\State{\textbf{E-step:} update posterior covariance $\Sigmam_{\widetilde{\uv}}^{(\ell)} \gets$ \eqref{eq:poster_cov}} 
		\State{\textbf{E-step:} update posterior mean $\muv_{\widetilde{\rhov}}^{(\ell)} \gets$ \eqref{eq:poster_mean}}
		\State{\textbf{M-step:} update hyperparameters $\widehat{\gammav}^{(\ell+1)} \gets$ \eqref{eq:M_step_gamma}}
		\State{\textbf{M-step:} update correlation $\widehat{\Psim}^{(\ell+1)} \gets$ \eqref{eq:M_step_psi}}
		\If{$\|\widehat{\gammav}^{(\ell+1)} - \widehat{\gammav}^{(\ell)}\|/{\|\widehat{\gammav}^{(\ell)}\|} < \epsilon$} 
		\State {Break for-loop}
		\EndIf
		\EndFor
		\State{\Return$\widehat{\rhov}_n = \left[\text{mat}(\muv_{\widetilde{\uv}}^{(\ell)}, N, Q)\right]_{n, :}$}
	\end{algorithmic}
\end{algorithm}

\section{Numerical Results}\label{section_simulation}
In this section, we present numerical results to validate the effectiveness of the proposed algorithms. We first consider the indoor scenario where the objective is to image a small target with high resolution, for which the RMA-based algorithm is used. Then, we consider the outdoor scenario where the objective is to reconstruct the large-scale 3D environment with coarse resolution, for which the SBL-based algorithm is used.

\subsection{Indoor Small-Scale Imaging}
\begin{figure*}[t]
	\centering
	\begin{minipage}{0.24\linewidth}
		\centering
		\includegraphics[width=\linewidth]{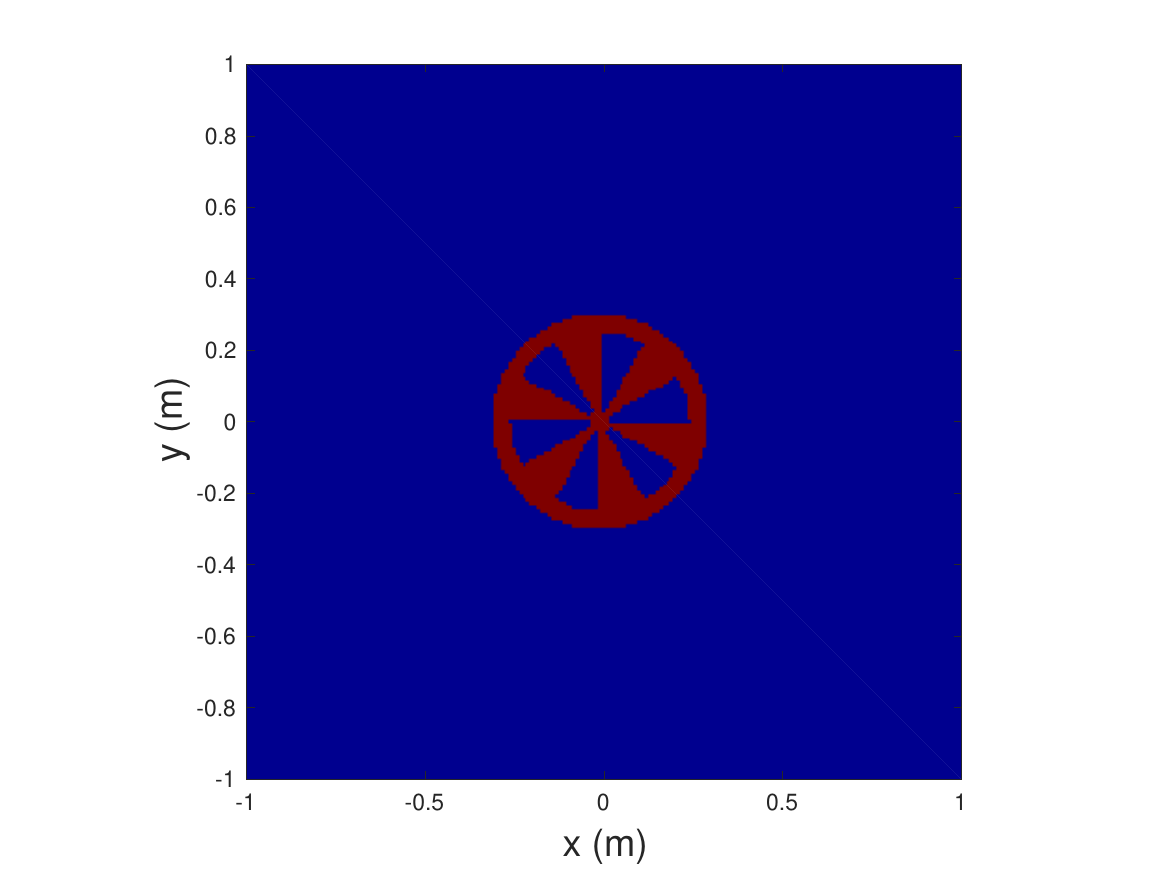}
		\centerline{(a) Spatial Domain}
	\end{minipage}
	\begin{minipage}{0.25\linewidth}
	\centering
	\includegraphics[width=\linewidth]{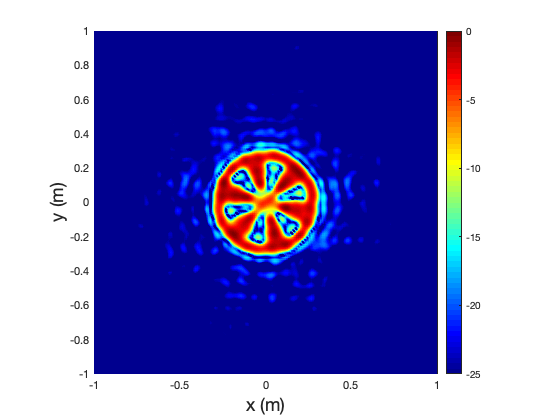}
	\centerline{(c) $P=30$ dBm}
\end{minipage}
		\begin{minipage}{0.25\linewidth}
		\centering
		\includegraphics[width=\linewidth]{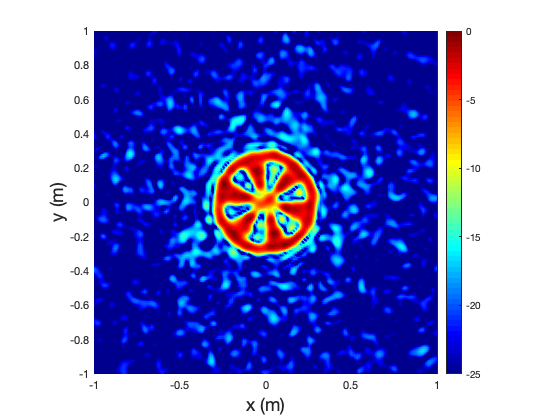}
		\centerline{(e) $P=20$ dBm}
	\end{minipage}
	\begin{minipage}{0.24\linewidth}
	\centering
	\includegraphics[width=\linewidth]{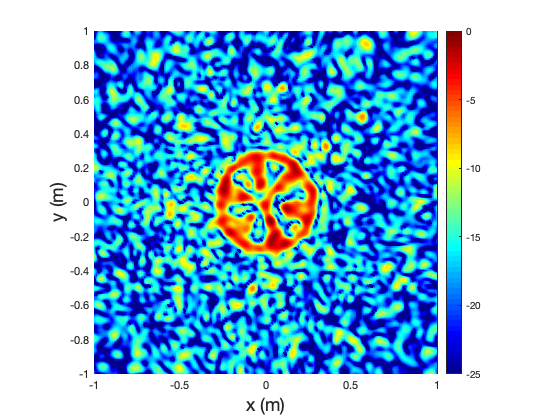}
	\centerline{(g) $P=10$ dBm}
\end{minipage}
	\begin{minipage}{0.24\linewidth}
		\centering
		\includegraphics[width=\linewidth]{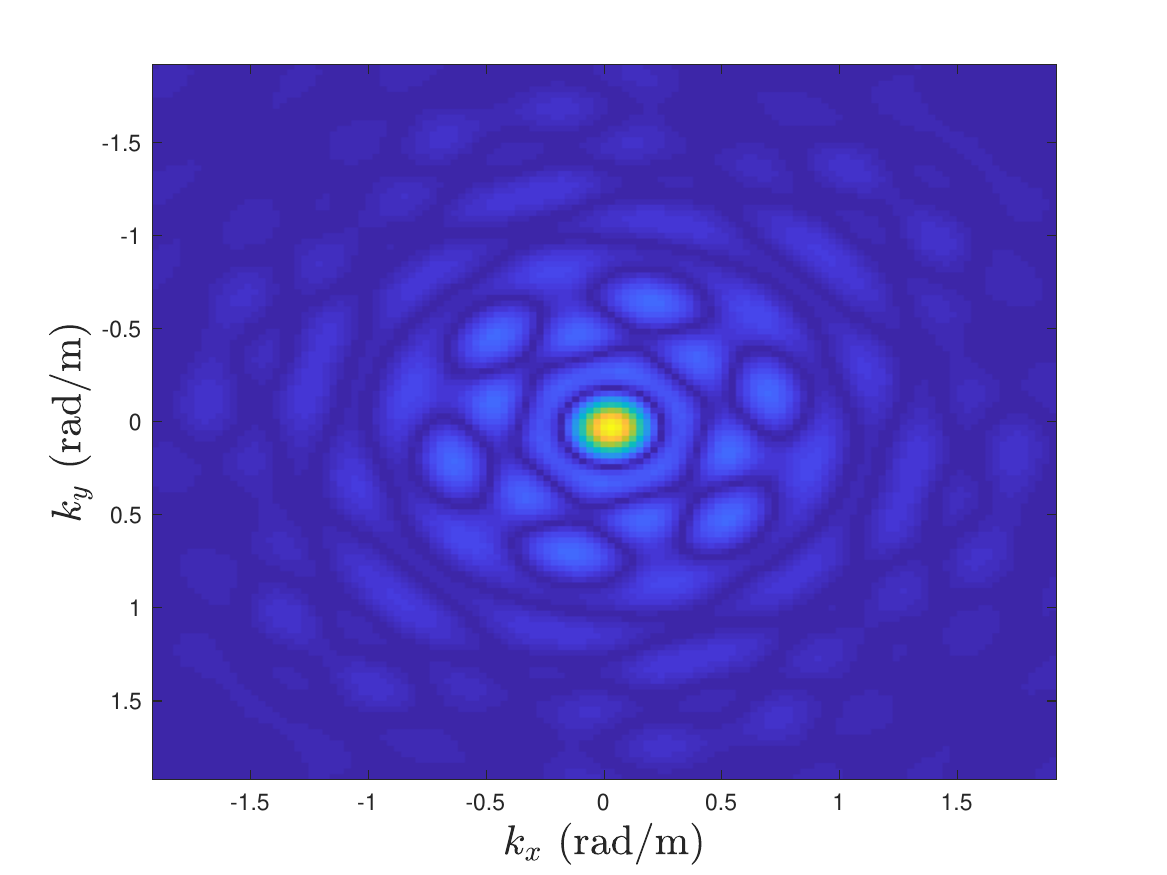}
		\centerline{(b) Wavenumber Domain}
	\end{minipage}
	\begin{minipage}{0.24\linewidth}
		\centering
		\includegraphics[width=\linewidth]{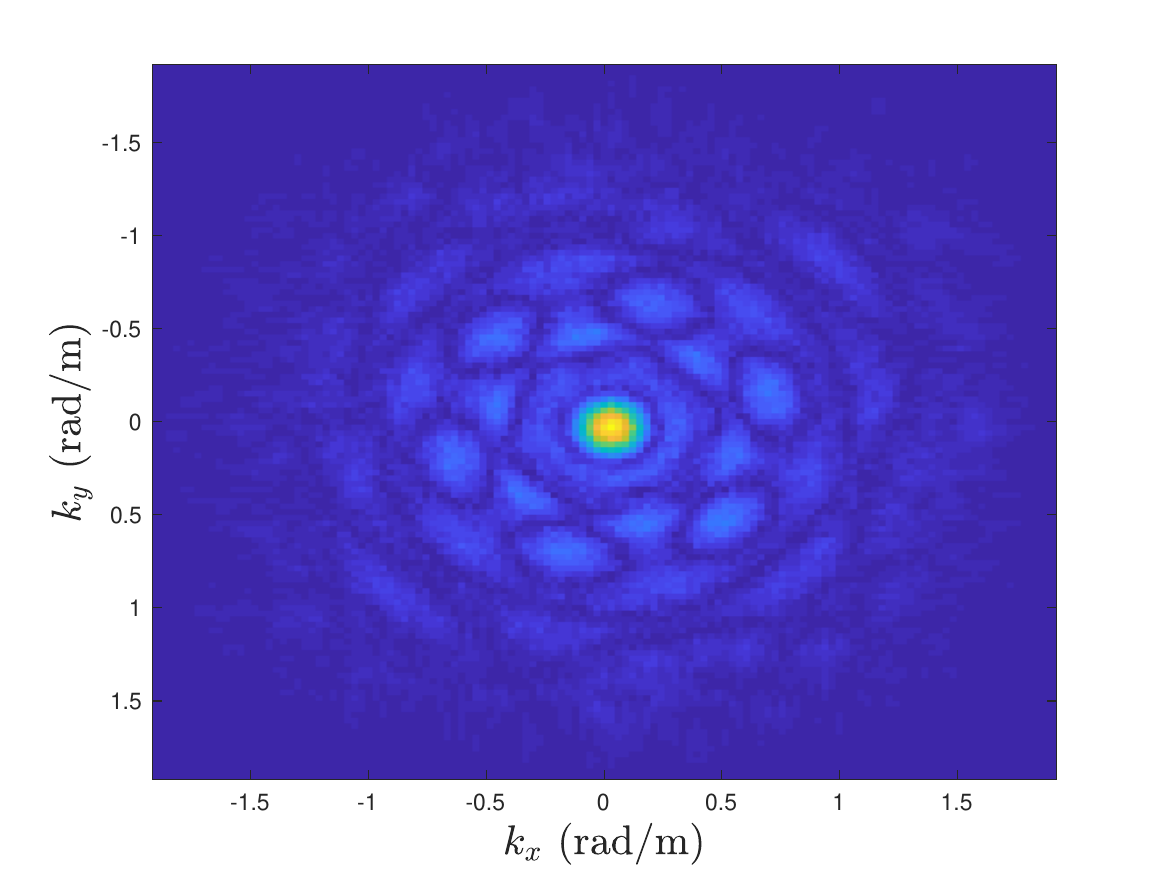}
		\centerline{(d) $P=30$ dBm}
	\end{minipage}
	\begin{minipage}{0.24\linewidth}
		\centering
		\includegraphics[width=\linewidth]{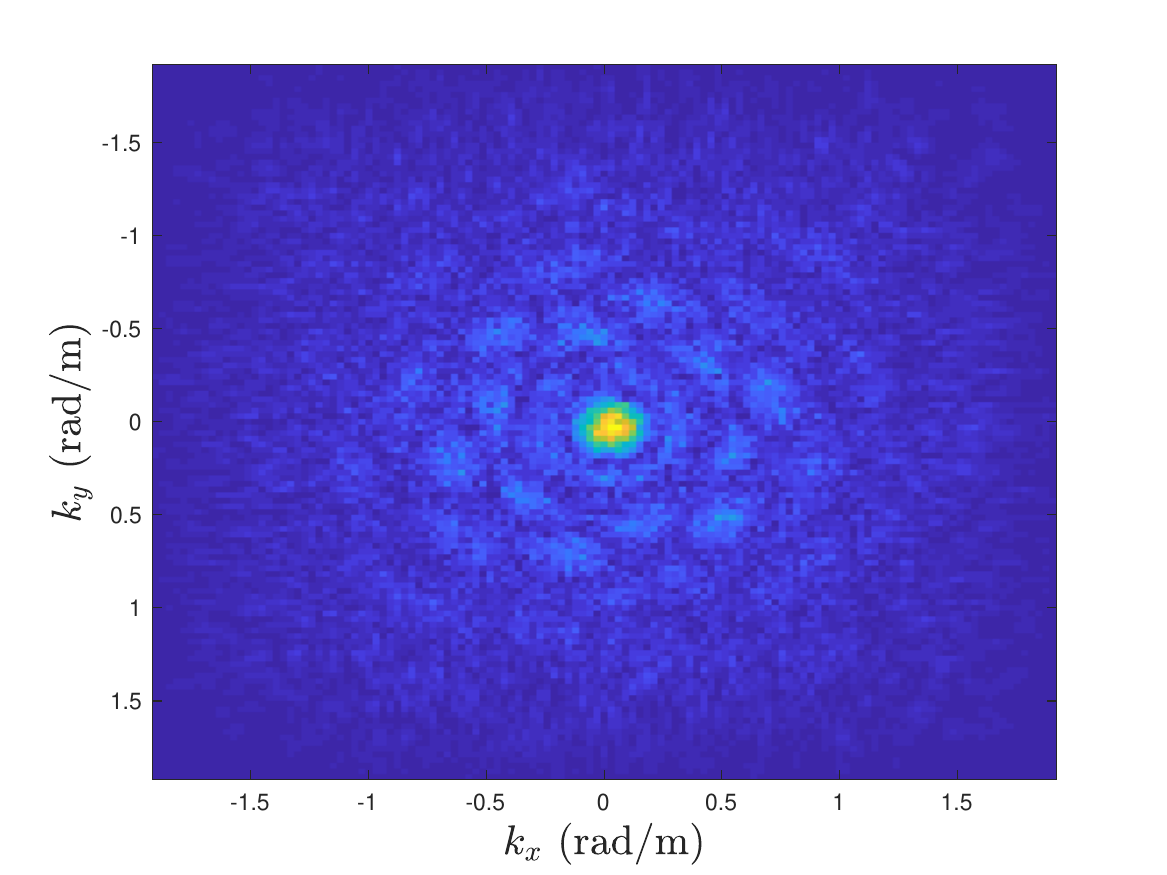}
		\centerline{(f) $P=20$ dBm}
	\end{minipage}
	\begin{minipage}{0.24\linewidth}
		\centering
		\includegraphics[width=\linewidth]{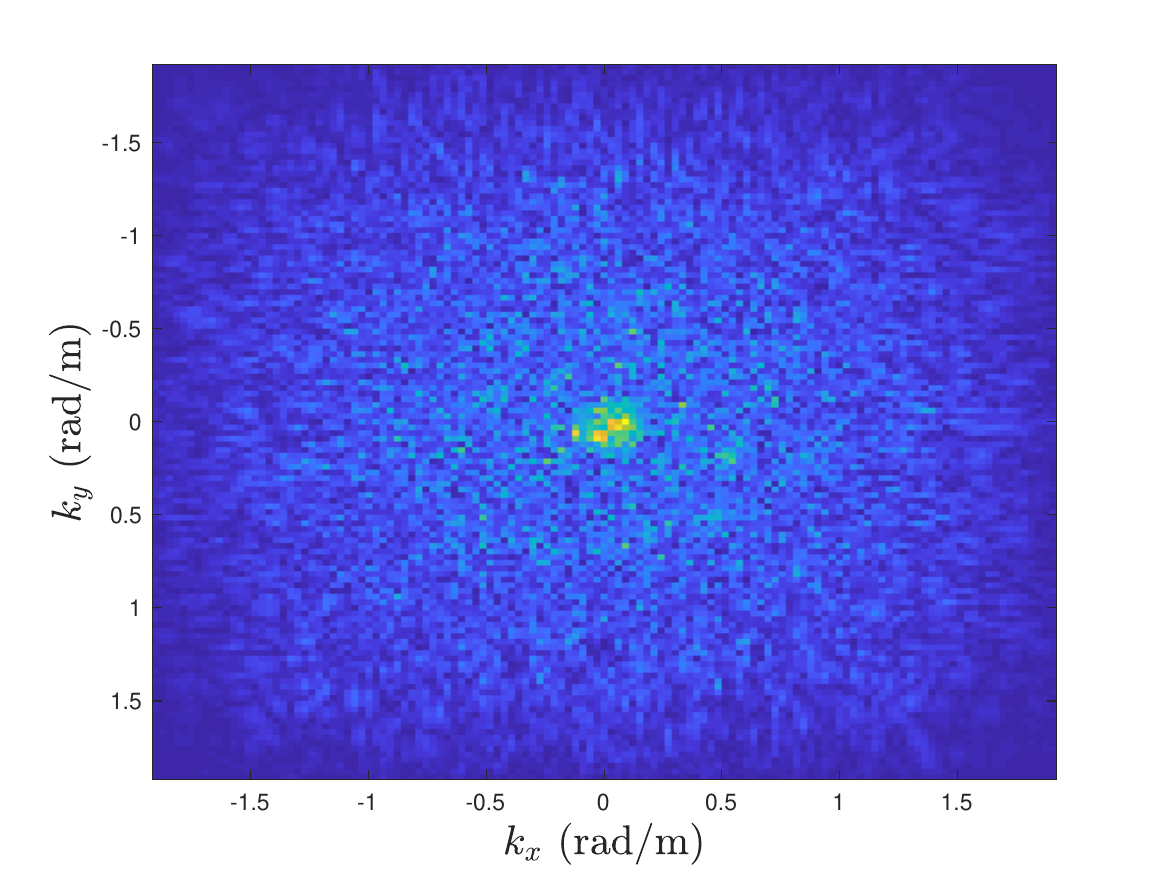}
		\centerline{(h) $P=10$ dBm}
	\end{minipage}
	\caption{Imaging for Siemens star pattern with boundary array under $M_l=60$, $M_w=4$, $\xi=4$. (a) True space-domain Image. (b) True wavenumber-domain image. (c)  Space-domain imaging result, $P=30$ dBm (d) Wavenumber-domain imaging result, $P=30$ dBm. (e)  Space-domain imaging result, $P=20$ dBm.   (f)    Wavenumber-domain imaging result, $P=20$ dBm.  (g)  Space-domain imaging result, $P=10$ dBm.     (h)    Wavenumber-domain imaging result, $P=10$ dBm.     }
	\label{figi1}
\end{figure*}

For indoor small-scale imaging, we first consider a Siemens star pattern as shown in Fig. \ref{figi1} (a), which is traditional in testing the imaging resolution. The object is comprised of pixels of size $1$ cm $\times$ $1$ cm. The $x$- and $y$-axes length of the star pattern is $0.8$ m. Unless otherwise stated, the boundary array is considered with $M_l=60$ and $M_w=4$ and with antenna spacing $\xi   \times  {\frac{\lambda_c}{2}}$ where $\xi = 4$. The variance of noise is set as $-50$ dBm, and the transmit power is $P=30$ dBm. The central carrier frequency is $f_c=10$ GHz. The imaging distance is $z=10$ m.

\subsubsection{Impact of SNR} In Fig. \ref{figi1}, we illustrate the imaging results based on RMA for a Siemens star pattern under different SNRs.  In (a) and (b), the ground truth images of the target in the spatial domain and wavenumber domain are shown, respectively. Then, at high SNR, it can be seen that the RMA-based algorithm with distributed MIMO effectively reconstructs the image with clear edges and hollow regions. The estimated shape and strength are quite similar to the true target, although we notice the appearance of some weak artifacts. Meanwhile, the data matrix obtained in the wavenumber domain involves rich information, which guarantees the effective reconstruction in the spatial domain. By reducing the power of the transmitted signals, it can be seen that more artifacts appear in the imaging result and the wavenumber-domain image becomes more and more noisy. Nevertheless, the scheme still yields an image of reasonably good quality and good agreement with the original shape of the object.

\begin{figure}[t]
	\centering
	\begin{minipage}[b]{0.49\linewidth}
		\centering
		\includegraphics[width=\linewidth]{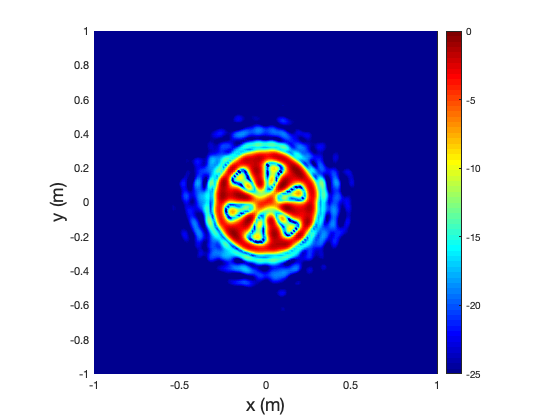}
		\centerline{(a) Boundary Array $M_w=8$}
	\end{minipage}
	\begin{minipage}[b]{0.49\linewidth}
		\centering
		\includegraphics[width=\linewidth]{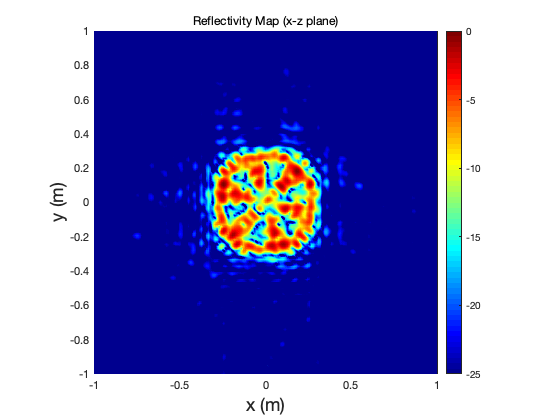}
		\centerline{(b) Boundary Array $M_w=2$}
	\end{minipage}
	\vspace{1mm}
	\begin{minipage}[b]{0.49\linewidth}
		\centering
		\includegraphics[width=\linewidth]{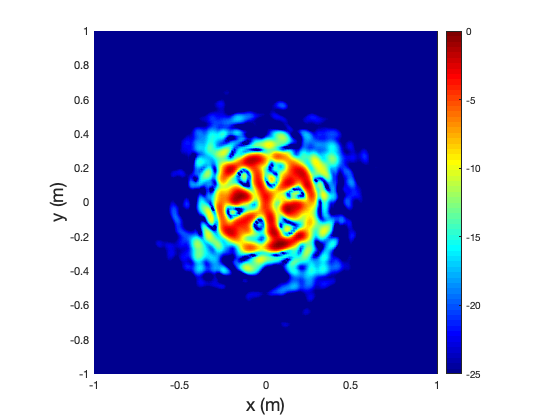}
		\centerline{(c) Distributed-Boundary Array}
	\end{minipage}
	\begin{minipage}[b]{0.49\linewidth}
		\centering
		\includegraphics[width=\linewidth]{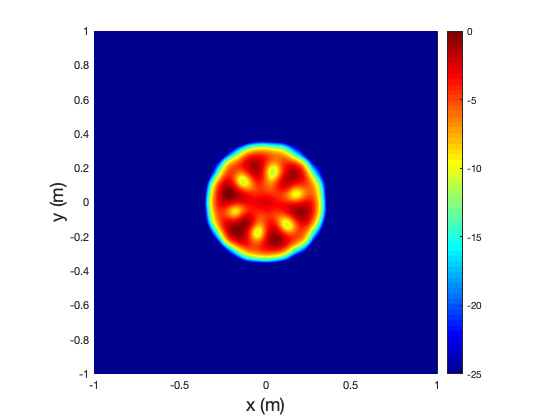}
		\centerline{(d) Full Array}
	\end{minipage}
	\vspace{2mm}
	\caption{Imaging with different array architectures. (a) Boundary array, $M_l=56$, $M_w=8$, $\xi = 4$. (b)  Boundary array, $M_l=62$, $M_w=2$, $\xi = 4$. (c)   Distributed-boundary array, $M_l=56$, $M_w=8$, $\xi = 4$, $\tau=1/3$. (d) Full array, $M_t=M_r=32\times 32$, $\xi = 8$.}
	\label{figi2}
\end{figure}

\subsubsection{Impact of array architecture}  In Fig. \ref{figi2}, we investigate the imaging performance using different array architectures. It can be seen that compared with the boundary array in (a) with wide UPAs, the boundary array in (b) with narrow UPAs yields an image of some speckle-like artifacts, due to the limited number of sampling antennas. For (c), it can be seen that the proposed distributed boundary array could also achieve a good imaging result that is consistent with the true shape, but with some stronger artifacts surrounding the true pattern. Finally, in (d), a full array with $32\times32$ transmitting and receiving antennas is adopted. In this case, it can be seen that with a large number of spatial samples, the imaging result is very clear with almost zero artifacts. However, it has lower resolution even though it has the same aperture as the boundary array by increasing the antenna spacing from $\xi=4$ to $\xi=8$. These observations demonstrate the effectiveness of the proposed imaging algorithm based on sparse distributed  MIMO.

\begin{figure*}[t]
	\centering
	\begin{minipage}{0.24\linewidth}
		\centering
		\includegraphics[width=\linewidth]{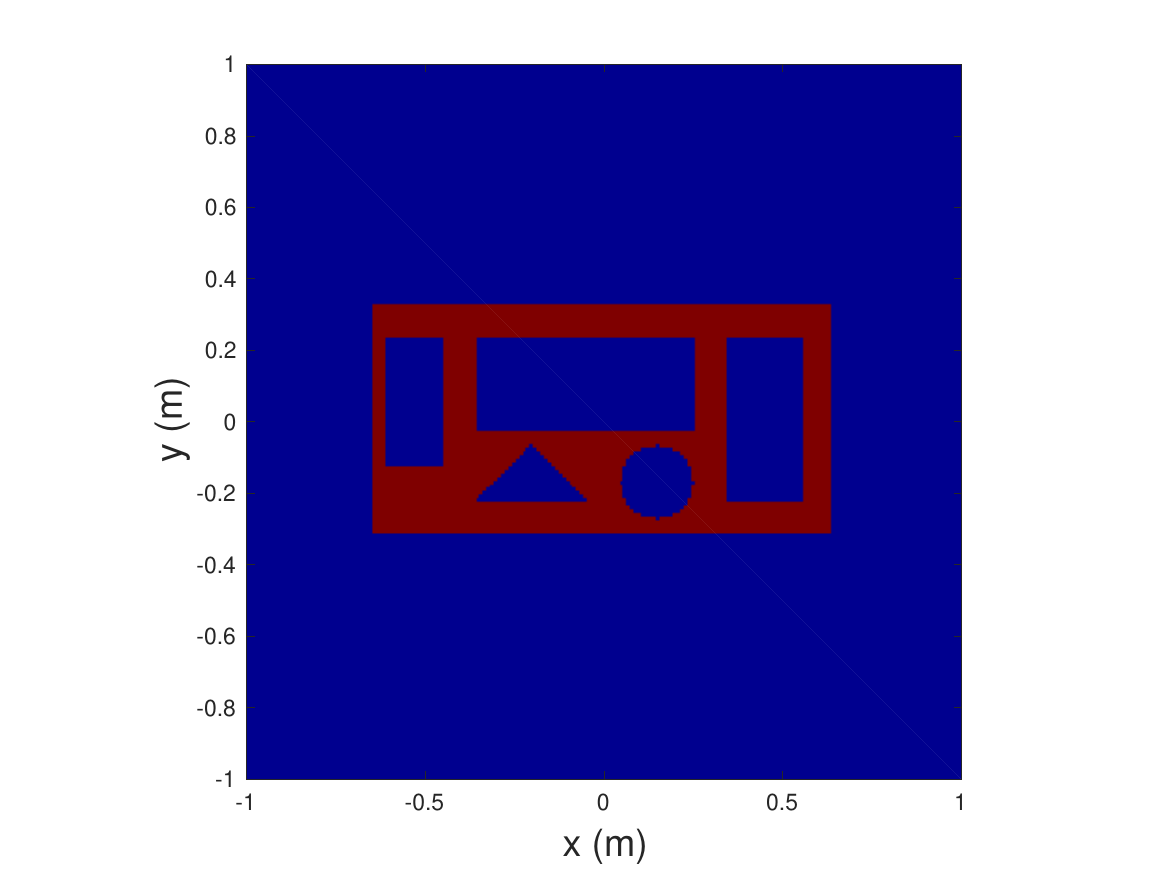}
		\centerline{(a)}
	\end{minipage}
	\begin{minipage}{0.25\linewidth}
	\centering
	\includegraphics[width=\linewidth]{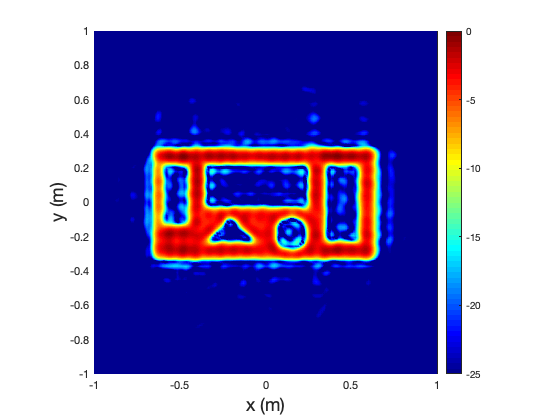}
	\centerline{(c) $z=10$ m }
\end{minipage}
	\begin{minipage}{0.25\linewidth}
	\centering
	\includegraphics[width=\linewidth]{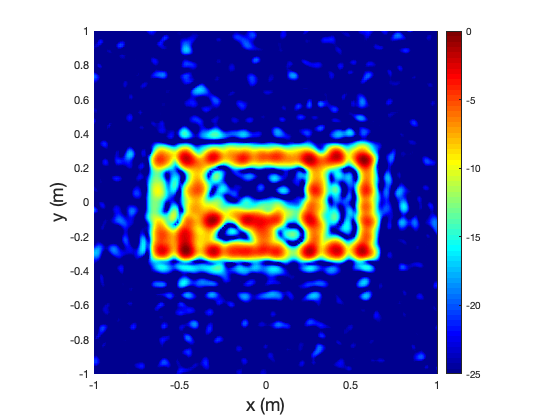}
	\centerline{(e) $z=20$ m }
\end{minipage}
	\begin{minipage}{0.24\linewidth}
	\centering
	\includegraphics[width=\linewidth]{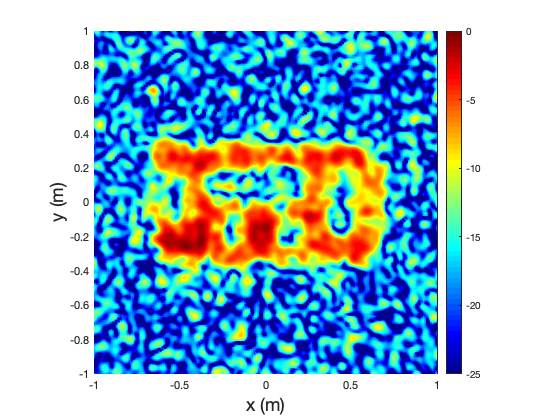}
	\centerline{(g) $z=30$ m }
\end{minipage}
	\begin{minipage}{0.24\linewidth}
		\centering
		\includegraphics[width=\linewidth]{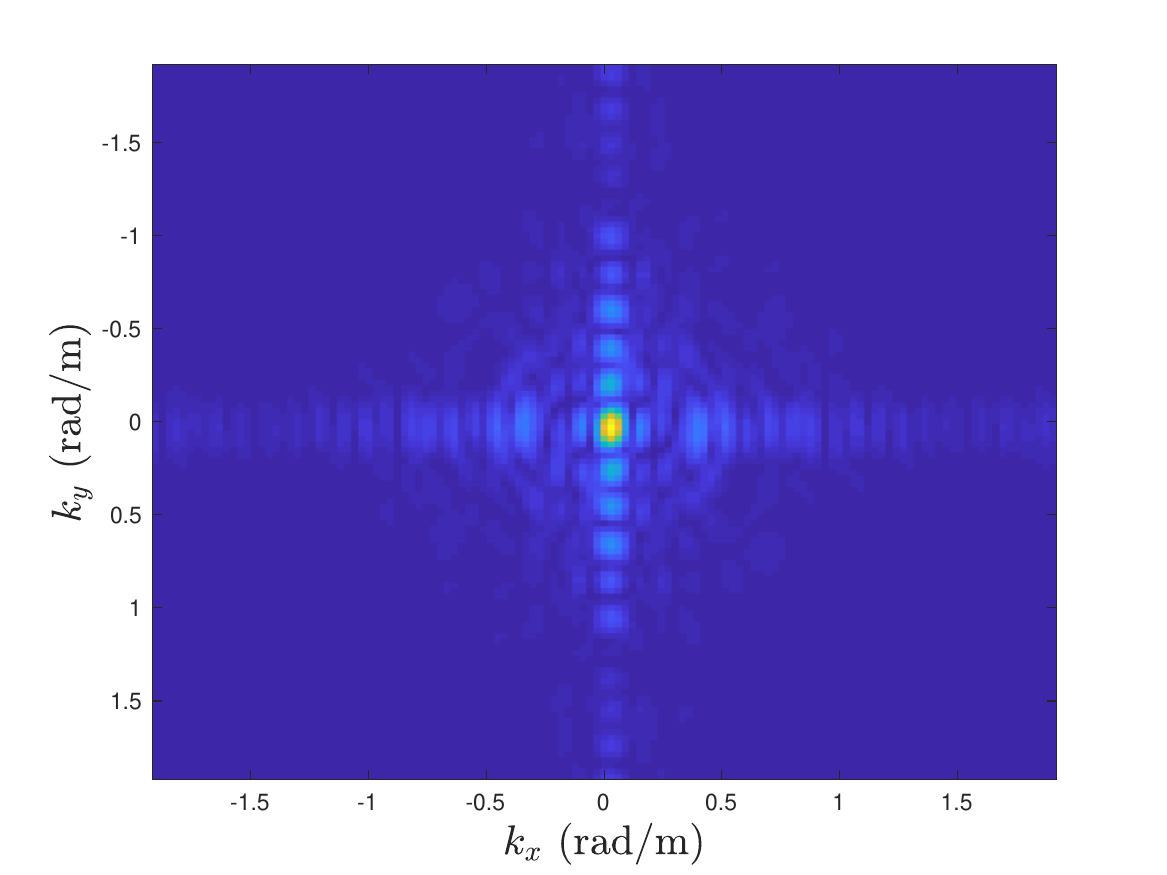}
		\centerline{(b)}
	\end{minipage}
	\begin{minipage}{0.24\linewidth}
		\centering
		\includegraphics[width=\linewidth]{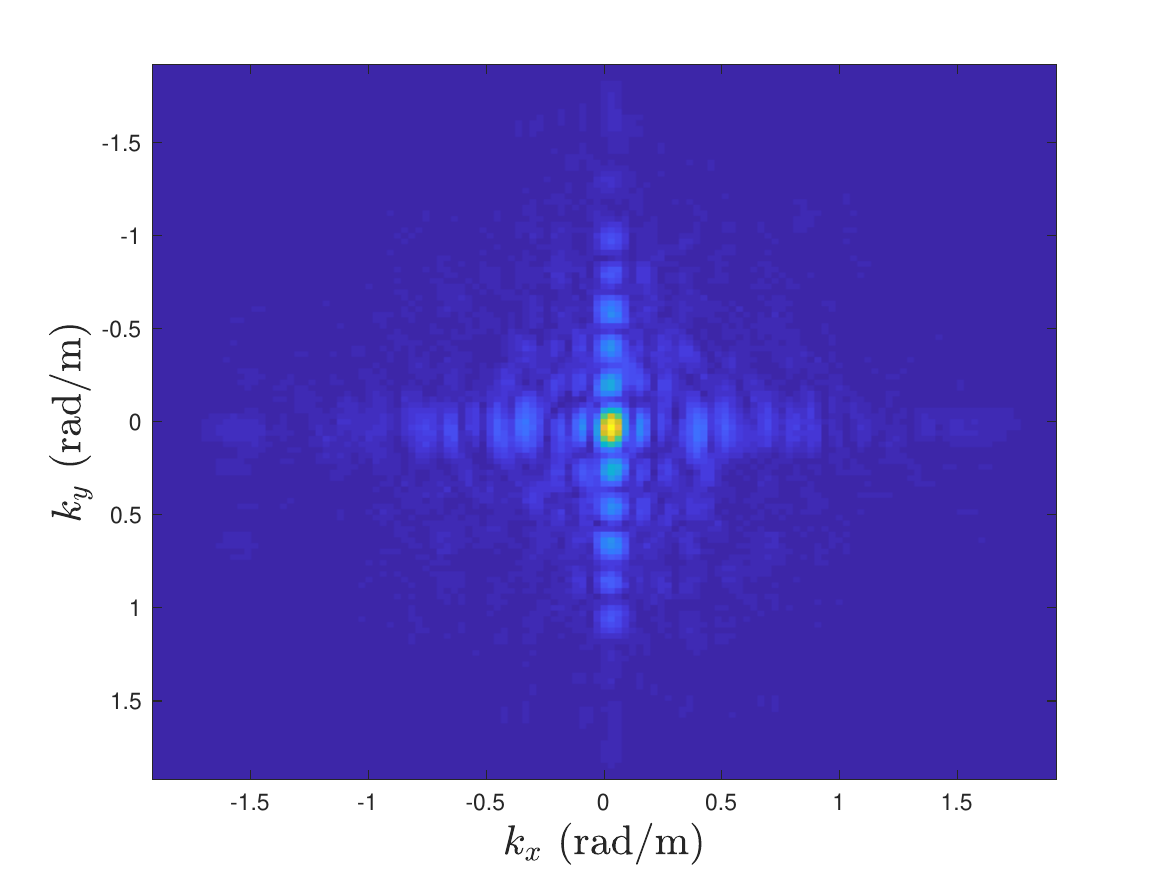}
		\centerline{(d) $z=10$ m }
	\end{minipage}
	\begin{minipage}{0.24\linewidth}
		\centering
		\includegraphics[width=\linewidth]{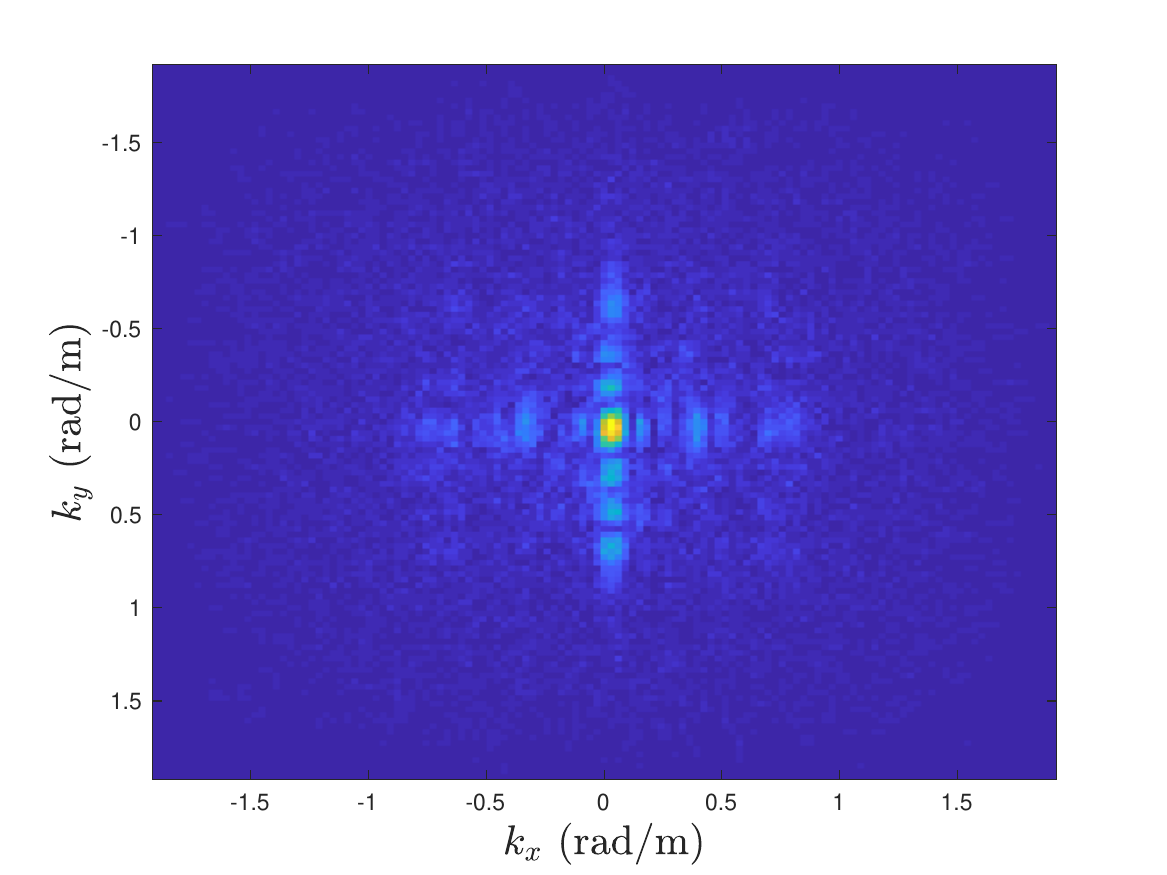}
		\centerline{(f)  $z=20$ m  }
	\end{minipage}
	\begin{minipage}{0.24\linewidth}
		\centering
		\includegraphics[width=\linewidth]{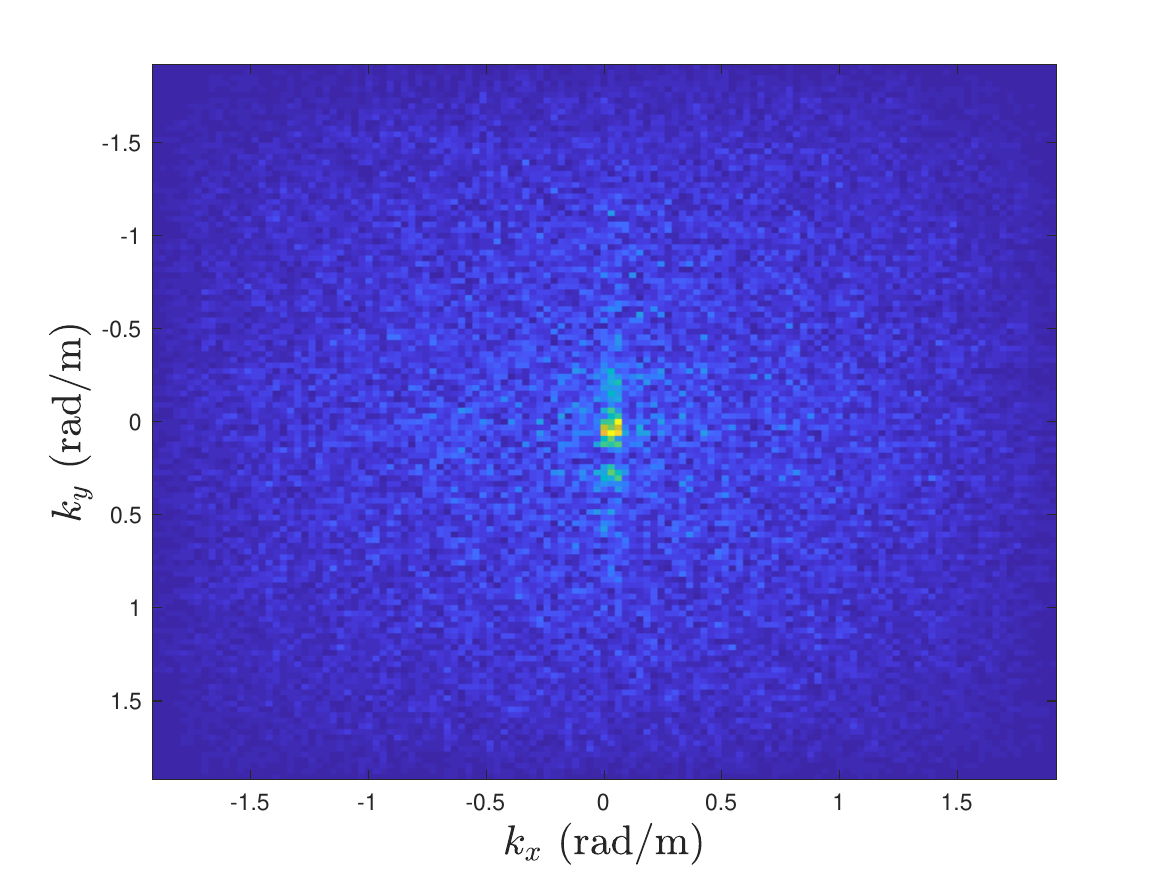}
		\centerline{(h) $z=30$ m }
	\end{minipage}
	\caption{Imaging for a rectangular-shape target with boundary array under $M_l=60$, $M_w=4$, $\xi=4$. (a) 2-D True space-domain Image. (b) True wavenumber-domain image. (c)  Space-domain imaging result, $z=10$ m (d) Wavenumber-domain imaging result, $z=10$ m. (e)  Space-domain imaging result, $z=20$ m.   (f)    Wavenumber-domain imaging result, $z=20$ m.  (g)  Space-domain imaging result, $z=30$ m.     (h)    Wavenumber-domain imaging result, $z=30$ m.     }
	\label{figi3}
\end{figure*}

\subsubsection{Impact of imaging distance} In Fig. \ref{figi3}, we investigate the impact of imaging distance on the RMA algorithm. We consider a rectangular-shaped target with some hollowing, as shown in Fig. \ref{figi3} (a). It is comprised of pixels of size $1$ cm $\times$ $1$ cm. The number of pixels of this target is $128$ and $64$ in the $x$- and $y$-axies, respectively. Thus, its length is $128$ cm and its width is $64$ cm. It can be seen from Fig. \ref{figi3} that, by increasing the distance from $z=10$ m to $z=30$ m, the imaging quality becomes worse. The reason is two-fold. First, as the distance increases, the path loss of the channel becomes more and more severe, thus reducing the  imaging SNR. Second, as the distance increases, the near-field effect becomes weaker and weaker and the diversity of parameter features of different channel elements becomes lower. This can be validated by observing the estimated image in the wavenumber domain. As the distance increases, the rich information in the wavenumber domain gradually collapses into a single point, since the near-field spherical wavefront tends to become a   far-field planar wavefront \cite{zhi2024performance}, thereby degrading the imaging quality.
These results also coincide with the theoretical resolution (\ref{reso}), which is proportional to the imaging distances.

\begin{figure*}
	\begin{minipage}[t]{0.33\linewidth}
	\centering
	\includegraphics[width=2.6in]{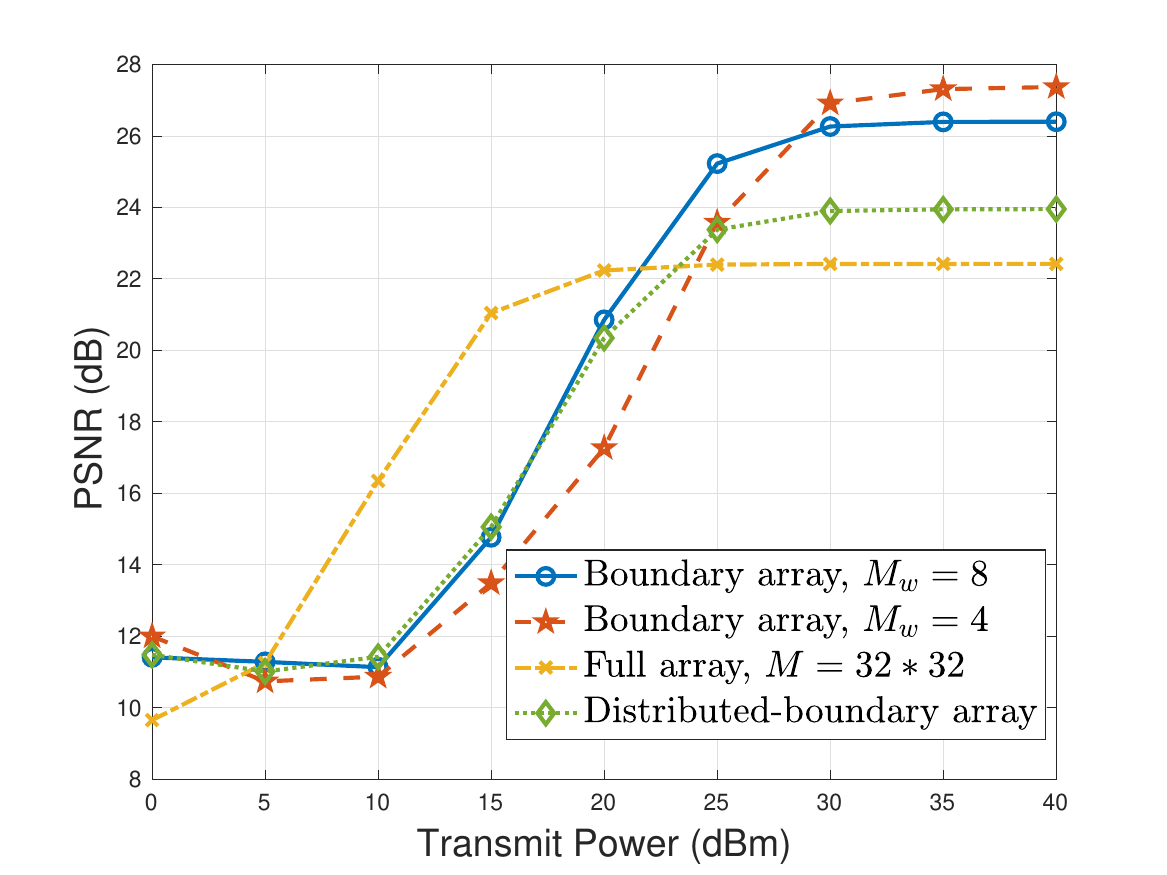}
			\captionsetup{justification=centering}
	\caption{PSNR performance versus $P$.}
	\label{figi5}
	\end{minipage}%
	\begin{minipage}[t]{0.33\linewidth}
	\centering
	\includegraphics[width=2.6in]{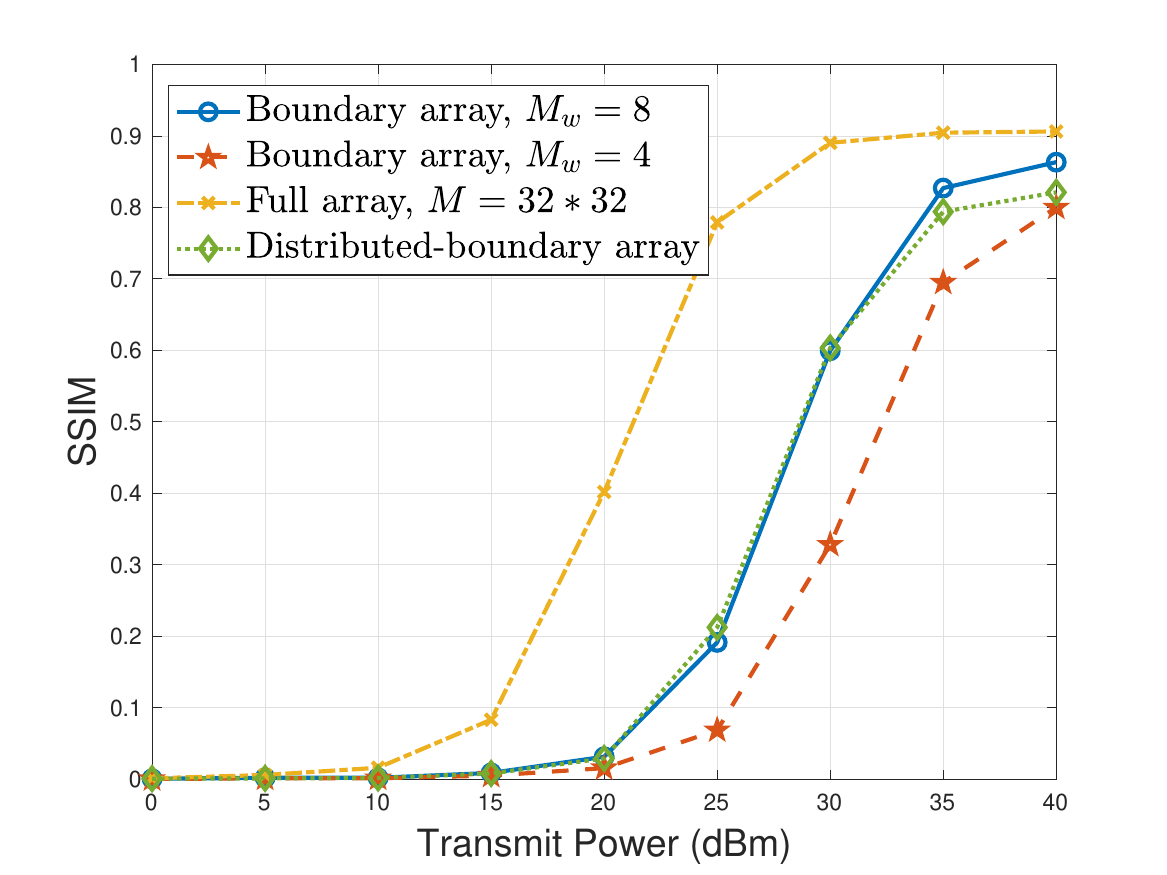}
			\captionsetup{justification=centering}
	\caption{SSIM performance versus $P$.}
	\label{figi6}
	\end{minipage}%
	\begin{minipage}[t]{0.33\linewidth}
	\centering
	\includegraphics[width=2.6in]{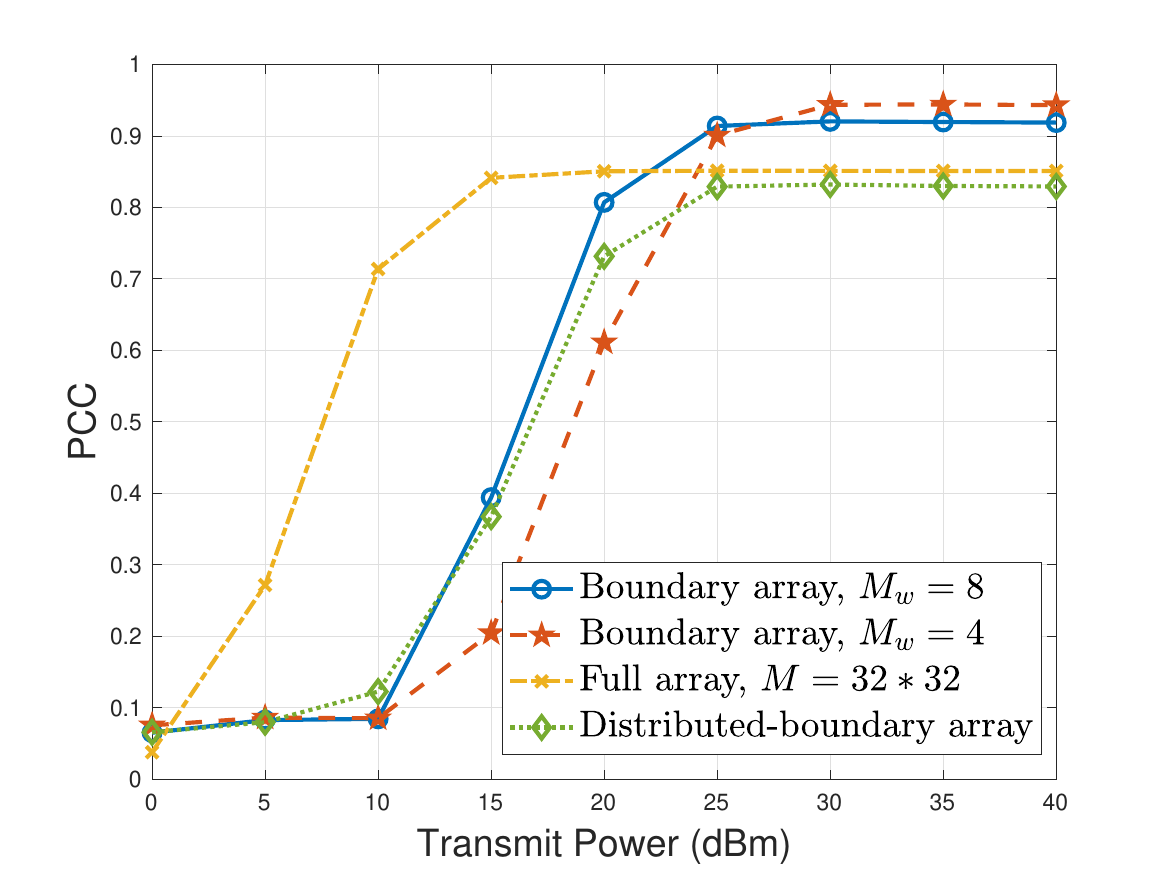}
	\captionsetup{justification=centering}
	\caption{PCC performance versus $P$.}
	\label{figi4}
\end{minipage}%
\end{figure*}

%
%
%
%
%

\subsubsection{Quantitative evaluation} Finally, we showcase the quantitative evaluation of imaging results in terms of  PSNR,  SSIM, and PCC in Figs. \ref{figi5},   \ref{figi6},  and \ref{figi4}, respectively. It can be seen that the full array has advantages in the low SNR regime, while the boundary array performs well in the high SNR regime. Meanwhile, it can be seen that the proposed distributed-boundary array, which has a small number of  antennas, can still achieve good performance, especially for SSIM. Besides, all arrays realize high PCCs, which implies that the imaging results have a similar shape to the true target. Furthermore, it can be observed that the boundary array with narrow UPAs (i.e., $M_w=4$) is sufficient to achieve a good imaging performance, which demonstrates its potential in reducing hardware cost.

\subsection{Outdoor Large-Scale Imaging}
\begin{figure*}[t]
	\centering
	\begin{minipage}[b]{0.24\linewidth}
		\centering
		\includegraphics[width=\linewidth]{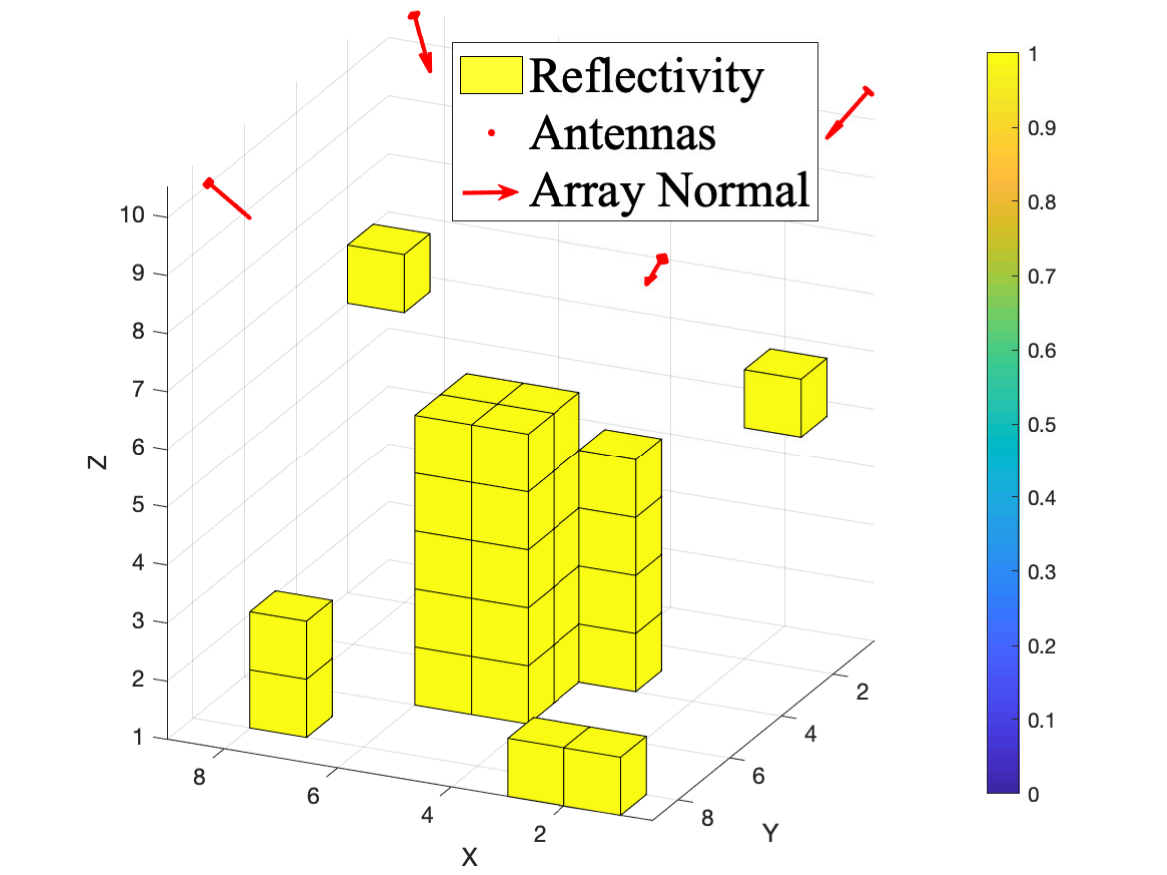}
		{(a) Ground Truth,  $n=1$}
	\end{minipage}
	\begin{minipage}[b]{0.24\linewidth}
		\centering
		\includegraphics[width=\linewidth]{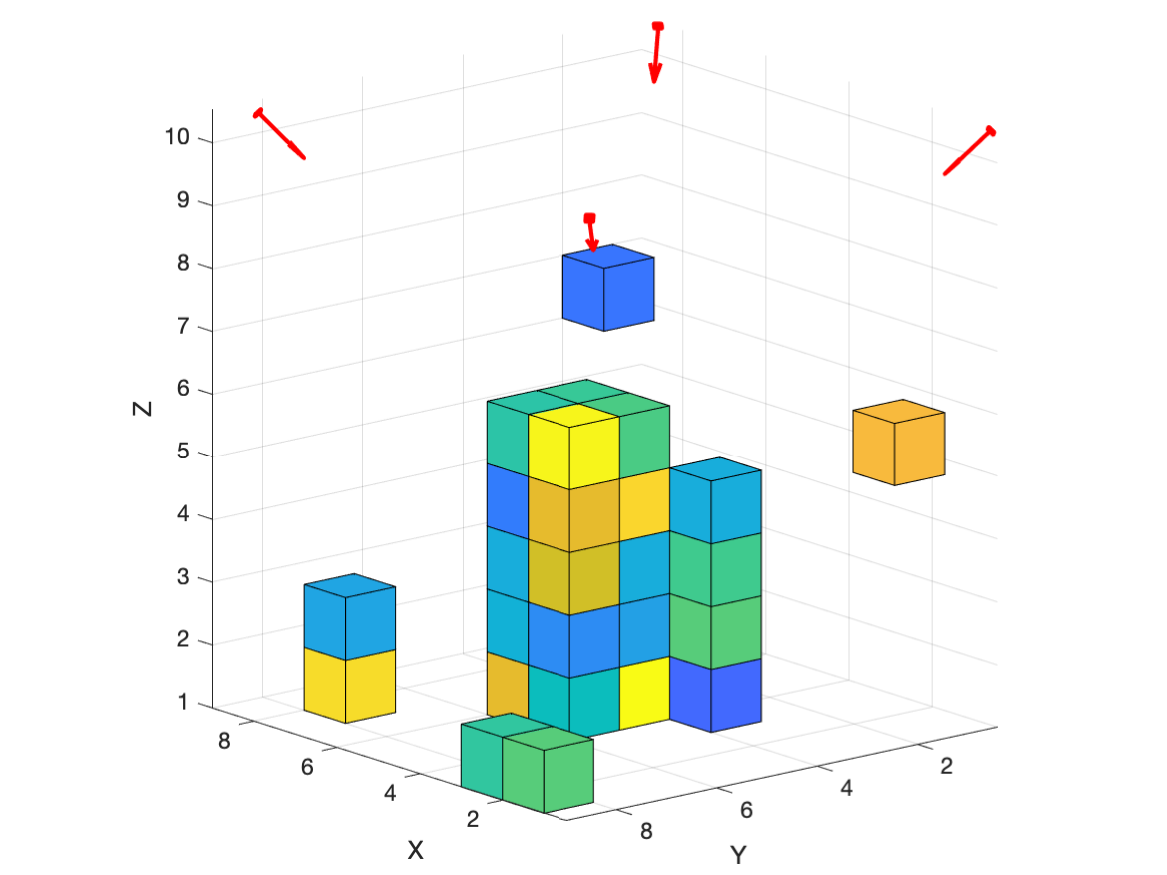} 
		{(c) Ground Truth,  $n=2$}
	\end{minipage}
	\begin{minipage}[b]{0.24\linewidth}
		\centering
		\includegraphics[width=\linewidth]{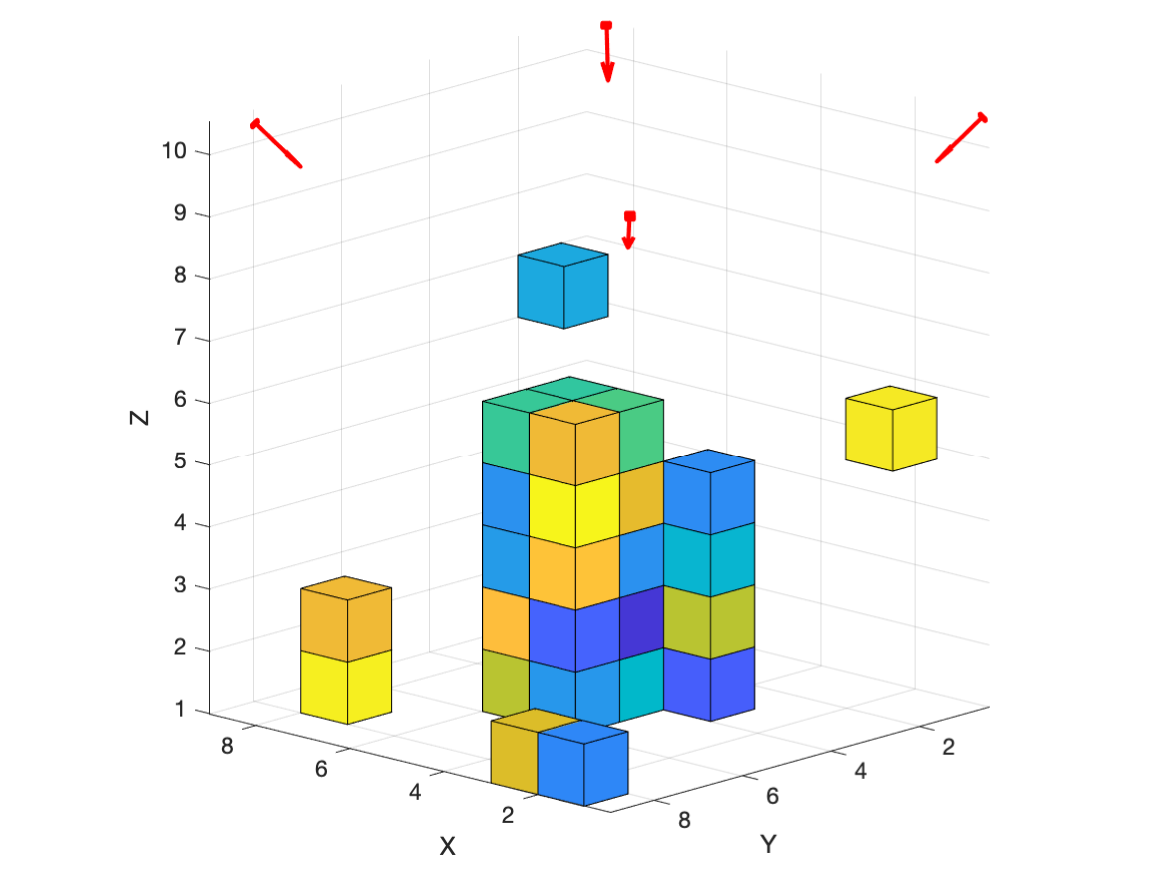} 
		(e) Ground Truth,  $n=4$
%
	\end{minipage}
	\begin{minipage}[b]{0.24\linewidth}
		\centering
		\includegraphics[width=\linewidth]{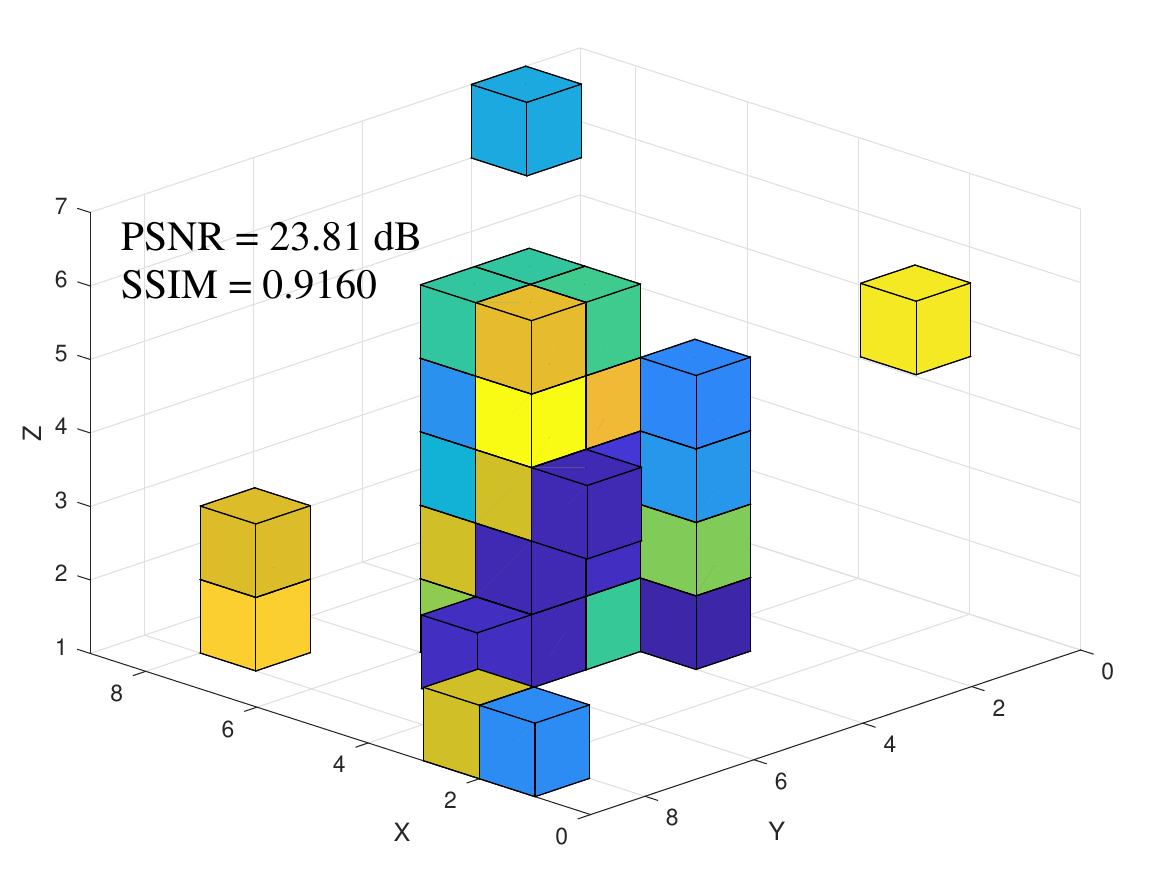}  
		{(g) Est.,  $n=4$,  $M=16$ }
	\end{minipage}
	\begin{minipage}[b]{0.24\linewidth}
		\centering
		\includegraphics[width=\linewidth]{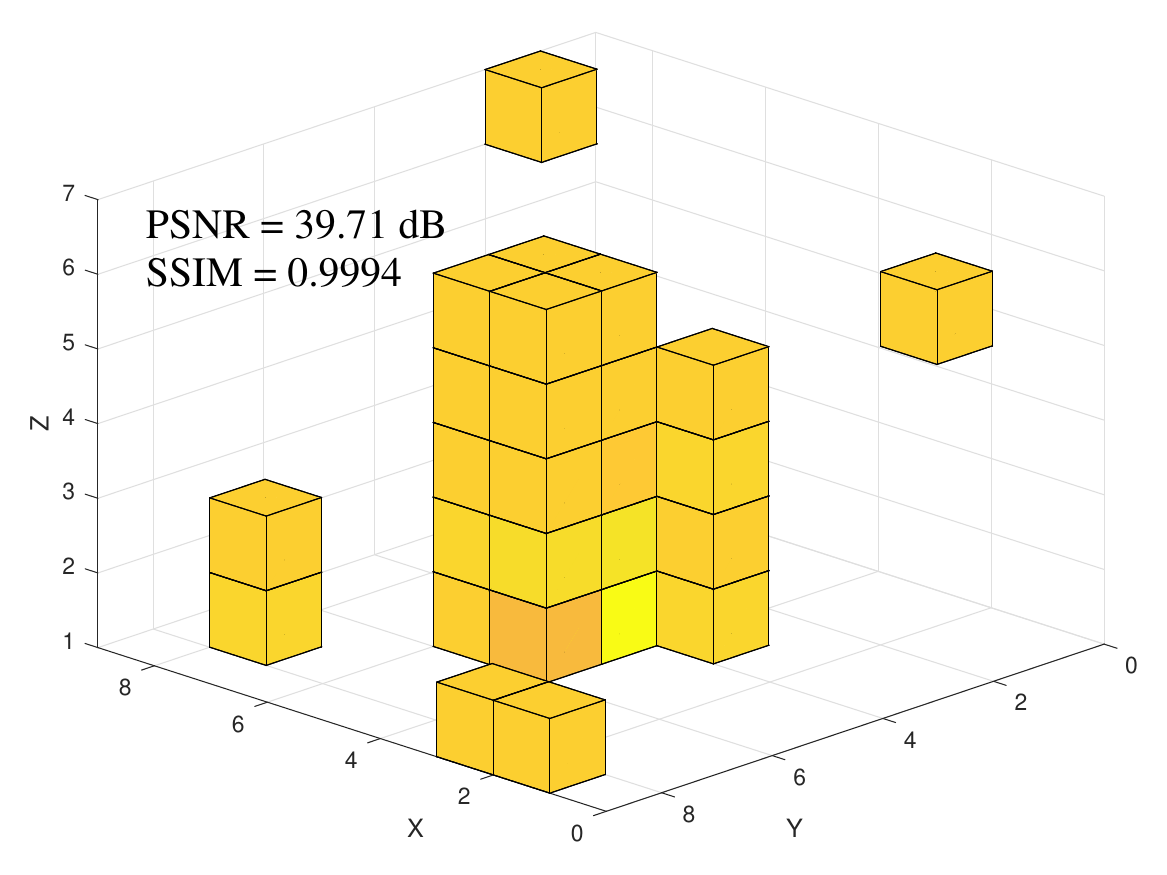}  
		 {(b) Est.,  $n=1$}
	\end{minipage}
	\begin{minipage}[b]{0.24\linewidth}
		\centering
		\includegraphics[width=\linewidth]{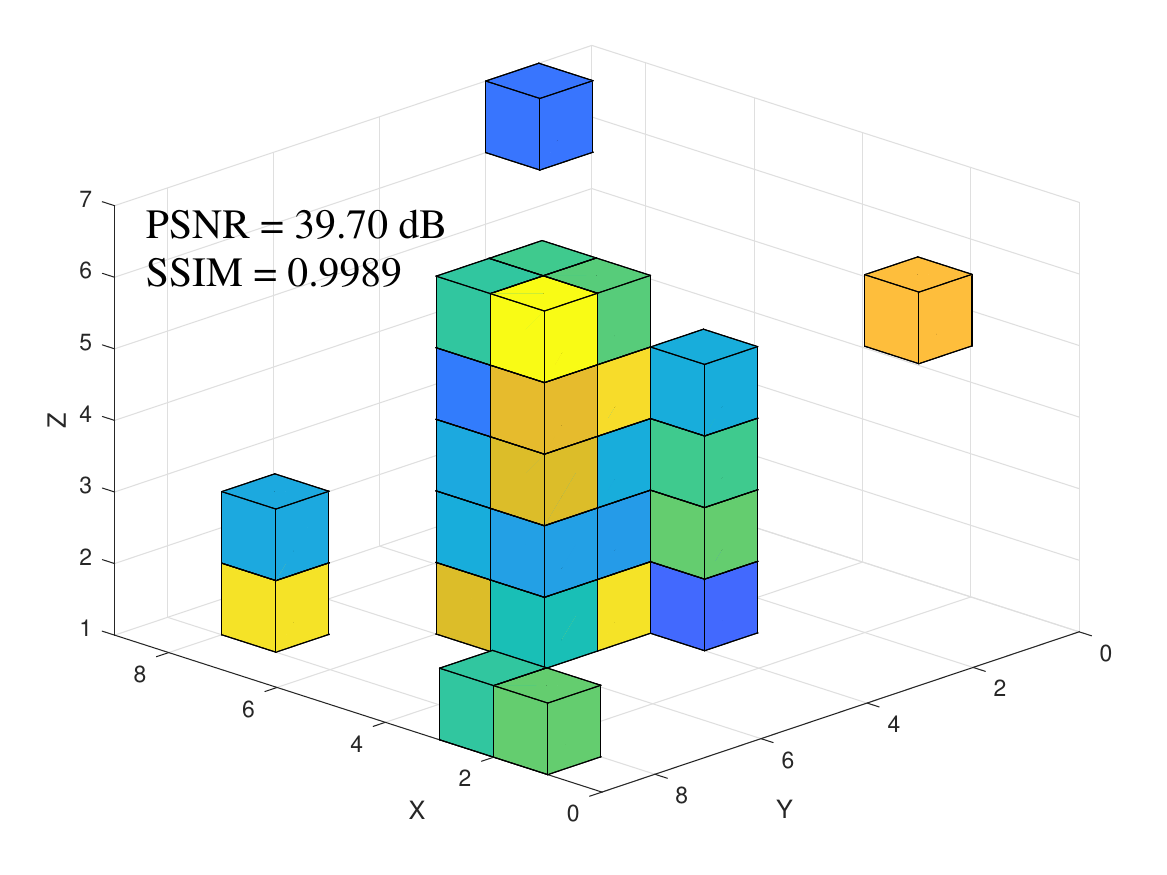}
		{(d) Est.,  $n=2$}
	\end{minipage}
	\begin{minipage}[b]{0.24\linewidth}
	\centering
	\includegraphics[width=\linewidth]{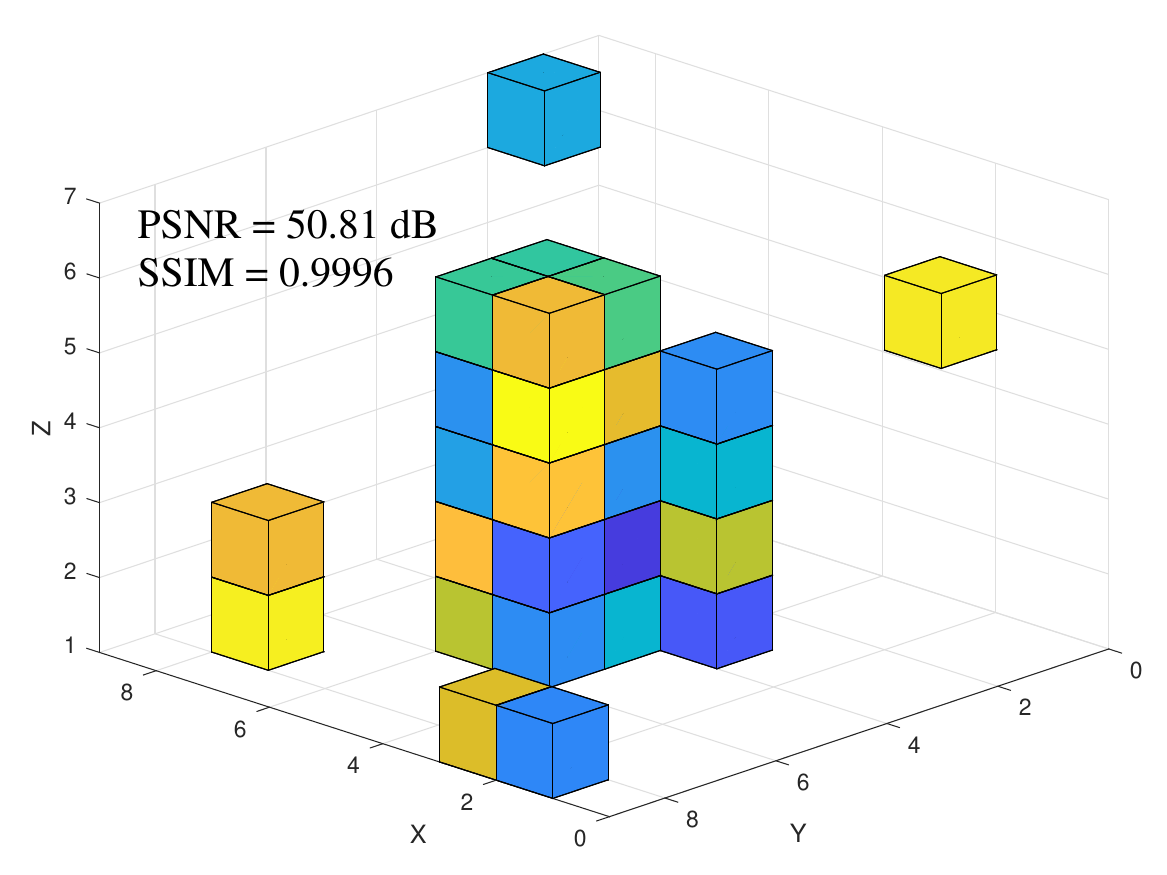}  
	{(f) Est.,  $n=4$}
\end{minipage}
\begin{minipage}[b]{0.24\linewidth}
	\centering
	\includegraphics[width=\linewidth]{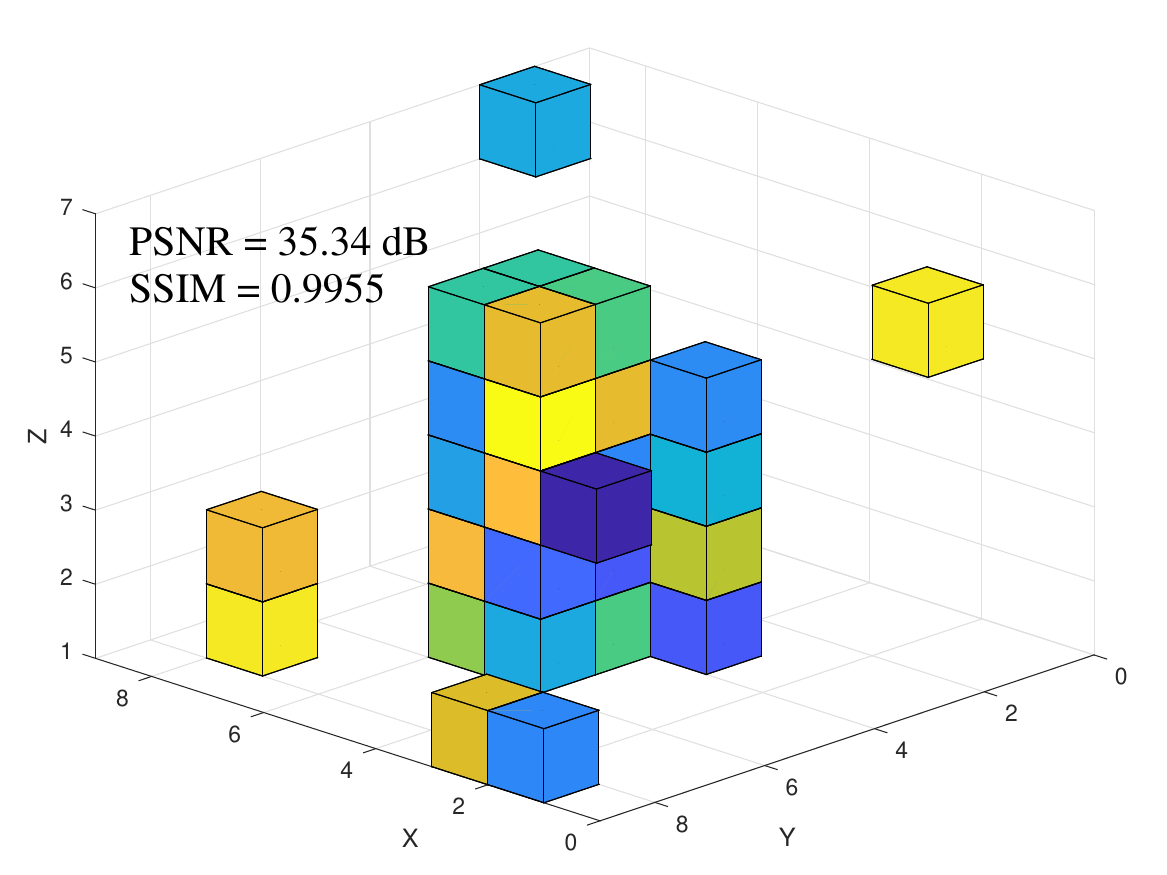}
	\centerline{(h) Est.,  $n=4$, $M=16$, $\xi=2$}
\end{minipage}
\vspace{2mm}
	\caption{Imaging under reflectivity correlation among subcarriers. 
		(a) 3-D True Image in the $n=1$ subcarrier. (b) Imaging result in subcarrier $1$ under $M=64$, $\xi=1$. 
		(c)  3-D True Image in the $n=2$ subcarrier. (d)  Imaging result in subcarrier $2$ under $M=64$, $\xi=1$. 
	   (e)  3-D True Image in the $n=4$ subcarrier. (f)  Imaging result in subcarrier $4$ under $M=64$, $\xi=1$.
	   (g)    Imaging result in subcarrier $4$ under $M=16$, $\xi=1$. (h)   Imaging result in subcarrier $4$ based on sparse array, with $M=16$, $\xi=2$.
	}
	\label{figo1}
\end{figure*}

In this subsection, we consider the outdoor scenario with large-scale 3D targets. Unless otherwise stated, each voxel is of size $1$ m $\times$ $1$ m $\times$ $1$ m. The area of interest is of $10$ m $\times$ $10$ m $\times$ $10$ m, which is discretized into $10$   $\times$ $10$   $\times$ $10$ grids. We consider $4$ distributed UPAs with a downtilt of $  45^{o}$ located at the four vertices of the interested area with a height of $z=10$ m, as illustrated in Fig. \ref{figo1} (a). The number of antennas on each UPA is $8\times 8$ and the antenna spacing is $\xi=1$ times of half wavelength. The central carrier frequency is $10$ GHz and the number of subcarriers is $N=10$ with subcarrier spacing $2$ MHz. The transmitting power is $P=30$ dBm and the covariance of noise is $\sigma^2=-50$ dBm. We consider $S=4$ time slots, in which the four  RUs  take turns acting as receivers, while the remaining three RUs transmit random pilot signals in each slot and on each subcarrier under the power constraint (\ref{power_constraint}).

In Fig. \ref{figo1}, we examine the SBL-based imaging algorithm for the frequency-selective 3D image. We use the first-order autoregressive (AR-1) process to model the reflection coefficients at different subcarriers, resulting in a Toeplitz correlation matrix $\Psim = {\sf Toeplitz}([1, \psi, \dots,  \psi^{N-1}])$, where $\psi = 0.9$ is the correlation coefficient between every two neighboring subcarriers.  In this case, we consider $N=4$ subcarriers where the true 3D images on the first, second, and fourth subcarriers are shown in Figs. \ref{figo1} (a), (c), and (e), respectively. It can be seen that the proposed SBL algorithm effectively solves the MMV problems, yielding very accurate image estimation (i.e., (b), (d), and (f)) for all subcarriers. Their high values of PSNR and SSIM also coincide with the subjective feeling. In (g), we reduce the number of antennas on each UPA from $8\times 8$ to $4\times 4$. Then, it can be seen that the imaging of the 3D reflectivity shows deviation from the true image, although the estimated shape is still of high precision. In (h), we increase the antenna spacing from $\frac{\lambda_c}{2}$ to $\frac{2\lambda_c}{2}$ and form a sparse array with the same aperture as that used in (f). In this case, it can be seen that the estimation quality of image reflectivity is improved significantly, compared to the small compact array used in (g). This is because the sparse array has a larger resolution, as discussed in (\ref{reso}), compared to the compact array. Nevertheless, due to the sparse antenna spacing, slight artifacts exist in (h) caused by the aliasing effect.

\begin{figure*}[t]
	\centering
	\begin{minipage}[b]{0.24\linewidth}
		\centering
		\includegraphics[width=\linewidth]{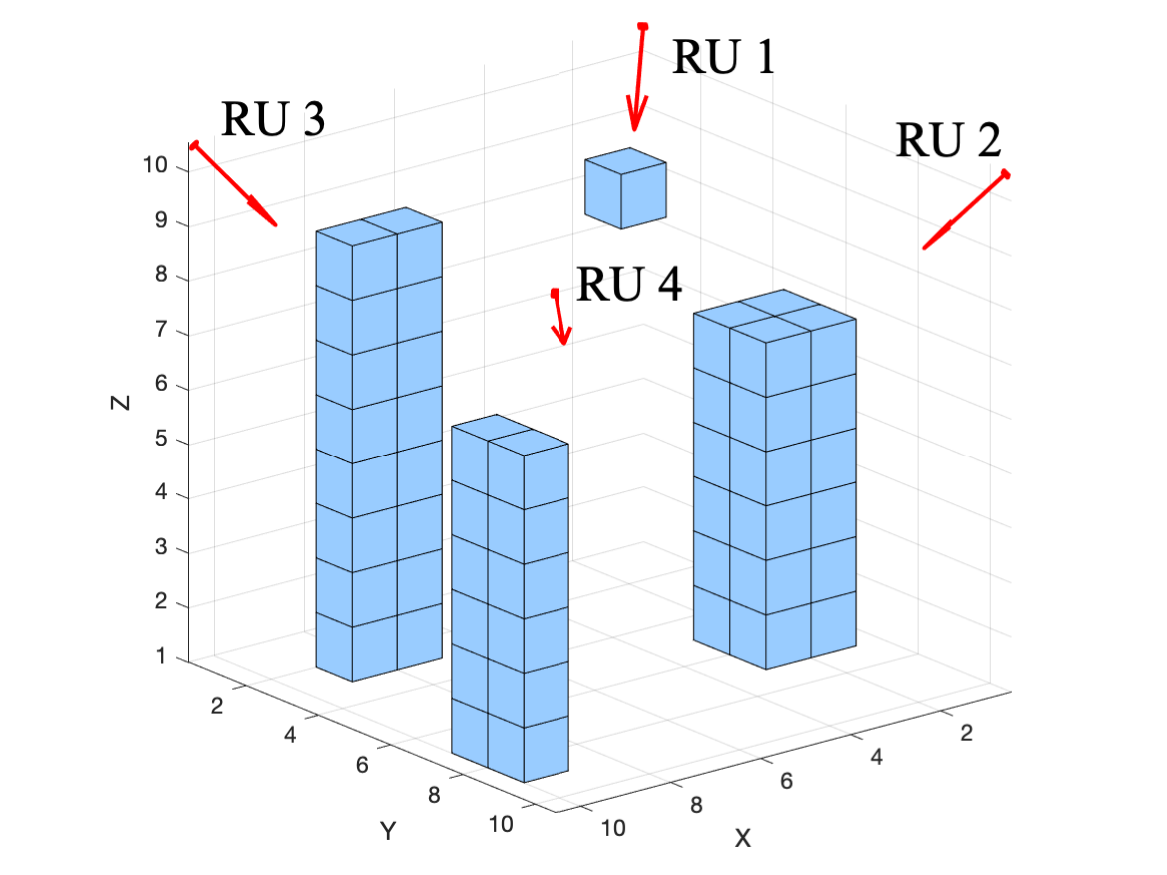}
		 {(a) Ground Truth}
	\end{minipage}
	\begin{minipage}[b]{0.24\linewidth}
		\centering
		\includegraphics[width=\linewidth]{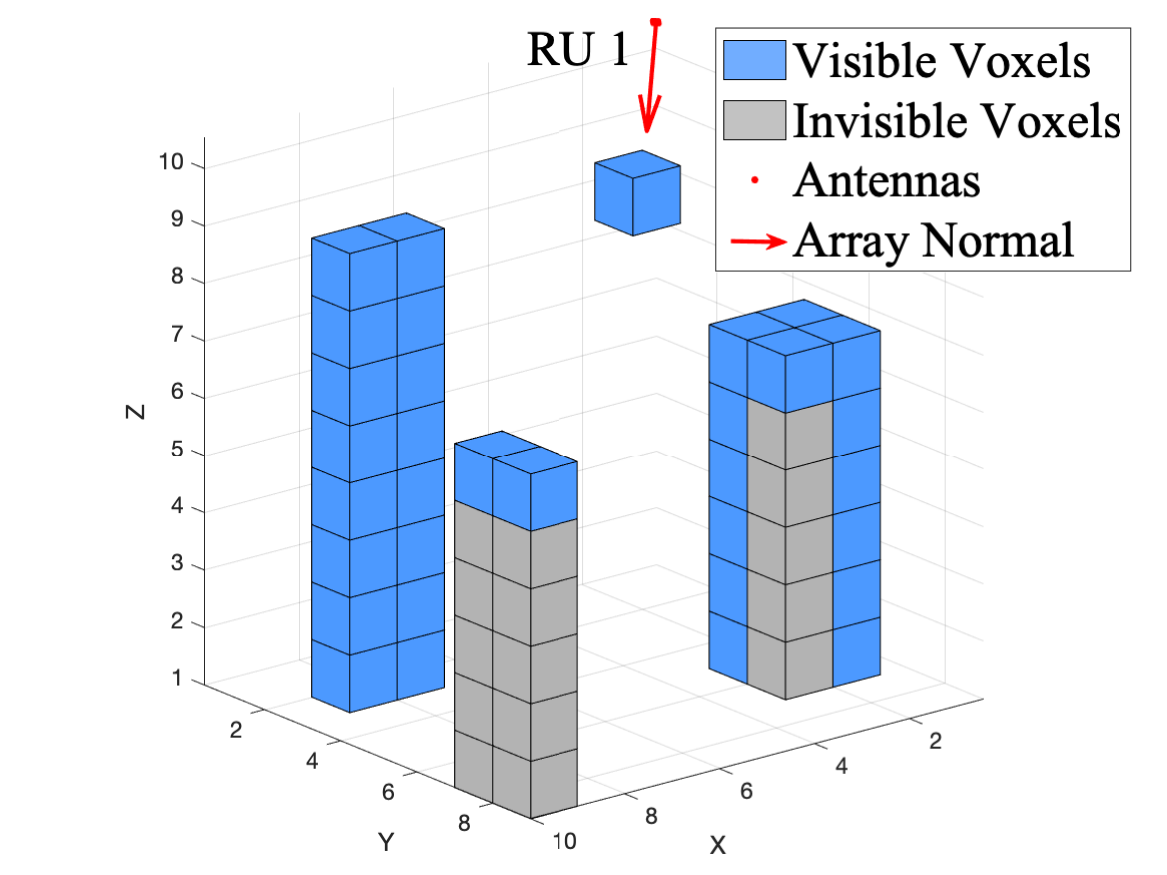} 
		\centerline{(c) Single view from RU $1$}
	\end{minipage}
	\begin{minipage}[b]{0.24\linewidth}
		\centering
		\includegraphics[width=\linewidth]{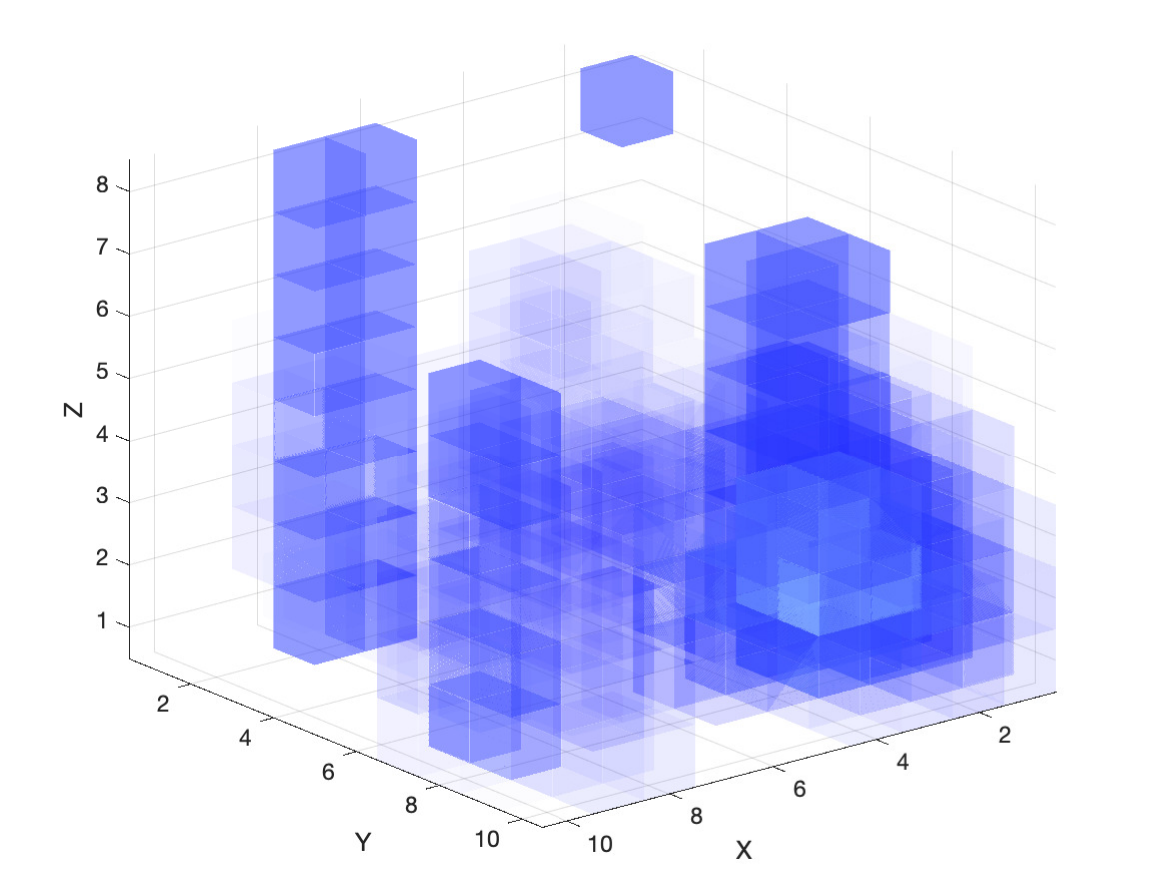} 
		 {(e) Multi-view LS}
	\end{minipage}
	\begin{minipage}[b]{0.24\linewidth}
		\centering
		\includegraphics[width=\linewidth]{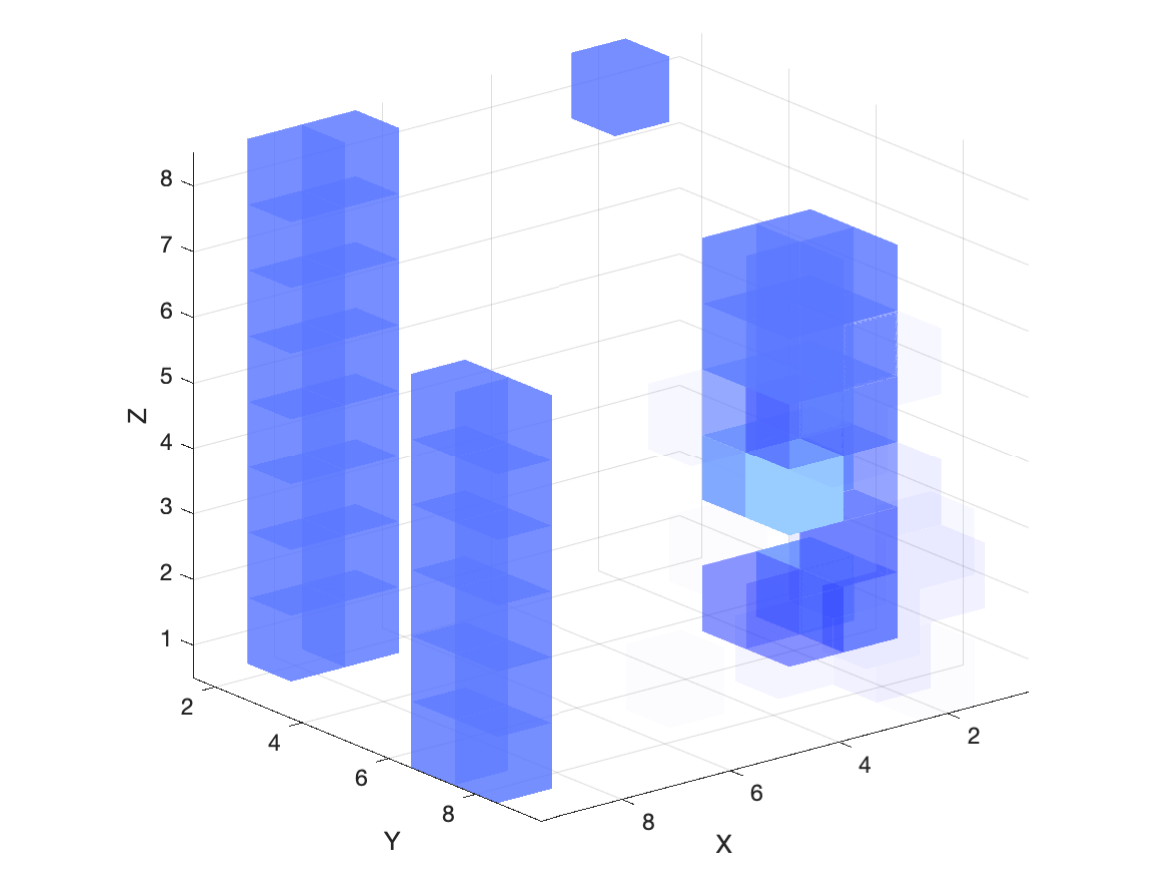}  
		 {(g) Multi-view OMP}
	\end{minipage}
	\begin{minipage}[b]{0.24\linewidth}
	\centering
	\includegraphics[width=\linewidth]{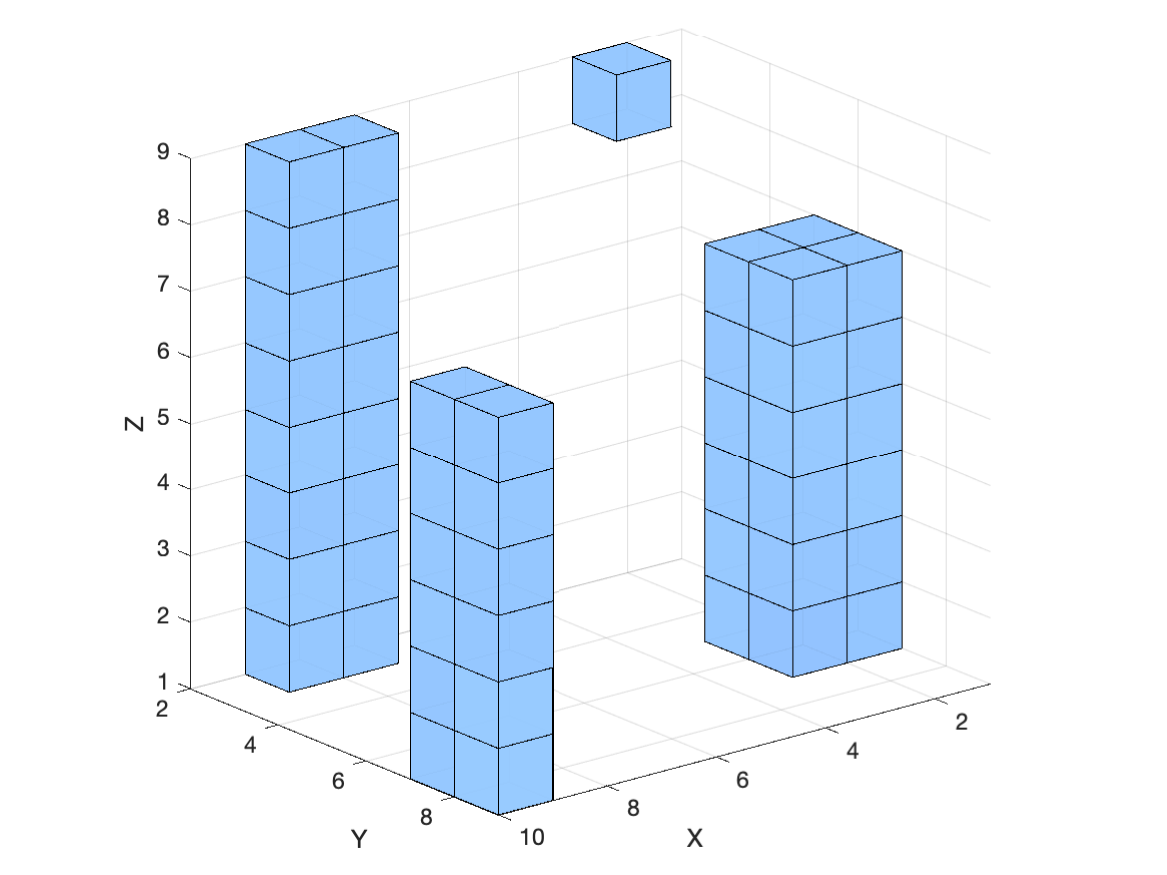}  
	 {(b) Multi-view SBL }
\end{minipage}
\begin{minipage}[b]{0.24\linewidth}
	\centering
	\includegraphics[width=\linewidth]{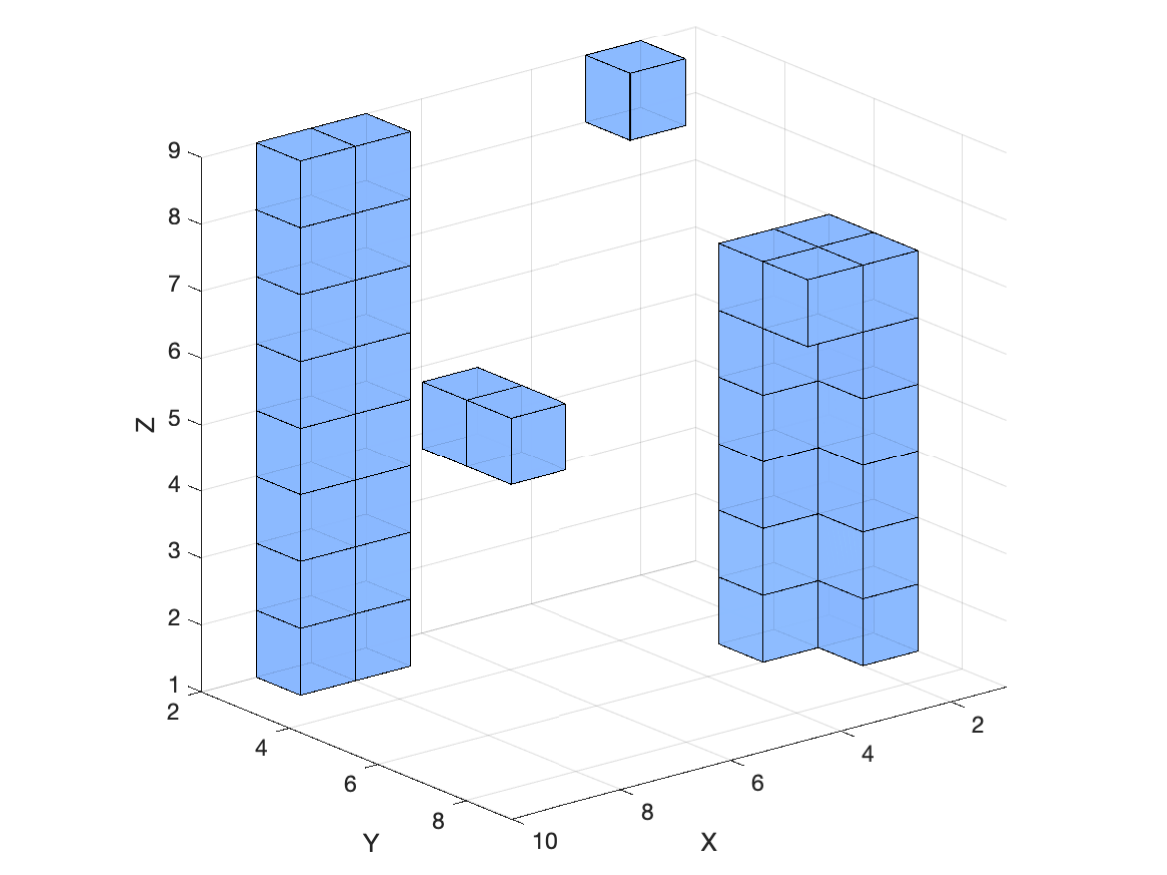}
	\centerline{(d)   Single-view SBL}
\end{minipage}
	\begin{minipage}[b]{0.24\linewidth}
	\centering
	\includegraphics[width=\linewidth]{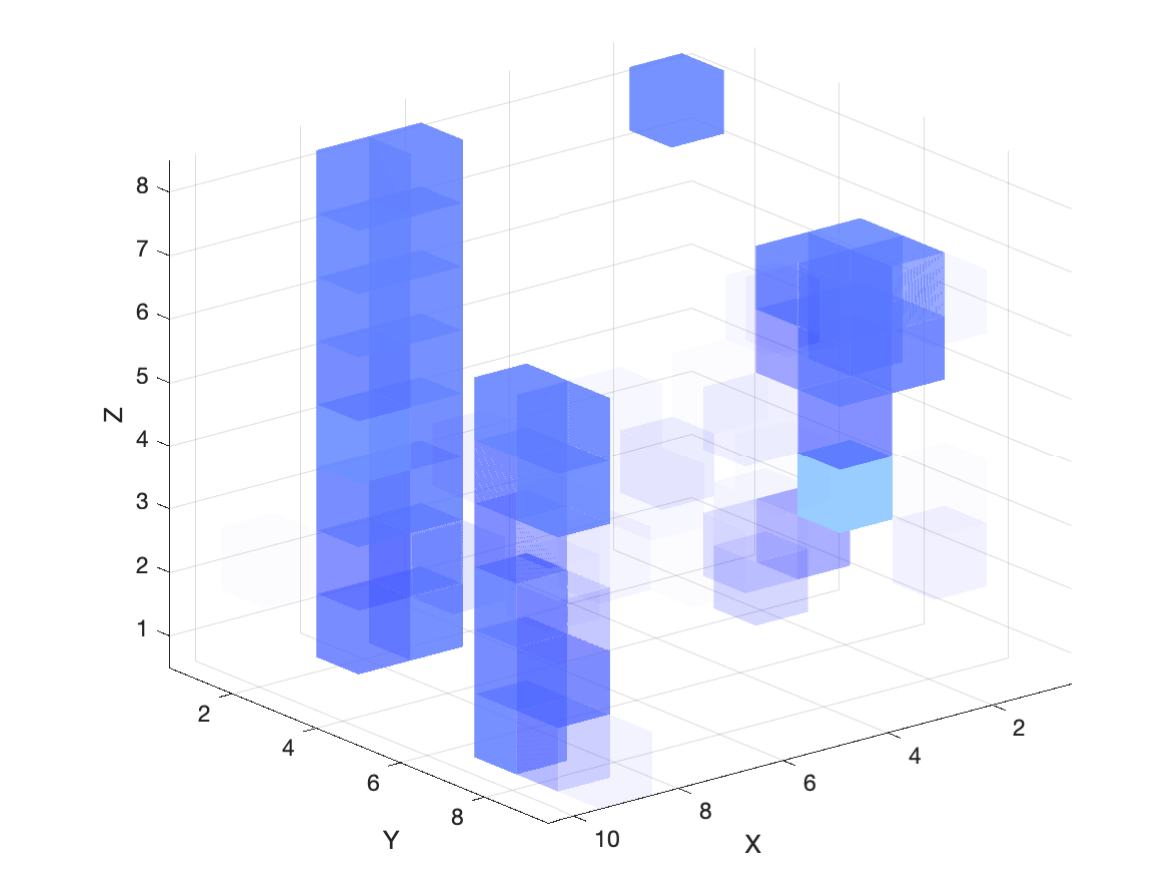}  
	 {(f) Multi-view ISTA}
\end{minipage}
\begin{minipage}[b]{0.24\linewidth}
	\centering
	\includegraphics[width=\linewidth]{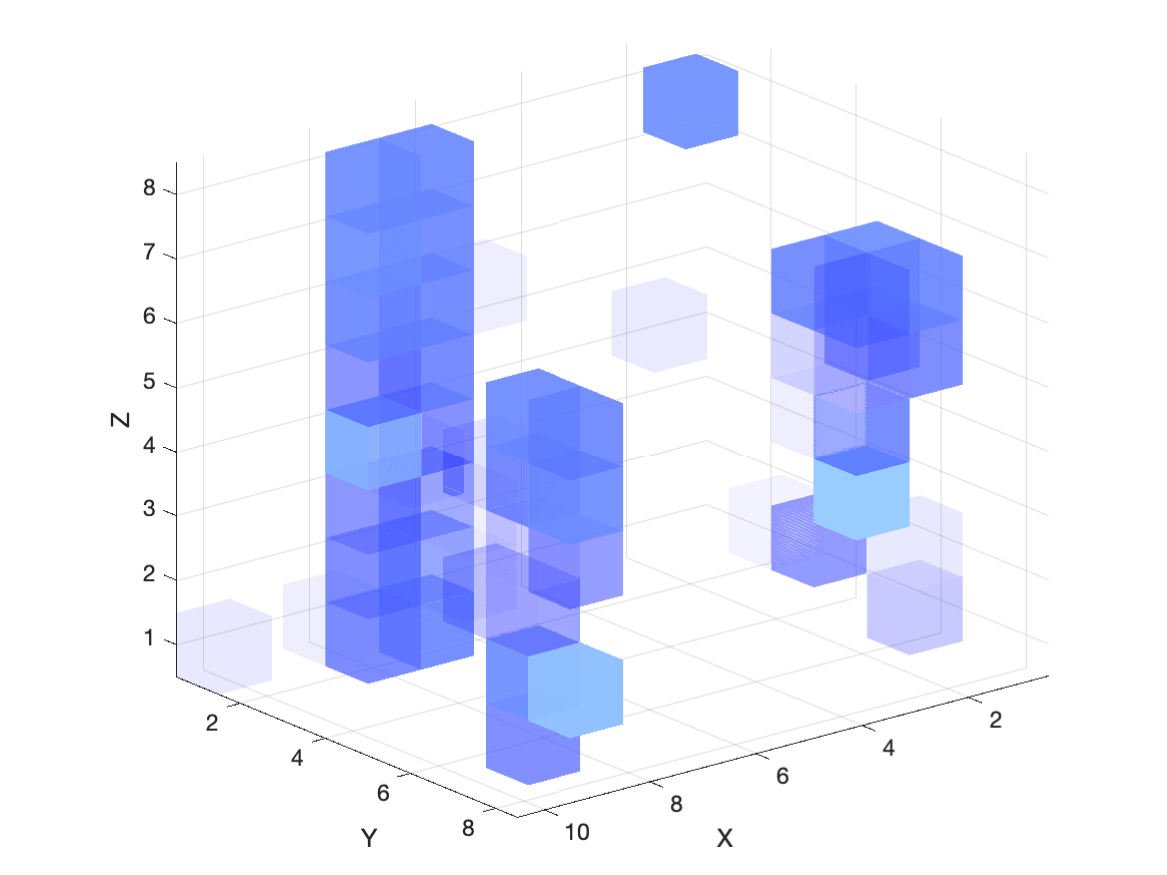}
	\centerline{(h) Multi-view LASSO}
\end{minipage}
	\vspace{2mm}
	\caption{Imaging results under different algorithms. (a) 3-D True Image. (b) Multi-view SBL. (c) Limited views from RU $1$. (d) SBL imaging based on a single view from RU $1$. (e) Multi-view least-squares (LS). (f) Multi-view iterative shrinkage-thresholding algorithm (ISTA) with a regularization parameter of $0.0002$.    (g) Multi-view orthogonal matching pursuit (OMP) algorithm. (h)  Multi-view least absolute shrinkage and selection operator (LASSO) algorithm with a regularization parameter of $0.0002$.   }
	\label{figo2}
\end{figure*}

In the following, for clarity, we simulate the cases where the images are fully correlated among different subcarriers (i.e., $\rhov_1 = \rhov_2 = \ldots =\rhov_N$) due to the small bandwidth.

In Fig. \ref{figo2}, we compare the proposed SBL-based multi-view imaging algorithm with some benchmarks. The considered 3D ground truth environment is shown in Fig. \ref{figo2} (a). It can be seen from (b) that with the multi-view information in distributed MIMO networks, the proposed SBL-based algorithm can reconstruct the 3D environment accurately. By contrast, in (c), we consider the imaging based on a single view from RU $1$. The gray voxel indicates that it is not visible to most of the antennas on RU $ 1 $. In this case, we assume that the RU $1$ keeps acting as the receiver for all the $S=4$ time slots while the other three RUs continue to serve as transmitters with different transmitting signals. From (d), it can be seen that due to the obstacle, single-view imaging can not effectively reconstruct the overall environment, which demonstrates the motivation of the proposed multi-view imaging algorithms. Then, in (e), (f), (g), and (h), we compare the proposed algorithms with four benchmark algorithms using the same multi-view setup as (a). It can be seen that the least-square algorithm (e)  estimates the required voxels, but at the same time produces severe artifacts. For the iterative shrinkage-thresholding algorithm (ISTA) and the least absolute shrinkage and selection operator (LASSO), they cannot reconstruct the complete image, although they have relatively fewer artifacts. The OMP algorithm cannot accurately estimate the most complicated part of the images. Besides the unsatisfactory performance, OMP algorithms require the knowledge of the number of voxels, and ISTA/LASSO require a sophisticated selection of regularization parameters. These observations demonstrate the effectiveness and robustness of the proposed SBL-based algorithm.

\begin{figure}[t]
	\centering
	\includegraphics[width=\linewidth]{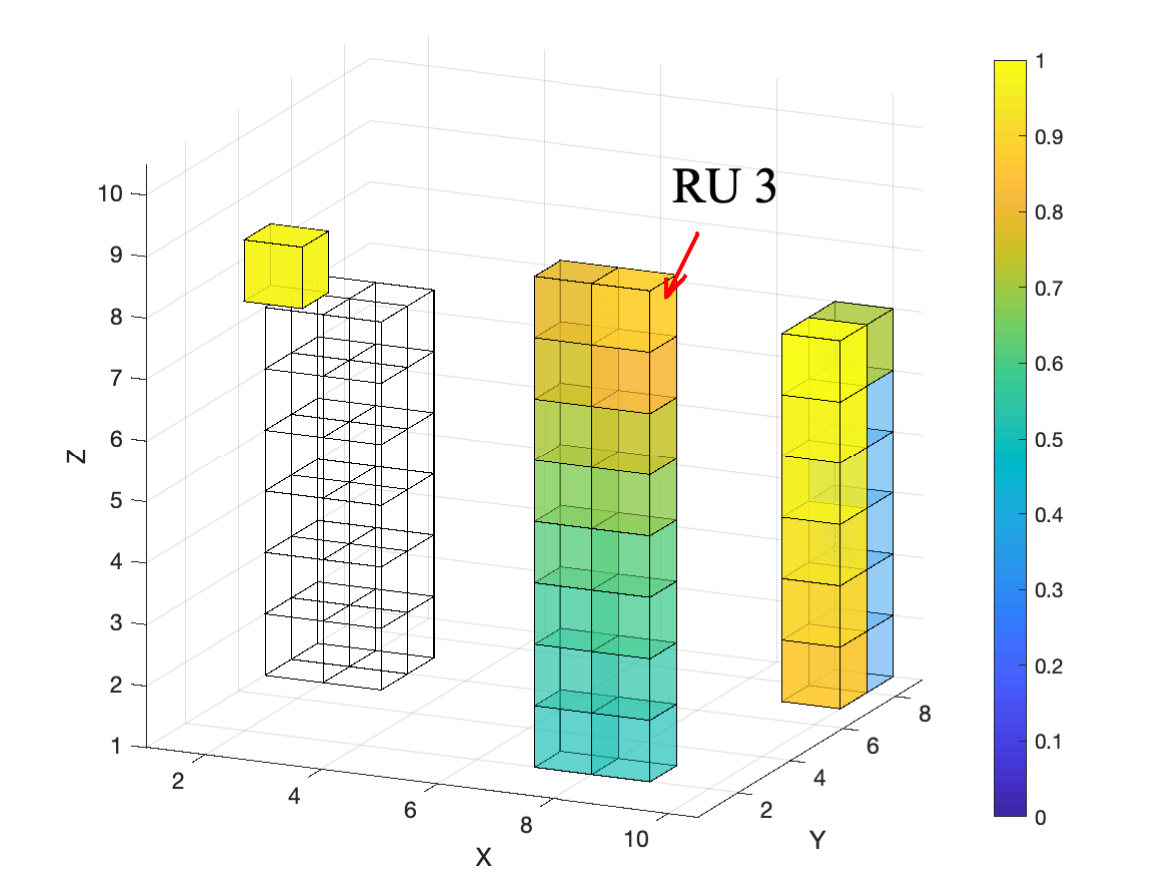}
	\caption{Projected channel strength  $\sqrt{\cos\theta_{\mathbf{p}}^{\ell_t,m_t} \cos\phi_{\mathbf{p}}^{\ell_t,m_t} }  $ defined in (\ref{ht}) for the central antenna of RU $3$.}
	\label{figo7}
\end{figure}

\begin{figure}[t]
	\centering
\includegraphics[width=\linewidth]{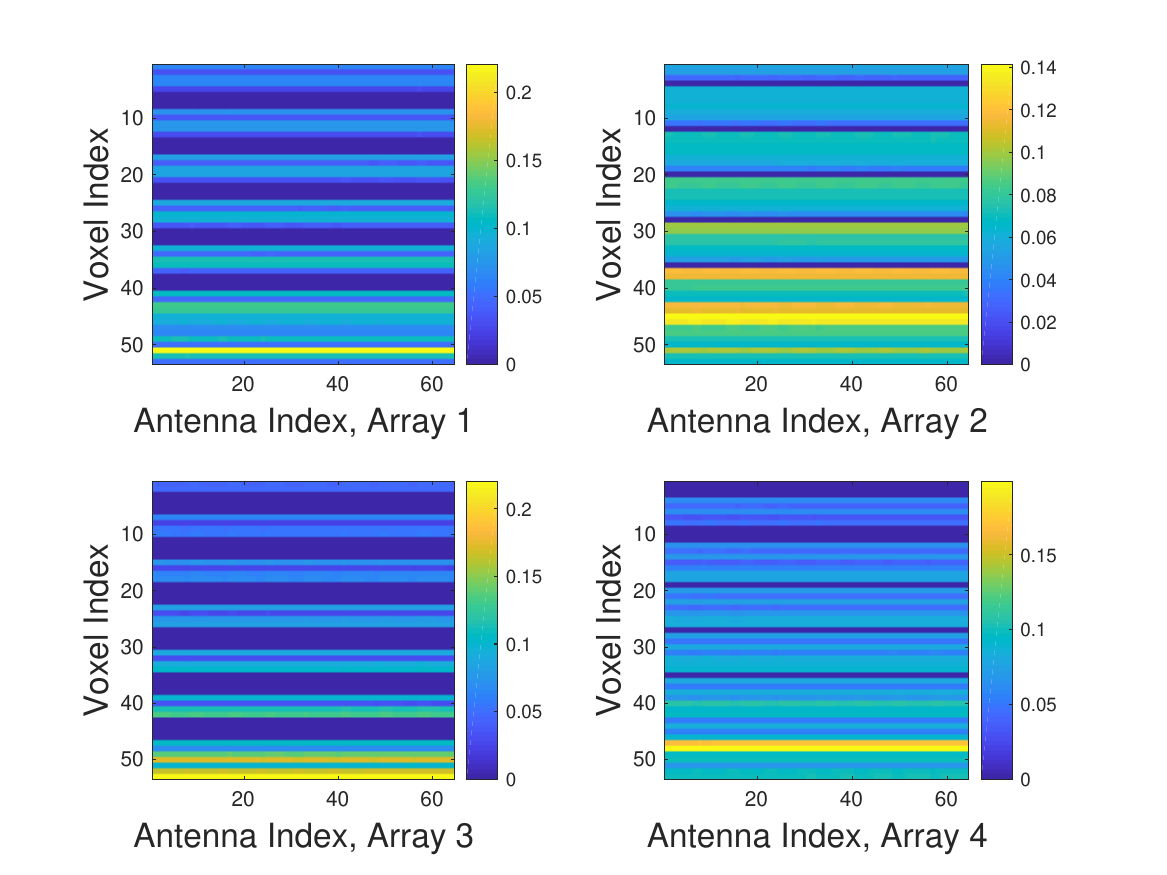}
	\caption{Non-isotropic channel strengths for antennas and arrays.}
\label{figo3}
\end{figure}

In Figs. \ref{figo7} and \ref{figo3}, we illustrate the effect of visibility and non-isotropic views in the considered channel modeling. In Fig. \ref{figo7}, we visualize the impact of non-isotropic projection angle on the channel strength, by plotting the non-stationary strength of $\sqrt{\cos\theta_{\mathbf{p}}^{\ell_t,m_t} \cos\phi_{\mathbf{p}}^{\ell_t,m_t} }  $ from the central antenna of RU $3$ to each voxel. It can be seen that for xovels with lower heights that own large signal incident angles, the power losses caused by signal projection are severe. Besides, for voxels nearly obstructed by other voxels, due to the narrow view, the projection loss is also larger, and the channel is weaker. If the voxels are fully blocked, the corresponding channel will have no power. This effect can be further validated in Fig. \ref{figo3}, by observing the channel strength of four RUs. It can be seen that due to the obstacle, all RUs have blind spots, especially for RU $1$ and $3$. Nevertheless, by fusing the views from all RUs, a complete view of the imaging area can be obtained, which ensures the accurate reconstruction of the environment. 

\begin{figure}[t]
	\centering
	\includegraphics[width=\linewidth]{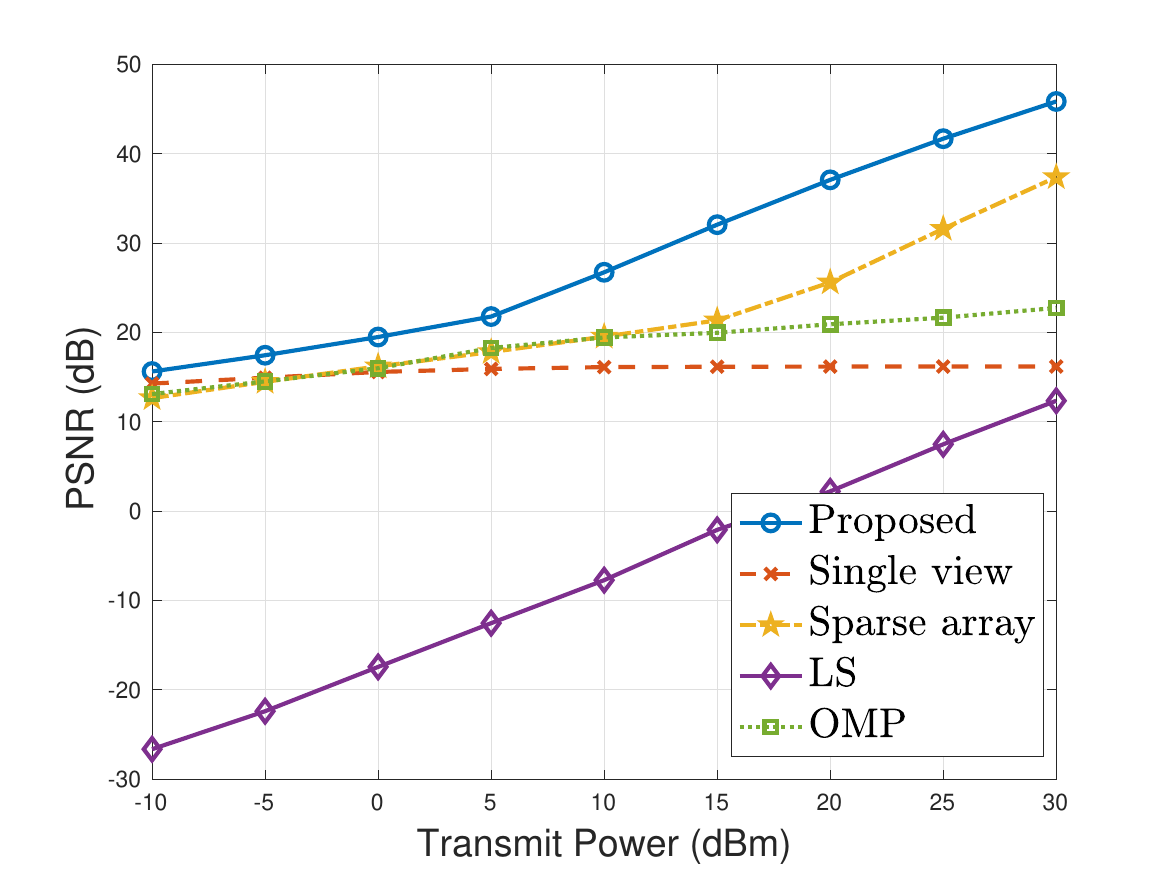}
	\caption{PSNR performance versus transmit power}
	\label{figo4}
\end{figure}

\begin{figure}[t]
	\centering
	\includegraphics[width=\linewidth]{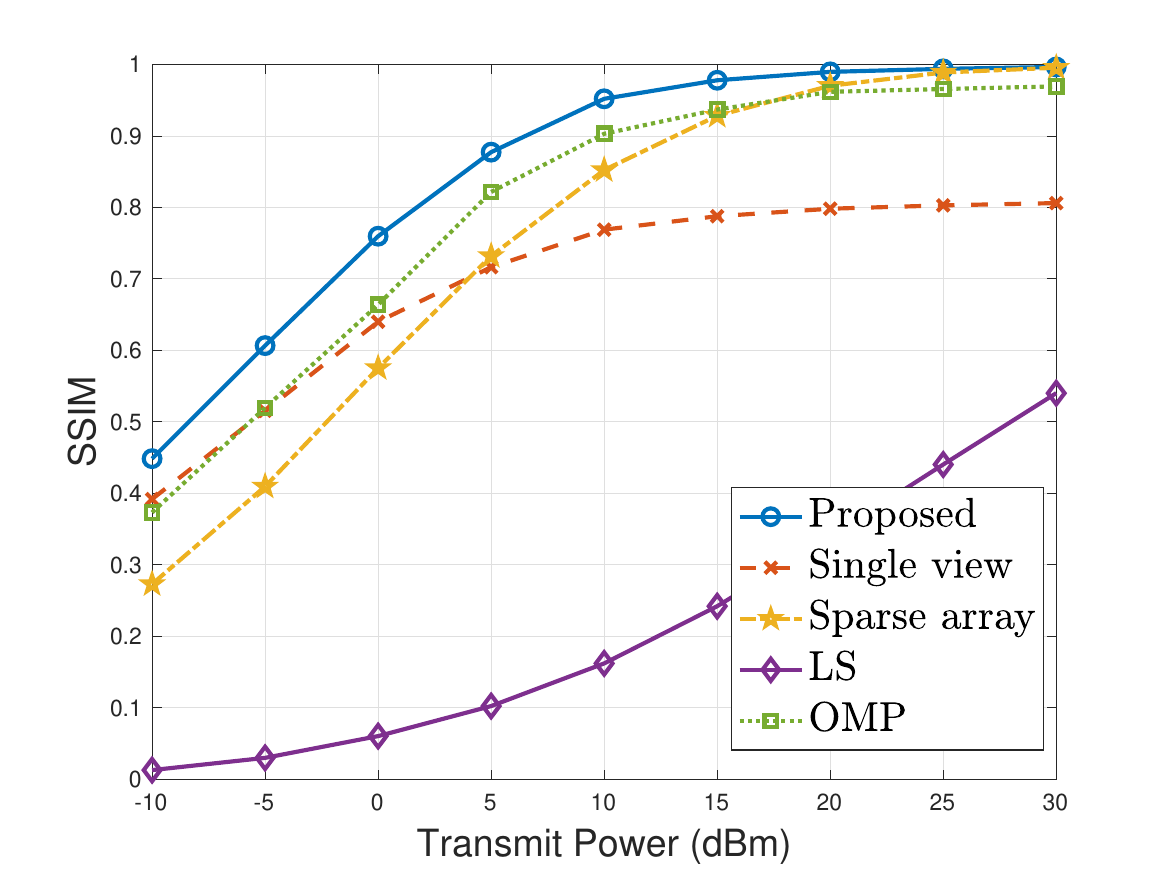}
	\caption{SSIM performance versus transmit power}
	\label{figo5}
\end{figure}

Finally, in Figs. \ref{figo4} and \ref{figo5}, we investigate the quantitative performance of the imaging tasks in terms of PSNR and SSIM, following the same scenario of Fig. \ref{figo2} (a). It can be seen that the proposed multi-view imaging algorithm realizes a superior imaging quality than all the other benchmarks. The proposed scheme ($M=64$) achieves a high PSNR and achieves the optimal SSIM at the high SNR regime. Besides, it can be seen that when using sparse arrays ($M=16$, $\xi=2$), satisfactory estimation quality can also be achieved in the high SNR regime, which demonstrates the feasibility of using sparse XL-MIMO with reduced cost. 
By contrast, the performance of the single-view imaging scheme saturates at a moderate value as SNR increases, due to the unavoidable loss caused by limited views.  
Obvious advantages of the proposed scheme compared to benchmark algorithms of LS and OMP can also be observed. Also, the sparse array scheme using the proposed algorithm outperforms the OMP-based scheme, showcasing the promising potential of the proposed algorithm.

\section{Summary}\label{section_summary}
This paper investigates the radio imaging based on distributed MIMO communication arrays, taking into consideration of non-isotropic signal observations and the obstruction of visibility from different views. We first propose an RMA-based algorithm based on three kinds of array architectures, i.e., the full array, the boundary array, and the distributed-boundary array. Then, we propose an SBL-based algorithm by discretizing the 3D space into voxels and exploiting the sparsity. Simulation demonstrates the effectiveness of the two proposed algorithms in acquiring high-quality images in indoor and outdoor environments.

\bibliographystyle{IEEEtran}
\bibliography{reference}

\begin{thebibliography}{10}
\providecommand{\url}[1]{#1}
\csname url@samestyle\endcsname
\providecommand{\newblock}{\relax}
\providecommand{\bibinfo}[2]{#2}
\providecommand{\BIBentrySTDinterwordspacing}{\spaceskip=0pt\relax}
\providecommand{\BIBentryALTinterwordstretchfactor}{4}
\providecommand{\BIBentryALTinterwordspacing}{\spaceskip=\fontdimen2\font plus
\BIBentryALTinterwordstretchfactor\fontdimen3\font minus
  \fontdimen4\font\relax}
\providecommand{\BIBforeignlanguage}[2]{{%
\expandafter\ifx\csname l@#1\endcsname\relax
\typeout{** WARNING: IEEEtran.bst: No hyphenation pattern has been}%
\typeout{** loaded for the language `#1'. Using the pattern for}%
\typeout{** the default language instead.}%
\else
\language=\csname l@#1\endcsname
\fi
#2}}
\providecommand{\BIBdecl}{\relax}
\BIBdecl

\bibitem{wp5d2023m}
I.~WP5D, ``M. 2160: framework and overall objectives of the future development
  of {IMT} for {2030} and beyond,'' \emph{ITU Radiocommunication Sector
  (ITU-R), ITU-R recommendations}, 2023.

\bibitem{FengLAE2025}
Y.~Feng, C.~Zhao, H.~Luo, F.~Gao, F.~Liu, and S.~Jin, ``Networked {ISAC} based
  {UAV} tracking and handover towards low-altitude economy,'' \emph{IEEE
  Transactions on Wireless Communications}, pp. 1--1, 2025.

\bibitem{TangLAE2025}
J.~Tang, Y.~Yu, C.~Pan, H.~Ren, D.~Wang, J.~Wang, and X.~You, ``Cooperative
  {ISAC}-empowered low-altitude economy,'' \emph{IEEE Transactions on Wireless
  Communications}, vol.~24, no.~5, pp. 3837--3853, 2025.

\bibitem{lyu2025empowering}
Z.~Lyu, Y.~Gao, J.~Chen, H.~Du, J.~Xu, K.~Huang, and D.~I. Kim, ``Empowering
  intelligent low-altitude economy with large {AI} model deployment,''
  \emph{arXiv preprint arXiv:2505.22343}, 2025.

\bibitem{lyu2022joint}
Z.~Lyu, G.~Zhu, and J.~Xu, ``Joint maneuver and beamforming design for
  {UAV}-enabled integrated sensing and communication,'' \emph{IEEE Trans.
  Wireless Commun.}, vol.~22, no.~4, pp. 2424--2440, 2022.

\bibitem{xiong2023fundamental}
Y.~Xiong, F.~Liu, Y.~Cui, W.~Yuan, T.~X. Han, and G.~Caire, ``On the
  fundamental tradeoff of integrated sensing and communications under gaussian
  channels,'' \emph{IEEE Transactions on Information Theory}, vol.~69, no.~9,
  pp. 5723--5751, 2023.

\bibitem{hua20243d}
M.~Hua, G.~Chen, K.~Meng, S.~Ma, C.~Yuen, and H.~C. So, ``{3D} multi-target
  localization via intelligent reflecting surface: Protocol and analysis,''
  \emph{IEEE Transactions on Wireless Communications}, 2024.

\bibitem{10769538}
K.~Meng, C.~Masouros, A.~P. Petropulu, and L.~Hanzo, ``Cooperative {ISAC}
  networks: Performance analysis, scaling laws, and optimization,'' \emph{IEEE
  Transactions on Wireless Communications}, vol.~24, no.~2, pp. 877--892, 2025.

\bibitem{yu2023active}
Z.~Yu, H.~Ren, C.~Pan, G.~Zhou, B.~Wang, M.~Dong, and J.~Wang, ``Active
  {RIS}-aided {ISAC} systems: Beamforming design and performance analysis,''
  \emph{IEEE Transactions on Communications}, vol.~72, no.~3, pp. 1578--1595,
  2023.

\bibitem{Meng2024SG}
K.~Meng, C.~Masouros, G.~Chen, and F.~Liu, ``Network-level integrated sensing
  and communication: Interference management and {BS} coordination using
  stochastic geometry,'' \emph{IEEE Transactions on Wireless Communications},
  vol.~23, no.~12, pp. 19\,365--19\,381, 2024.

\bibitem{songperformance}
Y.~Song, K.~Zhi, T.~Yang, S.~Li, P.~Ciblat, and G.~Caire, ``Performance
  analysis of network sensing in the distributed {MIMO} radar system,'' in
  \emph{ICC 2022 - IEEE International Conference on Communications}, 2025.

\bibitem{yang2025cooperative}
T.~Yang, S.~Li, Y.~Song, K.~Zhi, and G.~Caire, ``Cooperative multistatic target
  detection in cell-free communication networks,'' in \emph{2025 IEEE Wireless
  Communications and Networking Conference (WCNC)}.\hskip 1em plus 0.5em minus
  0.4em\relax IEEE, 2025, pp. 1--6.

\bibitem{babu2024precoding}
N.~Babu, C.~Masouros, C.~B. Papadias, and Y.~C. Eldar, ``Precoding for
  multi-cell {ISAC}: from coordinated beamforming to coordinated multipoint and
  bi-static sensing,'' \emph{IEEE Transactions on Wireless Communications},
  2024.

\bibitem{10726912}
K.~Meng, C.~Masouros, A.~P. Petropulu, and L.~Hanzo, ``Cooperative {ISAC}
  networks: Opportunities and challenges,'' \emph{IEEE Wireless
  Communications}, vol.~32, no.~3, pp. 212--219, 2025.

\bibitem{wang2025wideband}
X.~Wang, W.~Zhai, X.~Wang, M.~Amin, and A.~Zoubir, ``Wideband near-field
  integrated sensing and communications: A hybrid precoding perspective,''
  \emph{IEEE Signal Processing Magazine}, vol.~42, no.~1, pp. 88--105, 2025.

\bibitem{10791452}
J.~Wu, W.~Yuan, Z.~Wei, K.~Zhang, F.~Liu, and D.~Wing Kwan~Ng, ``Low-complexity
  minimum {BER} precoder design for {ISAC} systems: A delay-doppler
  perspective,'' \emph{IEEE Transactions on Wireless Communications}, vol.~24,
  no.~2, pp. 1526--1540, 2025.

\bibitem{tao2025survey}
Q.~Tao, Z.~Li, K.~Zhi, S.~Li, W.~Yuan, L.~Zaniboni, S.~Stanczak, E.~Viterbo,
  and X.~Wang, ``A survey on reconfigurable intelligent surface-assisted
  orthogonal time frequency space systems,'' \emph{IEEE Open Journal of
  Vehicular Technology}, 2025.

\bibitem{wu2024exploit}
T.~Wu, C.~Pan, K.~Zhi, H.~Ren, M.~Elkashlan, C.-X. Wang, R.~Schober, and
  X.~You, ``Exploit high-dimensional {RIS} information to localization: What is
  the impact of faulty element?'' \emph{IEEE Journal on Selected Areas in
  Communications}, 2024.

\bibitem{pan2022overview}
C.~Pan, G.~Zhou, K.~Zhi, S.~Hong, T.~Wu, Y.~Pan, H.~Ren, M.~Di~Renzo, A.~L.
  Swindlehurst, R.~Zhang \emph{et~al.}, ``An overview of signal processing
  techniques for {RIS}/{IRS}-aided wireless systems,'' \emph{IEEE Journal of
  Selected Topics in Signal Processing}, vol.~16, no.~5, pp. 883--917, 2022.

\bibitem{wu2024employing}
T.~Wu, C.~Pan, K.~Zhi, H.~Ren, M.~Elkashlan, J.~Wang, and C.~Yuen, ``Employing
  high-dimensional {RIS} information for {RIS}-aided localization systems,''
  \emph{IEEE Communications Letters}, 2024.

\bibitem{zhou2024fluid}
L.~Zhou, J.~Yao, M.~Jin, T.~Wu, and K.-K. Wong, ``Fluid antenna-assisted {ISAC}
  systems,'' \emph{IEEE Wireless Communications Letters}, 2024.

\bibitem{chen2025multi}
G.~Chen, Q.~Wu, S.~Lu, M.~Hua, and W.~Chen, ``Multi-{IRS} aided {ISAC} system:
  Multi-path exploitation versus reduction,'' \emph{arXiv preprint
  arXiv:2506.21968}, 2025.

\bibitem{hua2023secure}
M.~Hua, Q.~Wu, W.~Chen, O.~A. Dobre, and A.~L. Swindlehurst, ``Secure
  intelligent reflecting surface-aided integrated sensing and communication,''
  \emph{IEEE Transactions on Wireless Communications}, vol.~23, no.~1, pp.
  575--591, 2023.

\bibitem{yang2025towards}
S.~Yang, J.~Yao, J.~Tang, T.~Wu, M.~Elkashlan, C.~Yuen, M.~Debbah, H.~Shin, and
  M.~Valenti, ``Towards intelligent antenna positioning: Leveraging {DRL} for
  {FAS}-aided {ISAC} systems,'' \emph{IEEE Internet of Things Journal}, 2025.

\bibitem{10596930}
S.~Lu, F.~Liu, F.~Dong, Y.~Xiong, J.~Xu, Y.-F. Liu, and S.~Jin, ``Random {ISAC}
  signals deserve dedicated precoding,'' \emph{IEEE Transactions on Signal
  Processing}, vol.~72, pp. 3453--3469, 2024.

\bibitem{liu2024joint}
M.~Liu, H.~Ren, C.~Pan, B.~Wang, Z.~Yu, R.~Weng, K.~Zhi, and Y.~He, ``Joint
  beamforming design for double active {RIS}-assisted radar-communication
  coexistence systems,'' \emph{IEEE Transactions on Cognitive Communications
  and Networking}, 2024.

\bibitem{hua2024integrated}
M.~Hua, Q.~Wu, W.~Chen, A.~Jamalipour, C.~Wu, and O.~A. Dobre, ``Integrated
  sensing and communication: Joint pilot and transmission design,'' \emph{IEEE
  Transactions on Wireless Communications}, 2024.

\bibitem{manzoni2024wavefield}
M.~Manzoni, D.~Tagliaferri, S.~Tebaldini, M.~Mizmizi, A.~V. Monti-Guarnieri,
  C.~M. Prati, and U.~Spagnolini, ``Wavefield networked sensing: Principles,
  algorithms and applications,'' \emph{IEEE Open Journal of the Communications
  Society}, 2024.

\bibitem{li2021lightweight}
X.~Li and Y.~Chen, ``Lightweight 2d imaging for integrated imaging and
  communication applications,'' \emph{IEEE Signal Processing Letters}, vol.~28,
  pp. 528--532, 2021.

\bibitem{yang2025illumination}
Q.~Yang, H.~Zhang, C.~Li, R.~Liu, and B.~Wang, ``Illumination design for near
  field joint imaging and wireless power transfer systems,'' \emph{IEEE
  Internet of Things Journal}, 2025.

\bibitem{huang2024fourier}
Y.~Huang, J.~Yang, W.~Tang, C.-K. Wen, and S.~Jin, ``Fourier transform-based
  wavenumber domain {3D} imaging in {RIS}-aided communication systems,''
  \emph{IEEE Transactions on Wireless Communications}, 2024.

\bibitem{huang2024ris}
Y.~Huang, J.~Yang, C.-K. Wen, and S.~Jin, ``{RIS}-aided single-frequency {3D}
  imaging by exploiting multi-view image correlations,'' \emph{IEEE
  Transactions on Communications}, 2024.

\bibitem{li2024networked}
J.~Li, X.~Shao, F.~Chen, S.~Wan, C.~Liu, Z.~Wei, and D.~W.~K. Ng, ``Networked
  integrated sensing and communications for {6G} wireless systems,'' \emph{IEEE
  Internet of Things Journal}, 2024.

\bibitem{zheng2024random}
B.~Zheng and F.~Liu, ``Random signal design for joint communication and {SAR}
  imaging towards low-altitude economy,'' \emph{IEEE Wireless Communications
  Letters}, 2024.

\bibitem{jiang2024electromagnetic}
Y.~Jiang, F.~Gao, S.~Jin, and T.~J. Cui, ``Electromagnetic property sensing
  based on diffusion model in {ISAC} system,'' \emph{IEEE Transactions on
  Wireless Communications}, 2024.

\bibitem{tong2022environment}
X.~Tong, Z.~Zhang, Y.~Zhang, Z.~Yang, C.~Huang, K.-K. Wong, and M.~Debbah,
  ``Environment sensing considering the occlusion effect: A multi-view
  approach,'' \emph{IEEE Transactions on Signal Processing}, vol.~70, pp.
  3598--3615, 2022.

\bibitem{lu2024deep}
B.~Lu, Z.~Wei, H.~Wu, X.~Zeng, L.~Wang, X.~Lu, D.~Mei, and Z.~Feng, ``Deep
  learning based multi-node {ISAC} {4D} environmental reconstruction with
  uplink-downlink cooperation,'' \emph{IEEE Internet of Things Journal}, 2024.

\bibitem{tong2025computational}
X.~Tong, Z.~Zhang, Z.~Yang, Y.~Ge, and H.~Wymeersch, ``Computational
  imaging-based {ISAC} method with large pixel division,'' \emph{arXiv preprint
  arXiv:2505.07355}, 2025.

\bibitem{torcolacci2024holographic}
G.~Torcolacci, A.~Guerra, H.~Zhang, F.~Guidi, Q.~Yang, Y.~C. Eldar, and
  D.~Dardari, ``Holographic imaging with {XL}-{MIMO} and {RIS}: Illumination
  and reflection design,'' \emph{IEEE Journal of Selected Topics in Signal
  Processing}, 2024.

\bibitem{ahmed2012advanced}
S.~S. Ahmed, A.~Schiessl, F.~Gumbmann, M.~Tiebout, S.~Methfessel, and L.-P.
  Schmidt, ``Advanced microwave imaging,'' \emph{IEEE microwave magazine},
  vol.~13, no.~6, pp. 26--43, 2012.

\bibitem{shao2020advances}
W.~Shao and T.~McCollough, ``Advances in microwave near-field imaging:
  Prototypes, systems, and applications,'' \emph{IEEE microwave magazine},
  vol.~21, no.~5, pp. 94--119, 2020.

\bibitem{wang2019review}
Z.~Wang, T.~Chang, and H.-L. Cui, ``Review of active millimeter wave imaging
  techniques for personnel security screening,'' \emph{IEEE Access}, vol.~7,
  pp. 148\,336--148\,350, 2019.

\bibitem{wu1987diffraction}
R.-S. Wu and M.~N. Toks{\"o}z, ``Diffraction tomography and multisource
  holography applied to seismic imaging,'' \emph{Geophysics}, vol.~52, no.~1,
  pp. 11--25, 1987.

\bibitem{bolomey1990microwave}
J.-C. Bolomey and C.~Pichot, ``Microwave tomography: from theory to practical
  imaging systems,'' \emph{International Journal of Imaging Systems and
  Technology}, vol.~2, no.~2, pp. 144--156, 1990.

\bibitem{ren20183}
K.~Ren, J.~Chen, and R.~J. Burkholder, ``A 3-{D} uniform diffraction
  tomographic algorithm for near-field microwave imaging through stratified
  media,'' \emph{IEEE Transactions on Antennas and Propagation}, vol.~66,
  no.~6, pp. 3034--3045, 2018.

\bibitem{moreira2013tutorial}
A.~Moreira, P.~Prats-Iraola, M.~Younis, G.~Krieger, I.~Hajnsek, and K.~P.
  Papathanassiou, ``A tutorial on synthetic aperture radar,'' \emph{IEEE
  Geoscience and remote sensing magazine}, vol.~1, no.~1, pp. 6--43, 2013.

\bibitem{krieger2013mimo}
G.~Krieger, ``{MIMO}-{SAR}: Opportunities and pitfalls,'' \emph{IEEE
  transactions on geoscience and remote sensing}, vol.~52, no.~5, pp.
  2628--2645, 2013.

\bibitem{fang2013fast}
J.~Fang, Z.~Xu, B.~Zhang, W.~Hong, and Y.~Wu, ``Fast compressed sensing {SAR}
  imaging based on approximated observation,'' \emph{IEEE Journal of Selected
  Topics in Applied Earth Observations and Remote Sensing}, vol.~7, no.~1, pp.
  352--363, 2013.

\bibitem{xu2011bayesian}
G.~Xu, M.~Xing, L.~Zhang, Y.~Liu, and Y.~Li, ``Bayesian inverse synthetic
  aperture radar imaging,'' \emph{IEEE Geoscience and Remote Sensing Letters},
  vol.~8, no.~6, pp. 1150--1154, 2011.

\bibitem{vehmas2021inverse}
R.~Vehmas and N.~Neuberger, ``Inverse synthetic aperture radar imaging: A
  historical perspective and state-of-the-art survey,'' \emph{IEEE access},
  vol.~9, pp. 113\,917--113\,943, 2021.

\bibitem{yanik2020development}
M.~E. Yanik, D.~Wang, and M.~Torlak, ``Development and demonstration of
  mimo-sar mmwave imaging testbeds,'' \emph{IEEE Access}, vol.~8, pp.
  126\,019--126\,038, 2020.

\bibitem{soumekh1998wide}
M.~Soumekh, ``Wide-bandwidth continuous-wave monostatic/bistatic synthetic
  aperture radar imaging,'' in \emph{Proceedings 1998 International Conference
  on Image Processing. ICIP98 (Cat. No. 98CB36269)}.\hskip 1em plus 0.5em minus
  0.4em\relax IEEE, 1998, pp. 361--365.

\bibitem{grebner2023probabilistic}
T.~Grebner, A.~Grathwohl, P.~Schoeder, V.~Janoudi, and C.~Waldschmidt,
  ``Probabilistic sar processing for high-resolution mapping using
  millimeter-wave radar sensors,'' \emph{IEEE Transactions on Aerospace and
  Electronic Systems}, vol.~59, no.~5, pp. 4800--4814, 2023.

\bibitem{desai1992convolution}
M.~D. Desai and W.~K. Jenkins, ``Convolution backprojection image
  reconstruction for spotlight mode synthetic aperture radar,'' \emph{IEEE
  Transactions on Image Processing}, vol.~1, no.~4, pp. 505--517, 1992.

\bibitem{ren2019fast}
K.~Ren, Q.~Wang, and R.~J. Burkholder, ``A fast back-projection approach to
  diffraction tomography for near-field microwave imaging,'' \emph{IEEE
  Antennas and Wireless Propagation Letters}, vol.~18, no.~10, pp. 2170--2174,
  2019.

\bibitem{ge2024efficient}
S.~Ge, S.~Song, D.~Feng, J.~Wang, L.~Chen, J.~Zhu, and X.~Huang, ``Efficient
  near-field millimeter-wave sparse imaging technique utilizing one-bit
  measurements,'' \emph{IEEE Transactions on Microwave Theory and Techniques},
  2024.

\bibitem{sheen2001three}
D.~M. Sheen, D.~L. McMakin, and T.~E. Hall, ``Three-dimensional millimeter-wave
  imaging for concealed weapon detection,'' \emph{IEEE Transactions on
  microwave theory and techniques}, vol.~49, no.~9, pp. 1581--1592, 2001.

\bibitem{lopez20003}
J.~M. Lopez-Sanchez and J.~Fortuny-Guasch, ``3-d radar imaging using range
  migration techniques,'' \emph{IEEE Transactions on antennas and propagation},
  vol.~48, no.~5, pp. 728--737, 2000.

\bibitem{zhu2016frequency}
R.~Zhu, J.~Zhou, L.~Tang, Y.~Kan, and Q.~Fu, ``Frequency-domain imaging
  algorithm for single-input--multiple-output array,'' \emph{IEEE Geoscience
  and Remote Sensing Letters}, vol.~13, no.~12, pp. 1747--1751, 2016.

\bibitem{moulder2016development}
W.~F. Moulder, J.~D. Krieger, J.~J. Majewski, C.~M. Coldwell, H.~T. Nguyen,
  D.~T. Maurais-Galejs, T.~L. Anderson, P.~Dufilie, and J.~S. Herd,
  ``Development of a high-throughput microwave imaging system for concealed
  weapons detection,'' in \emph{2016 IEEE International Symposium on Phased
  Array Systems and Technology (PAST)}.\hskip 1em plus 0.5em minus 0.4em\relax
  IEEE, 2016, pp. 1--6.

\bibitem{gao2018novel}
J.~Gao, Y.~Qin, B.~Deng, H.~Wang, and X.~Li, ``Novel efficient 3d short-range
  imaging algorithms for a scanning 1d-mimo array,'' \emph{IEEE Transactions on
  Image Processing}, vol.~27, no.~7, pp. 3631--3643, 2018.

\bibitem{zhu2019sequential}
R.~Zhu, J.~Zhou, B.~Cheng, Q.~Fu, and G.~Jiang, ``Sequential frequency-domain
  imaging algorithm for near-field mimo-sar with arbitrary scanning paths,''
  \emph{IEEE Journal of Selected Topics in Applied Earth Observations and
  Remote Sensing}, vol.~12, no.~8, pp. 2967--2975, 2019.

\bibitem{zhuge2012three}
X.~Zhuge and A.~G. Yarovoy, ``Three-dimensional near-field mimo array imaging
  using range migration techniques,'' \emph{IEEE Transactions on Image
  Processing}, vol.~21, no.~6, pp. 3026--3033, 2012.

\bibitem{fromenteze2019transverse}
T.~Fromenteze, O.~Yurduseven, F.~Berland, C.~Decroze, D.~R. Smith, and A.~G.
  Yarovoy, ``A transverse spectrum deconvolution technique for mimo short-range
  fourier imaging,'' \emph{IEEE Transactions on Geoscience and Remote Sensing},
  vol.~57, no.~9, pp. 6311--6324, 2019.

\bibitem{yanik2019near}
M.~E. Yanik and M.~Torlak, ``Near-field mimo-sar millimeter-wave imaging with
  sparsely sampled aperture data,'' \emph{Ieee Access}, vol.~7, pp.
  31\,801--31\,819, 2019.

\bibitem{wang20203}
J.~Wang, P.~Aubry, and A.~Yarovoy, ``3-d short-range imaging with irregular
  mimo arrays using nufft-based range migration algorithm,'' \emph{IEEE
  Transactions on Geoscience and Remote Sensing}, vol.~58, no.~7, pp.
  4730--4742, 2020.

\bibitem{zhang2022fast}
W.~Zhang, Y.~Ji, W.~Shao, B.~Lin, C.~Li, and G.~Fang, ``A fast 3-d chirp
  scaling imaging technique for millimeter-wave near-field imaging,''
  \emph{IEEE Transactions on Microwave Theory and Techniques}, vol.~71, no.~2,
  pp. 827--841, 2022.

\bibitem{hu2018beyond}
S.~Hu, F.~Rusek, and O.~Edfors, ``Beyond massive {MIMO}: The potential of data
  transmission with large intelligent surfaces,'' \emph{IEEE Transactions on
  Signal Processing}, vol.~66, no.~10, pp. 2746--2758, 2018.

\bibitem{zhang2015generalized}
W.~Zhang and A.~Hoorfar, ``A generalized approach for sar and mimo radar
  imaging of building interior targets with compressive sensing,'' \emph{IEEE
  Antennas and Wireless Propagation Letters}, vol.~14, pp. 1052--1055, 2015.

\bibitem{wang2017fast}
L.-G. Wang, L.~Li, J.~Ding, and T.~J. Cui, ``A fast patches-based imaging
  algorithm for 3-d multistatic imaging,'' \emph{IEEE geoscience and remote
  sensing letters}, vol.~14, no.~6, pp. 941--945, 2017.

\bibitem{wang2022efficient}
M.~Wang, S.~Wei, Z.~Zhou, J.~Shi, and X.~Zhang, ``Efficient admm framework
  based on functional measurement model for mmw {3-D} {SAR} imaging,''
  \emph{IEEE Transactions on Geoscience and Remote Sensing}, vol.~60, pp.
  1--17, 2022.

\bibitem{gurbuz2009compressive}
A.~C. Gurbuz, J.~H. McClellan, and W.~R. Scott, ``A compressive sensing data
  acquisition and imaging method for stepped frequency gprs,'' \emph{IEEE
  Transactions on Signal Processing}, vol.~57, no.~7, pp. 2640--2650, 2009.

\bibitem{li2018compressive}
S.~Li, G.~Zhao, H.~Sun, and M.~Amin, ``Compressive sensing imaging of 3-{D}
  object by a holographic algorithm,'' \emph{IEEE Transactions on Antennas and
  Propagation}, vol.~66, no.~12, pp. 7295--7304, 2018.

\bibitem{ma2014mimo}
C.~Ma, T.~S. Yeo, Y.~Zhao, and J.~Feng, ``Mimo radar {3D} imaging based on
  combined amplitude and total variation cost function with sequential order
  one negative exponential form,'' \emph{IEEE Transactions on Image
  Processing}, vol.~23, no.~5, pp. 2168--2183, 2014.

\bibitem{bi2017l_}
H.~Bi, B.~Zhang, X.~X. Zhu, W.~Hong, J.~Sun, and Y.~Wu, ``{$
  L_1$}-regularization-based {SAR} imaging and {CFAR} detection via complex
  approximated message passing,'' \emph{IEEE Transactions on Geoscience and
  Remote Sensing}, vol.~55, no.~6, pp. 3426--3440, 2017.

\bibitem{yang2013sparse}
J.~Yang, T.~Jin, X.~Huang, J.~Thompson, and Z.~Zhou, ``Sparse mimo array
  forward-looking gpr imaging based on compressed sensing in clutter
  environment,'' \emph{IEEE Transactions on Geoscience and Remote Sensing},
  vol.~52, no.~7, pp. 4480--4494, 2013.

\bibitem{li2025compressive}
J.~Li, D.~Bi, X.~Li, L.~Peng, and Y.~Xie, ``A compressive super-resolution
  imaging algorithm for multi-frequency near-field millimeter-wave,''
  \emph{IEEE Transactions on Instrumentation and Measurement}, 2025.

\bibitem{chen2023adaptive}
X.~Chen, Q.~Yang, H.~Wang, Y.~Zeng, and B.~Deng, ``Adaptive {ADMM}-based
  high-quality fast imaging algorithm for short-range {MMW} {MIMO}-{SAR}
  systems,'' \emph{IEEE Transactions on Antennas and Propagation}, vol.~71,
  no.~11, pp. 8925--8935, 2023.

\bibitem{ongie2020deep}
G.~Ongie, A.~Jalal, C.~A. Metzler, R.~G. Baraniuk, A.~G. Dimakis, and
  R.~Willett, ``Deep learning techniques for inverse problems in imaging,''
  \emph{IEEE Journal on Selected Areas in Information Theory}, vol.~1, no.~1,
  pp. 39--56, 2020.

\bibitem{xiong2020spb}
K.~Xiong, G.~Zhao, Y.~Wang, and G.~Shi, ``Spb-net: A deep network for sar
  imaging and despeckling with downsampled data,'' \emph{IEEE Transactions on
  Geoscience and Remote Sensing}, vol.~59, no.~11, pp. 9238--9256, 2020.

\bibitem{shi20256g}
Q.~Shi, S.~Zhang, and L.~Liu, ``A {6G}-based multi-view reconstruction
  approach,'' in \emph{2025 IEEE Wireless Communications and Networking
  Conference (WCNC)}.\hskip 1em plus 0.5em minus 0.4em\relax IEEE, 2025, pp.
  1--6.

\bibitem{huang2024image}
N.~Huang, C.~Dou, Y.~Wu, L.~Qian, S.~Zhou, and R.~Lu, ``Image analysis oriented
  integrated sensing and communication via intelligent reflecting surface,''
  \emph{IEEE Transactions on Cognitive Communications and Networking}, 2024.

\bibitem{tahira2024irs}
S.~Tahira, T.~Fujihashi, T.~Takahashi, S.~Saruwatari, and T.~Watanabe,
  ``{IRS}-aided over-the-air image processing: Single antenna imaging,'' in
  \emph{2024 IEEE 35th International Symposium on Personal, Indoor and Mobile
  Radio Communications (PIMRC)}.\hskip 1em plus 0.5em minus 0.4em\relax IEEE,
  2024, pp. 1--7.

\bibitem{zhi2024performance}
K.~Zhi, C.~Pan, H.~Ren, K.~K. Chai, C.-X. Wang, R.~Schober, and X.~You,
  ``Performance analysis and low-complexity design for {XL}-{MIMO} with
  near-field spatial non-stationarities,'' \emph{IEEE Journal on Selected Areas
  in Communications}, 2024.

\bibitem{bjornson2020power}
E.~Bj{\"o}rnson and L.~Sanguinetti, ``Power scaling laws and near-field
  behaviors of massive {MIMO} and intelligent reflecting surfaces,'' \emph{IEEE
  Open Journal of the Communications Society}, vol.~1, pp. 1306--1324, 2020.

\bibitem{wang2019beam}
B.~Wang, M.~Jian, F.~Gao, G.~Y. Li, and H.~Lin, ``Beam squint and channel
  estimation for wideband mmwave massive {MIMO}-{OFDM} systems,'' \emph{IEEE
  transactions on signal processing}, vol.~67, no.~23, pp. 5893--5908, 2019.

\bibitem{cui2022channel}
M.~Cui and L.~Dai, ``Channel estimation for extremely large-scale {MIMO}:
  Far-field or near-field?'' \emph{IEEE transactions on communications},
  vol.~70, no.~4, pp. 2663--2677, 2022.

\bibitem{wu2024fluid}
T.~Wu, K.~Zhi, J.~Yao, X.~Lai, J.~Zheng, H.~Niu, M.~Elkashlan, K.-K. Wong,
  C.-B. Chae, Z.~Ding \emph{et~al.}, ``Fluid antenna systems enabling {6G}:
  Principles, applications, and research directions,'' \emph{arXiv preprint
  arXiv:2412.03839}, 2024.

\bibitem{dong2024movable}
Z.~Dong, Z.~Zhou, Z.~Xiao, C.~Zhang, X.~Li, H.~Min, Y.~Zeng, S.~Jin, and
  R.~Zhang, ``Movable antenna for wireless communications: Prototyping and
  experimental results,'' \emph{arXiv preprint arXiv:2408.08588}, 2024.

\bibitem{yao2025framework}
J.~Yao, X.~Lai, K.~Zhi, T.~Wu, M.~Jin, C.~Pan, M.~Elkashlan, C.~Yuen, and K.-K.
  Wong, ``A framework of {FAS}-{RIS} systems: Performance analysis and
  throughput optimization,'' \emph{IEEE Transactions on Wireless
  Communications}, 2025.

\bibitem{zhu2023movable}
L.~Zhu, W.~Ma, and R.~Zhang, ``Movable antennas for wireless communication:
  Opportunities and challenges,'' \emph{IEEE Communications Magazine}, vol.~62,
  no.~6, pp. 114--120, 2023.

\bibitem{papoulis1968systems}
A.~Papoulis, ``Systems and transforms with applications in optics,''
  \emph{McGraw-Hill Series in System Science}, 1968.

\bibitem{chew1999waves}
W.~C. Chew, \emph{Waves and fields in inhomogenous media}.\hskip 1em plus 0.5em
  minus 0.4em\relax John Wiley \& Sons, 1999, vol.~16.

\bibitem{lu2024tutorial}
H.~Lu, Y.~Zeng, C.~You, Y.~Han, J.~Zhang, Z.~Wang, Z.~Dong, S.~Jin, C.-X. Wang,
  T.~Jiang \emph{et~al.}, ``A tutorial on near-field {XL-MIMO} communications
  toward {6G},'' \emph{IEEE Communications Surveys \& Tutorials}, vol.~26,
  no.~4, pp. 2213--2257, 2024.

\bibitem{wang2017associations}
J.~Wang, N.~Zheng, B.~Chen, and J.~C. Principe, ``Associations among image
  assessments as cost functions in linear decomposition: {MSE}, {SSIM}, and
  correlation coefficient,'' \emph{arXiv preprint arXiv:1708.01541}, 2017.

\bibitem{zhang2020amp}
Z.~Zhang, Y.~Liu, J.~Liu, F.~Wen, and C.~Zhu, ``{AMP}-{Net}: Denoising-based
  deep unfolding for compressive image sensing,'' \emph{IEEE Transactions on
  Image Processing}, vol.~30, pp. 1487--1500, 2020.

\bibitem{zhang2013extension}
Z.~Zhang and B.~D. Rao, ``Extension of {SBL} algorithms for the recovery of
  block sparse signals with intra-block correlation,'' \emph{IEEE Trans. Signal
  Process.}, vol.~61, no.~8, pp. 2009--2015, 2013.

\end{thebibliography}
\end{document}